\documentclass[nobibstyle,review]{twcccreport}
\reportnumber{2024-03}
\usepackage{amsmath}
\usepackage{amssymb}
\usepackage{hyperref}
\usepackage{graphics,graphicx}
\newcommand{\email}[1]{\protect\href{mailto:#1}{#1}}
\usepackage{tikz}
\usepackage[shortlabels]{enumitem}

\usepackage{mathtools}
\usepackage{upgreek}

\usepackage[authoryear,round]{natbib}
\usepackage[caption=false, font=footnotesize]{subfig}

\usepackage{ifthen}
\newboolean{LongVersion}
\setboolean{LongVersion}{true} 
\newboolean{OneColumn}
\setboolean{OneColumn}{true} 

\newcommand{\pubidadjcol}[1]{#1}

\RequirePackage[amsthm,amsmath,thmmarks,hyperref]{ntheorem}
\usepackage[capitalize,nameinlink]{cleveref}

\theoremstyle{definition}
\newtheorem{theorem}{Theorem}
\newtheorem{assumption}{Assumption}
\crefname{assumption}{Assumption}{Assumptions}
\newtheorem{proposition}{Proposition}

\newtheorem{definition}{Definition}
\newtheorem{remark}{Remark}
\newtheorem{example}{Example}

\usepackage{amsfonts}
\usepackage{amsopn}
\usepackage{braket}
\usepackage{comment}

\newcommand{\QED}{$\square$}

\newcommand{\defas}{:=}
\newcommand{\asdef}{=:}

\newcommand{\lev}{\textnormal{lev}}
\newcommand{\id}{\textsc{id}}

\newcommand{\real}{\mathbb{R}}
\newcommand{\posreal}{\mathbb{R}_{>0}}
\newcommand{\nnegreal}{\mathbb{R}_{\geq 0}}

\newcommand{\nnegereal}{\overline{\mathbb{R}}_{\geq 0}}
\newcommand{\realm}[2]{\mathbb{R}^{#1\times #2}}

\newcommand{\allint}{\mathbb{I}}
\newcommand{\nnegint}{\mathbb{I}_{\geq 0}}
\newcommand{\posint}{\mathbb{I}_{>0}}
\newcommand{\intinterval}[2]{\mathbb{I}_{#1:#2}}

\newcommand{\calPD}{\mathcal{PD}}
\newcommand{\calK}{\mathcal{K}}
\newcommand{\calKinf}{{\mathcal{K}}_{\infty}}

\newcommand{\calKL}{\mathcal{KL}}

\usetikzlibrary{matrix} 
\usetikzlibrary{shapes,arrows} 
\usetikzlibrary{calc,patterns,angles,quotes}
\usetikzlibrary{fit}

\tikzstyle{block} = [draw,rectangle,thick,minimum height=2em,minimum width=2em,
align=center]
\tikzstyle{branch} = [circle,inner sep=0pt,minimum
size=0mm,fill=black,draw=black]
\tikzstyle{input} = [circle,inner sep=0pt,minimum size=1mmPackage,draw=black]
\tikzstyle{arrow} = [->,thick]
\tikzstyle{line} = [thick]
\tikzstyle{circ} = [draw,circle,thick,inner sep=0pt,minimum height=1em,
minimum width=1em,align=center]

\begin{document}

\title{Beyond inherent robustness: strong stability of MPC despite plant-model
  mismatch%
  \thanks{This report is an extended version of a submitted paper. This work was
    supported by the National Science Foundation (NSF) under Grant 2138985.
    (e-mail: \email{skuntz@ucsb.edu}; \email{jbraw@ucsb.edu})}
  \thanks{Version 2 includes additional discussions, self-contained proofs of
    the main Lyapunov results, and a revision of the main differentiability
    assumption. The main technical results remain unchanged.}%
}
\author{Steven J.~Kuntz and James B.~Rawlings \\
  Department of Chemical Engineering \\
  University of California, Santa Barbara}

\maketitle

\begin{abstract}
  In this \ifthenelse{\boolean{LongVersion}}{technical report}{article}, we
  establish the asymptotic stability of MPC under plant-model mismatch for
  problems where the origin remains a steady state despite mismatch. This class
  of problems includes, but is not limited to, inventory management,
  path-planning, and control of systems in deviation variables. Our results
  differ from prior results on the inherent robustness of MPC, which guarantee
  only convergence to a neighborhood of the origin, the size of which scales
  with the magnitude of the mismatch. For MPC with quadratic costs, continuous
  differentiability of the system dynamics is sufficient to demonstrate
  exponential stability of the closed-loop system despite mismatch. For MPC with
  general costs, a joint comparison function bound and scaling condition
  guarantee asymptotic stability despite mismatch. The results are illustrated
  in numerical simulations, including the classic upright pendulum problem. The
  tools developed to establish these results can address the stability of
  offset-free MPC, an open and interesting question in the MPC research
  literature.
\end{abstract}

\section{Introduction}\label{sec:intro}
\ifthenelse{\boolean{OneColumn}}{%
  Plant-model %
}{%
  \IEEEPARstart{P}{lant-model} %
}%
mismatch is an ever-present challenge in model predictive control (MPC)
practice. In industrial implementations, the main driver of MPC performance is
model quality~\citep{qin:badgwell:2003,darby:nikolaou:2012}. There has been
recent progress on improving model quality and MPC performance through
disturbance modeling and estimator
tuning~\citep{kuntz:rawlings:2022,
  kuntz:rawlings:2024a,
  simpson:asprion:muntwiler:kohler:diehl:2024}, 
simultaneous state and parameter
estimation~\citep{baumgartner:reiter:diehl:2022,%
  muntwiler:kohler:zeilinger:2023a,%
  schiller:muller:2023}, %
and data-driven MPC design and analysis~\citep{%
  yin:iannelli:smith:2023,
  dorfler:coulson:markovsky:2022,
  berberich:kohler:muller:allgower:2021,
  berberich:kohler:muller:allgower:2022
} %
to name a few methods. However, there is not yet a sharp theoretical
understanding of the robustness of MPC to plant-model mismatch.

Before discussing MPC robustness, let us first define \emph{robustness}. In the
stability literature, \emph{robust asymptotic stability} has been used to refer
to both (i) input-to-state stability (ISS)~\citep{jiang:wang:2001} and (ii)
asymptotic stability despite disturbances~\citep{kellett:teel:2005}. To avoid
confusion, we reserve the term \emph{robust asymptotic stability} for (i) and
use \emph{strong asymptotic stability} to refer to (ii).\footnote{The latter
  term is borrowed from the differential inclusion
  literature~\citep{clarke:ledyaev:stern:1998}%
  \ifthenelse{\boolean{LongVersion}}{ %
    (see \cite{jiang:wang:2002,kellett:teel:2005} for discrete-time
    definitions). %
  }{.} %
  Some authors~\citep{jiang:wang:2001,jiang:wang:2002} use the term
  \emph{uniform asymptotic stability} to refer to (ii), but we wish to avoid
  confusion with the time-varying case.} %
When such properties are given by a nominal MPC,%
\footnote{\emph{Nominal} MPC refers to MPC designed without a disturbance
  model\ifthenelse{\boolean{LongVersion}}{, possibly admitting parameter errors.
    This includes not only standard nonlinear MPC, but also suboptimal,
    offset-free, and (some) data-driven MPC.}{.}} %
we call it \emph{inherently robust} or \emph{inherently strongly stabilizing}.
Robust and strong exponential stability are defined similarly.

It is well-known that MPC is stabilizing under certain assumptions on the
terminal ingredients (cf.~\cite[Ch.~2]{rawlings:mayne:diehl:2020}). To achieve
robust stability in the presence of parameter errors, estimation errors, and
exogenous perturbations, a disturbance model can be included%
\ifthenelse{\boolean{LongVersion}}{.}{ %
  (cf.~\cite[Ch.~1,~3]{rawlings:mayne:diehl:2020}). %
}\ifthenelse{\boolean{LongVersion}}{%
  The simplest manner of handling disturbances is with feedback. For MPC this
  would require future knowledge of the disturbance trajectory, or at least a
  forecast of it, to implement the controller. While this is a strong
  requirement, it would confer strong stability rather than robust stability.
  Alternatively, a disturbance model may be included. Several MPC variants
  include disturbance models in their design, such as
  offset-free~\citep{pannocchia:gabiccini:artoni:2015},
  stochastic~\citep{mcallister:2022}, tube-based
  (cf.~\cite[Ch.~3]{rawlings:mayne:diehl:2020}), and min-max
  MPC~\citep{limon:alamo:salas:camacho:2006}. For a survey of these methods,
  see~\cite[Ch.~1,~3]{rawlings:mayne:diehl:2020}. %

}{}%
Even in the absence of a disturbance model, a wide range of nominal MPC designs
are inherently robust to disturbances.
Continuity of the control law was first proven to be sufficient for inherent
robustness~\citep{denicolao:magni:scattolini:1996,%
  scokaert:rawlings:meadows:1997}. Later, \cite{grimm:messina:tuna:teel:2004}
proved continuity of the optimal value function is sufficient for inherent
robustness, and stated MPC examples with discontinuous optimal value functions
that are nominally stable but otherwise not robust to disturbances.
A special class of time-varying terminal constraints were proven to confer
robust stability to nominal MPC by \cite{grimm:messina:tuna:teel:2007}, and to
suboptimal MPC by \cite{lazar:heemels:2009}.
In \cite{pannocchia:rawlings:wright:2011,allan:bates:risbeck:rawlings:2017}, the
inherent robustness of optimal and suboptimal MPC, using a class of
time-invariant terminal constraints, was proven.
The inherent stochastic robustness (in probability, expectation, and
distribution) of nominal MPC was shown
by~\cite{mcallister:rawlings:2022f,mcallister:rawlings:2022d,mcallister:rawlings:2023d}.
Finally, direct data-driven MPC was shown to be inherently robust to noisy
data~\citep{berberich:kohler:muller:allgower:2021}.

\pubidadjcol{}

If the origin remains a steady state under mismatch, we might expect strong
asymptotic stability. While this assumption may seem strong, it includes a wide
class of problems, including inventory management, path-planning, and control of
systems that can be recast in deviation variables. %
\ifthenelse{\boolean{LongVersion}}{%
  In unconstrained linear optimal control problems (LQR/LQG), the margin of
  stability (maximum perturbation to the open-loop gain that still gives a
  closed-loop system) is always nonzero. However, it is important to note that
  there is no guaranteed relative value of this margin below which the closed
  loop is stable, save a few exceptional cases such as a single input, or with
  diagonally-weighted stage
  costs~\citep{doyle:1978,lehtomaki:sandell:athans:1981,zhang:fu:1996}. Examples
  are shown by \cite{doyle:1978,zhang:fu:1996} in which arbitrarily small
  perturbations to the gain matrix destabilize the system. These examples use
  \emph{multiplicative disturbances} that, while persistent in the
  aforementioned papers, do not need to be time-invariant for the results to
  hold. The disturbances treated in the MPC literature are typically
  \emph{additive disturbances} entering the states and measurements
  (cf.~\cite[Ch.~3]{rawlings:mayne:diehl:2020}). In the multiplicative case,
  borrowing from knowledge of linear systems, we should expect strong
  exponential stability. However, in the additive case, we should expect only
  robust exponential stability. %
}{%
  For linear systems, unconstrained optimal control stabilizes the origin
  despite bounded perturbations to the system
  gain~\citep{doyle:1978,lehtomaki:sandell:athans:1981,zhang:fu:1996}. In the
  nonlinear setting, we might expect similar behavior under such disturbances. %
}%
To the best of our knowledge, the inherent strong stability of nominal MPC to
plant-model mismatch has been discussed by
only~\cite{santos:biegler:1999,santos:biegler:castro:2008}. For unconstrained
systems with a sufficiently small bound on the mismatch, nominal MPC is shown to
stabilize the plant to the origin. While exact penalty functions are considered
for handling constraints, there is no guarantee of recursive feasibility.

In this \ifthenelse{\boolean{LongVersion}}{report}{article}, we extend the work
of~\cite{santos:biegler:castro:2008} to include input constraints and
stabilizing terminal constraints. %
We show in \Cref{cor:mpc:mismatch:exp} that MPC with quadratic costs achieves
strong exponential stability given (i) a fixed steady state, (ii) a mild
differentiability condition, and (iii) standard stabilizing terminal ingredients
(cf.~\cite{allan:bates:risbeck:rawlings:2017}). For MPC with general, positive
definite cost functions, a fixed steady state, and stabilizing terminal
ingredients, we show a \emph{joint \(\calK\)-function} bound holds on the
increase in the optimal value function (\Cref{prop:lyap}), but strong stability
is implied only if this bound decays sufficiently quickly near the origin
(\Cref{cor:mpc:mismatch}). A counterexample (\Cref{ssec:example:sqrt}) shows
this property does \emph{not} hold in general.

\ifthenelse{\boolean{LongVersion}}{%
  To illustrate the main results, we consider
  \ifthenelse{\boolean{LongVersion}}{three}{two} numerical examples. The first
  example is a continuous yet nondifferentiable system with a general cost MPC
  that is robust yet not strongly stable, demonstrating inherent strong
  stability is not a guaranteed property of nonlinear MPC.\@ %
  \ifthenelse{\boolean{LongVersion}}{%
    The second example is a nondifferentiable system for which the quadratic
    cost MPC is strongly stabilizing. %
  }{}%
  In the \ifthenelse{\boolean{LongVersion}}{third and final}{second} example, we
  use the upright pendulum problem to showcase several types of plant-model
  mismatch that are covered by the main results, namely, discretization errors,
  unmodeled dynamics, and errors in estimated parameters. %
}{}

The theory in this \ifthenelse{\boolean{LongVersion}}{report}{article} can be
extended to address the open problem of offset-free MPC
stability~\cite{kuntz:rawlings:2024f}. In offset-free MPC, an integrating
disturbance model is used to effectively estimate the steady states as a
function of the disturbances. This guarantees (in the absence of estimation
errors) the steady state is uniform in the parameters, and strong stability can
be established (for quadratic costs and differentiable plants).

For brevity, complete proofs of
\Cref{thm:lyap:robust,thm:lyap:strong,thm:mpc:robust,thm:mpc:robust:exp}, an
additional nondifferentiable example, and additional remarks throughout are
deferred to an extended technical report~\cite{kuntz:rawlings:2024d}.

\ifthenelse{\boolean{LongVersion}}{%
  We outline the \ifthenelse{\boolean{LongVersion}}{report}{article} as follows.
  In \Cref{sec:problem}, we state the MPC with plant-model mismatch problem,
  review nominal MPC stability, and present a motivating example exhibiting both
  robust and strong closed-loop stability. In \cref{sec:robust:strong}, we
  define and prove results on robust and strong stability. In
  \cref{sec:mpc:robust}, we review inherent robustness of MPC.\@ In
  \Cref{sec:mpc:strong}, we present the main results. In \Cref{sec:examples}, we
  present numerical examples. In \Cref{sec:conclusion}, we discuss extensions
  and future work, including the extension of our theory to the offset-free MPC
  stability problem (cf.~\cite{kuntz:rawlings:2024f}). %
}{}%

\paragraph*{Notation}
\ifthenelse{\boolean{LongVersion}}{%
  Let \(\real\), \(\nnegreal\), and \(\posreal\) denote the real, nonnegative
  real, and positive real numbers, respectively. Let \(\allint\), \(\nnegint\),
  \(\posint\), and \(\intinterval{m}{n}\) denote the integers, nonnegative
  integers, positive integers, and integers from \(m\) to \(n\) (inclusive),
  respectively. Let \(\real^n\) and \(\realm{n}{m}\) denote real \(n\)-vectors
  and \(n\times m\) matrices, respectively.
}{}%
Let 
\(\nnegereal\defas\nnegreal\cup\set{\infty}\) denote the extended nonnegative
reals.
For any function \(V:\real^n\rightarrow\nnegereal\) and finite \(\rho\geq 0\),
we define the sublevel set \(\textnormal{lev}_\rho V\defas\set{x\in\real^n |
  V(x)\leq\rho}\). We say \(V:\real^n\rightarrow\nnegereal\) is lower
semicontinuous (l.s.c.) if \(\textnormal{lev}_\rho V\) is closed for each
\(\rho\geq 0\).
We say a symmetric matrix \(P=P^\top\in\realm{n}{n}\) is positive definite if
\(x^\top Px>0\) for all \(x\in\real^n\setminus\set{0}\).
We define the Euclidean and \(Q\)-weighted norms by \(|x|\defas\sqrt{x^\top x}\)
and \(|x|_Q\defas\sqrt{x^\top Qx}\) for each \(x\in\real^n\), where \(Q\) is
positive definite. Moreover, \(|\cdot|_Q\) has the property
\(\underline{\sigma}(Q)|x|^2 \leq |x|_Q^2 \leq \overline{\sigma}(Q)|x|^2\) for
all \(x\in\real^n\), where \(\underline{\sigma}(Q)\) and
\(\overline{\sigma}(Q)\) denote the smallest and largest singular values of
\(Q\).
For any signal \(a(k)\), we denote both infinite and finite sequences in bold
font as \(\mathbf{a} \defas (a(0),\ldots,a(k))\) and \(\mathbf{a} \defas
(a(0),a(1),\ldots)\). We define the infinite and length-\(k\) signal norm as
\(\|\mathbf{a}\| \defas \sup_{k\geq 0} |a(k)|\) and \(\|\mathbf{a}\|_{0:k}
\defas \max_{0\leq i\leq k} |a(i)|\).
Let \(\calPD\) be the class of functions
\(\alpha:\nnegreal\rightarrow\nnegreal\) such that \(\alpha(0)=0\) and
\(\alpha(s)>0\) for all \(s>0\). Let \(\calK\) be the class of
\(\calPD\)-functions that are continuous and strictly increasing. Let
\(\calKinf\) be the class of \(\calK\)-functions that are unbounded. Let
\(\calKL\) be the set of functions
\(\beta:\nnegreal\times\nnegint\rightarrow\nnegreal\) such that
\(\beta(\cdot,k)\in\calK\), \(\beta(r,\cdot)\) is nonincreasing, and
\(\lim_{i\rightarrow\infty}\beta(r,i) = 0\) for all
\((r,k)\in\nnegreal\times\nnegint\). %
\ifthenelse{\boolean{LongVersion}}{%
  Let \(\id(\cdot) \defas (\cdot) \in \calKinf\) denote the identity map. %
}{}

\section{Problem statement}\label{sec:problem}
Consider the following discrete-time plant:
\begin{equation}\label{eq:plant}
  x^+ = f(x,u,\theta)
\end{equation}
where \(x\in\real^n\) is the plant state, \(u\in\real^m\) is the plant input,
and \(\theta\in\real^{n_\theta}\) is an \emph{unknown} parameter vector. We
denote the parameter estimate by \(\hat{\theta}\in\real^{n_\theta}\) and the
modeled system by
\begin{equation}\label{eq:model}
  x^+ = f(x,u,\hat{\theta}).
\end{equation}
We assume the parameter estimate is time-invariant, while the parameter vector
itself may be time-varying. For simplicity, let \(\hat{\theta}=0\) and denote
the model as
\begin{equation}\label{eq:nominal}
  x^+ = \hat f(x,u)\defas f(x,u,0).
\end{equation}

In this \ifthenelse{\boolean{LongVersion}}{report}{article}, we study the
behavior of an MPC designed with the model \cref{eq:model}, but applied to the
plant \cref{eq:plant}. %
\ifthenelse{\boolean{LongVersion}}{%
  We adopt a user-oriented perspective in this analysis: while the model is
  fixed (e.g., via system identification or prior knowledge), the plant behavior
  is unknown and possibly changing over time as equipment or the environment
  changes. %
}{} %
Under the assumption \(\hat\theta=0\), \(\theta\) takes the role of an estimate
residual. In the language of inherent robustness, the model \cref{eq:nominal} is
the nominal system, and the plant \cref{eq:plant} is the uncertain system.

\subsection{Nominal MPC and basic assumptions}
We consider an MPC problem with control constraints
\(u\in\mathbb{U}\subseteq\real^m\), a horizon length of \(N\in\posint\), a stage
cost \(\ell:\real^n\times\real^m\rightarrow\nnegreal\), a terminal constraint
\(\mathbb{X}_f\subseteq\real^n\), and a terminal cost
\(V_f:\real^n\rightarrow\nnegreal\). For an initial state \(x\in\real^n\), we
define the set of admissible \((x,\mathbf{u})\) pairs \cref{eq:mpc:admit},
admissible input sequences \cref{eq:mpc:admit:inputs}, and admissible initial
states \cref{eq:mpc:admit:states} by
\begin{align}
  \mathcal{Z}_N
  &\defas \set{ (x,\mathbf{u}) \in \real^n\times\mathbb{U}^N |
    \hat\phi(N; x, \mathbf{u}) \in \mathbb{X}_f } \label{eq:mpc:admit} \\
  \mathcal{U}_N(x)
  &\defas \set{ \mathbf{u}\in\mathbb{U}^N | (x,\mathbf{u})\in\mathcal{Z}_N }
    \label{eq:mpc:admit:inputs} \\
  \mathcal{X}_N
  &\defas \set{ x\in\real^n | \mathcal{U}_N(x) \; \textnormal{is nonempty} }
    \label{eq:mpc:admit:states}
\end{align}
where \(\hat\phi(k;x,\mathbf{u})\) denotes the solution to \cref{eq:nominal} at
time \(k\), given an initial state \(x\) and a sufficiently long input sequence
\(\mathbf{u}\). For each \((x,\mathbf{u})\in\real^{n+Nm}\), we define the MPC
objective by
\begin{equation}\label{eq:mpc:obj}
  V_N(x,\mathbf{u}) \defas \sum_{k=0}^{N-1} \ell(\hat\phi(k;x,\mathbf{u}),u(k))
  + V_f(\hat\phi(N;x,\mathbf{u}))
\end{equation}
and for each \(x\in\mathcal{X}_N\), we define the MPC problem by
\begin{equation}\label{eq:mpc}
  V_N^0(x) \defas \min_{\mathbf{u}\in\mathcal{U}_N(x)} V_N(x,\mathbf{u}).
\end{equation}
Using the convention of~\cite{rockafellar:wets:1998} for infeasible problems, we
take \(V_N^0(x)\defas\infty\) for all \(x\not\in\mathcal{X}_N\).

Throughout, we use the standard assumptions for inherent robustness of MPC
from~\cite{allan:bates:risbeck:rawlings:2017}.

\begin{assumption}[Continuity]\label{assum:cont}%
  The functions \(f:\real^n\times\real^m\times\real^{n_\theta} \rightarrow
  \real^n\), \(\ell:\real^n\times\real^m\rightarrow\nnegreal\), and
  \(V_f:\real^n\rightarrow\nnegreal\) are continuous and \(\hat f(0,0)=0\),
  \(\ell(0,0)=0\), and \(V_f(0)=0\).
\end{assumption}

\begin{assumption}[Constraint properties]\label{assum:cons}%
  The set \(\mathbb{U}\) is compact and contains the origin.
  The set \(\mathbb{X}_f\) is defined by \(\mathbb{X}_f \defas \lev_{c_f} V_f\)
  for some \(c_f>0\).
\end{assumption}

\begin{assumption}[Terminal control law]\label{assum:stabilizability}%
  There exists a terminal control law
  \(\kappa_f:\mathbb{X}_f\rightarrow\mathbb{U}\) such that
  \begin{align*}
    V_f(\hat f(x,\kappa_f(x))) &\leq V_f(x) - \ell(x,\kappa_f(x)),
    & \forall\; x&\in\mathbb{X}_f.
  \end{align*}
\end{assumption}

\begin{assumption}[Stage cost bound]\label{assum:posdef} There exists a function
  \(\alpha_1\in\calKinf\) such that
  \begin{align}\label{eq:bound:posdef}
    \ell(x,u) &\geq \alpha_1(|(x,u)|),
    & \forall \; (x,u)&\in\real^n\times\mathbb{U}.
  \end{align}
\end{assumption}

\begin{remark}
  \Cref{assum:cons,assum:stabilizability} imply \(V_f(\hat f(x,\kappa_f(x)))
  \leq V_f(x) \leq c_f\) for all \(x\in\mathbb{X}_f\) and therefore \(\mathbb{X}_f\)
  is positive invariant for \(x^+=\hat f(x,\kappa_f(x))\).
\end{remark}

Under \cref{assum:cont,assum:cons}, the existence of solutions to \cref{eq:mpc}
follows from~\cite[Prop.~2.4]{rawlings:mayne:diehl:2020}. To ensure uniqueness,
we assume some selection rule has been applied and denote the solution by
\(\mathbf{u}^0(x)=(u^0(0;x),\ldots,u^0(N-1;x))\), denote the corresponding
optimal state sequence by \(\hat x^0(k;x)\defas\hat\phi(k;x,\mathbf{u}^0(x))\)
for each \(k\in\intinterval{0}{N}\), and define the MPC control law
\(\kappa_N:\mathcal{X}_N\rightarrow\mathbb{U}\) by \(\kappa_N(x) \defas
u^0(0;x)\). Note that the subsequent analyses do not depend on the chosen
selection rule, so the results hold no matter what solutions are selected at a
particular time. It is also useful to define the following suboptimal input
sequence:
\[
  \tilde{\mathbf{u}}(x) \defas (u^0(1;x),\ldots,u^0(N-1;x),\kappa_f(\hat
  x^0(N;x))).
\]

Quadratic stage and terminal costs are of particular interest in this work.
Throughout, we call an MPC satisfying the following assumption a \emph{quadratic
  cost MPC}.
\begin{assumption}[Quadratic cost]\label{assum:quad}
  We have
  \begin{align}\label{eq:cost:quad}
    \ell(x,u) &\defas |x|_Q^2 + |u|_R^2,
    & V_f(x) &\defas |x|_{P_f}^2
  \end{align}
  for all \((x,u)\in\real^n\times\real^m\) and positive definite \(Q\), \(R\),
  and \(P_f\).
\end{assumption}

Consider the \emph{modeled} closed-loop system
\begin{equation}\label{eq:model:cl}
  x^+=\hat f_c(x)\defas\hat f(x,\kappa_N(x)).
\end{equation}
From \Cref{assum:cont,assum:cons,assum:stabilizability,assum:posdef}, it can be
shown \(x^+=\hat f_c(x)\) is asymptotically stable in \(\mathcal{X}_N\) with the
Lyapunov function \(V_N^0\)~\cite[Thm.~2.19]{rawlings:mayne:diehl:2020}. %
\ifthenelse{\boolean{LongVersion}}{%
  For completeness, we include a sketch of the proof in \Cref{app:mpc:stable}.
  \begin{theorem}[Thm.~2.19~of~\cite{rawlings:mayne:diehl:2020}]\label{thm:mpc:stable}
    Suppose \Cref{assum:cont,assum:cons,assum:stabilizability,assum:posdef}
    hold. Then, the following holds:
    \begin{enumerate}[(a)]
    \item \(\mathcal{X}_N\) is positive invariant for \(x^+=\hat f_c(x)\);
    \item there exists \(\alpha_2\in\calKinf\) such that, for each
      \(x\in\mathcal{X}_N\),
      \begin{subequations}\label{eq:mpc:lyap}
        \begin{align}
          \alpha_1(|x|) &\leq V_N^0(x) \leq \alpha_2(|x|) \label{eq:mpc:lyap:a} \\
          V_N^0(\hat f_c(x)) &\leq V_N^0(x) - \alpha_1(|x|); \label{eq:mpc:lyap:b}
        \end{align}
      \end{subequations}
    \item and \(x^+=\hat f_c(x)\) is asymptotically stable on
      \(\mathcal{X}_N\).
    \end{enumerate}
  \end{theorem}

}{} %
Similarly, it is shown in \cite[Sec.~2.5.5]{rawlings:mayne:diehl:2020} that,
under \Cref{assum:cont,assum:cons,assum:stabilizability,assum:quad}, the
quadratic cost MPC \emph{exponentially} stabilizes the closed-loop system
\cref{eq:model:cl} on any sublevel set of the optimal value function
\(\mathcal{S}\defas\textnormal{lev}_\rho V_N^0\). %
\ifthenelse{\boolean{LongVersion}}{%
  Note that, because \(V_N^0\) is only defined on \(\mathcal{X}_N\), we have
  \(\mathcal{S}\subseteq\mathcal{X}_N\) by the definition of the sublevel set.
  For completeness, we restate the conclusion of
  \cite[Sec.~2.5.5]{rawlings:mayne:diehl:2020} in the theorem below and include
  a sketch of the proof in \Cref{app:mpc:stable}.
  \begin{theorem}\label{thm:mpc:stable:exp}%
    Suppose \Cref{assum:cont,assum:cons,assum:stabilizability,assum:quad} hold.
    Let \(\rho>0\) and \(\mathcal{S}\defas\textnormal{lev}_\rho V_N^0\). Then,
    the following holds:
    \begin{enumerate}[(a)]
    \item \(\mathcal{S}\) is positive invariant for \(x^+=\hat f_c(x)\);
    \item there exists a constant \(c_2>0\) such that
      \begin{subequations}\label{eq:mpc:lyap:exp}
        \begin{align}
          c_1|x|^2 &\leq V_N^0(x) \leq c_2|x|^2 \label{eq:mpc:lyap:exp:a} \\
          V_N^0(\hat f_c(x)) &\leq V_N^0(x) - c_1|x|^2 \label{eq:mpc:lyap:exp:b}
        \end{align}
      \end{subequations}
      for each \(x\in\mathcal{S}\), where \(c_1\defas\underline{\sigma}(Q)\); and
    \item \(x^+=\hat f_c(x)\) is exponentially stable on \(\mathcal{S}\).
    \end{enumerate}
  \end{theorem}
}{%
  These facts are stated as special cases of the inherent robustness results in
  \Cref{sec:mpc:robust}. %
}

To show strong stability of the MPC with mismatch, we eventually require one or
both of the following assumptions.

\begin{assumption}[Steady state]\label{assum:steady-state}
  The origin is a steady state, uniformly in \(\theta\in\real^{n_\theta}\),
  i.e., \(f(0,0,\theta)=0\) for all \(\theta\in\real^{n_\theta}\).
\end{assumption}

\begin{assumption}[Differentiability]\label{assum:diff}
  The derivative \(\partial_{(x,u)} f\) exists and is continuous on
  \(\real^{n+m+n_\theta}\).
\end{assumption}

\begin{remark}\label{rem:statecons}
  State constraints were not considered, as there is no way to guarantee robust
  feasibility of state-constrained nominal
  MPC~\cite{allan:bates:risbeck:rawlings:2017}. Soft state constraints
  (cf.~\cite{santos:biegler:castro:2008,kuntz:rawlings:2024f}) are compatible
  with our general cost MPC assumptions, but using them in the quadratic cost
  MPC would require some modifications to our analysis.
\end{remark}

\begin{remark}\label{rem:steady-state}
  \Cref{assum:steady-state} limits our results to problems where the steady
  state is known and fixed (e.g., path-planning and inventory problems). If the
  steady state depends on \(\theta\), i.e., \(x_s(\theta) =
  f(x_s(\theta),u_s(\theta),\theta)\), we can still work with deviation
  variables \((\delta x,\delta u)\defas (x-x_s(\theta),u-u_s(\theta))\), but (i)
  we have to estimate the steady-state pair \((x_s(\theta),u_s(\theta))\) (e.g.,
  via an integrating disturbance model~\cite[Ch.~1]{rawlings:mayne:diehl:2020}),
  and (ii) strong stability is only achieve when the steady-state map is
  continuous, the parameters are asymptotically constant, and the estimation
  errors converge~\cite{kuntz:rawlings:2024f}.
\end{remark}

\begin{remark}
  \Cref{assum:diff} effectively requires the plant~\cref{eq:plant} to be
  continuous in \(\theta\). Continuous differentiability of \(f\) is sufficient,
  but not necessary, for guaranteeing \Cref{assum:diff}.
\end{remark}

\begin{remark}\label{ex:lpv}
  The linear parameter-varying (LPV) system \(x^+=A(\theta)x+B(\theta)u\), where
  \((A,B)\) are continuous in \(\theta\), is a simple example satisfying both
  \Cref{assum:steady-state,assum:diff}. If \((A,B)\) are known, one could treat
  \(\hat{\theta}\) as a fitted parameter estimate, construct the MPC from the
  data-driven surrogate model \(x^+=A(\hat{\theta})x + B(\hat{\theta})u\), and
  use the theory herein to demonstrate stability given sufficiently accurate
  estimates \(\hat{\theta}\).
\end{remark}

\subsection{Motivating example}\label{ssec:example:scalar}
\begin{figure}
  \centering
  \ifthenelse{\boolean{OneColumn}}{%
    \includegraphics[width=0.6\linewidth,page=2]{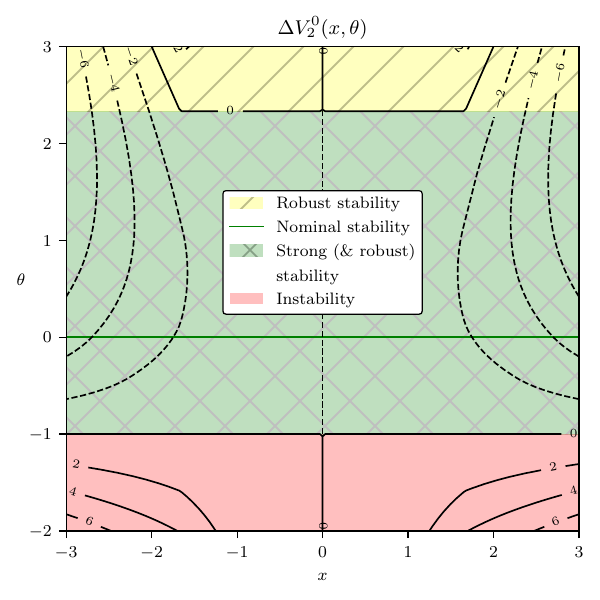}
  }{%
    \includegraphics[width=\linewidth,page=2]{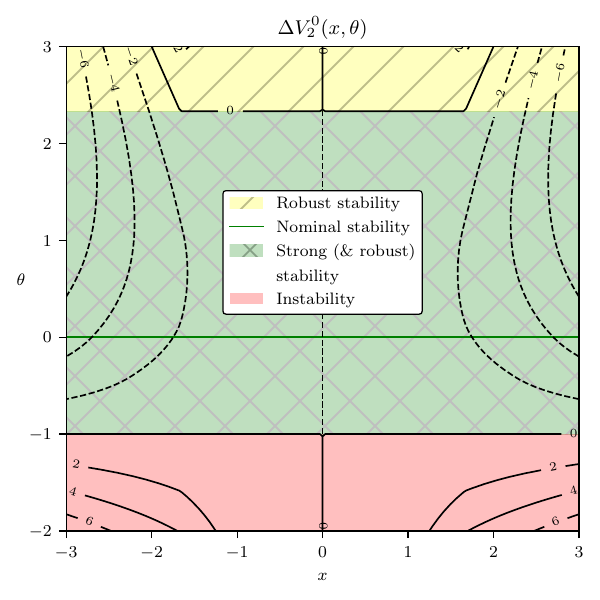}
  }%
  \caption{Contours of the cost difference as a function of the initial state
    \(x\) and the parameter \(\theta\).}%
  \label{fig:scalar:contour}
\end{figure}

\begin{figure}
  \centering
  \ifthenelse{\boolean{OneColumn}}{%
    \includegraphics[width=0.6\linewidth,page=3]{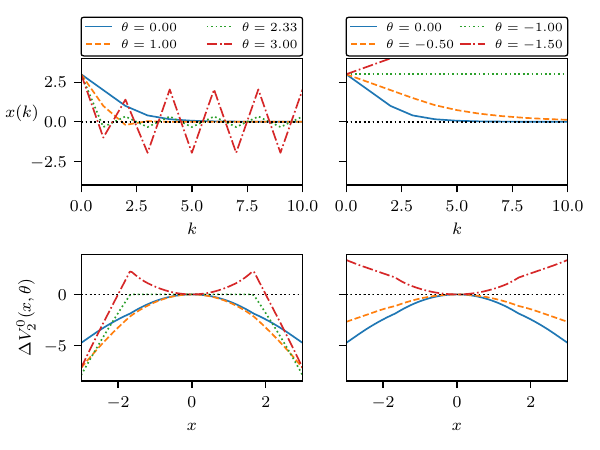}
  }{%
    \includegraphics[width=\linewidth,page=3]{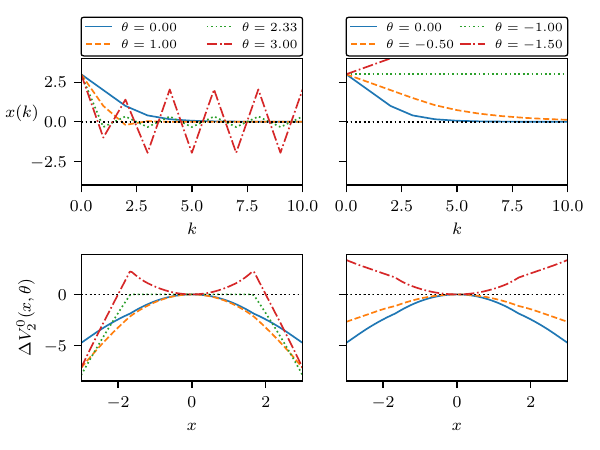}
  }%
  \caption{For (left) positive and (right) negative values of \(\theta\), the
    (top) closed-loop trajectories with initial state \(x=3\), and (bottom) cost
    differences as a function of \(x\), along with the nominal values.}%
  \label{fig:scalar:traj}
\end{figure}

We close this section with a motivating example exhibiting many types of
stability under persistent mismatch. Recall from the introduction we define
\emph{robust stability} as an ISS property for parameter errors, and
\emph{strong stability} as convergence to the origin despite mismatch. While
precise definitions are given in \Cref{sec:robust:strong}, these informal
definitions suffice for the example.

Consider the scalar linear system
\begin{equation}\label{eq:plant:scalar}
  x^+ = f(x,u,\theta) \defas x + (1+\theta)u.
\end{equation}
The plant \cref{eq:plant:scalar} is a prototypical integrating system, such as a
storage tank or vehicle on a track, with an uncertain input gain. As usual the
system is modeled with \(\hat\theta=0\),
\begin{equation}\label{eq:model:scalar}
  x^+ = \hat f(x,u) \defas f(x,u,0) = x + u.
\end{equation}

We define a nominal MPC with \(\mathbb{U}\defas[-1,1]\), \(\ell(x,u)\defas
(1/2)(x^2+u^2)\), \(V_f(x)\defas (1/2)x^2\), \(\mathbb{X}_f\defas[-1,1]\), and
\(N\defas 2\). Notice that the terminal set can be reached in \(N=2\) moves if
and only if \(|x|\leq 3\), so we have the set of admissible initial states
\(\mathcal{X}_2=[-3,3]\). \emph{Without} the terminal constraint (i.e.,
\(\mathbb{X}_f=\real\)), the optimal control sequence is
\[
  \mathbf{u}^0(x) = \begin{cases} (-3x/5, -x/5), & |x|\leq 5/3 \\
    (-\textrm{sgn}(x),-x/2+\textrm{sgn}(x)/2), & 5/3<|x|\leq 3 \end{cases}
\]
and the control law is \(\kappa_2(x) \defas
-\textrm{sat}(3x/5)\)~\cite[p.~104]{rawlings:mayne:diehl:2020}. However, the
optimal input sequence gives
\[
  \hat x^0(2;x) = \begin{cases} x/5, & |x|\leq 5/3 \\ x/2 - \textrm{sgn}(x)/2, &
    5/3<|x|\leq 3 \end{cases}
\]
so the terminal constraint \(\mathbb{X}_f=[-1,1]\) is automatically satisfied
for all \(|x|\leq 3\). Therefore \(\kappa_2(x)=-\textrm{sat}(3x/5)\) is also the
control law of the problem \emph{with} the terminal constraint.

In \Cref{fig:scalar:contour} we plot contours of the cost difference \(\Delta
V_2^0(x,\theta) \defas V_2^0(f(x,\kappa_2(x),\theta)) - V_2^0(x)\), and in
\Cref{fig:scalar:traj}, we plot closed-loop trajectories and the cost difference
curve \(\Delta V_2^0(\cdot,\theta)\) for several values of \(\theta\). %
\ifthenelse{\boolean{LongVersion}}{%
  The system is strongly stable for all \(-1<\theta<7/3\) as the cost difference
  curve is negative definite. When \(\theta<-1\), the entire cost difference
  curve is positive definite, so the trajectories become unbounded. This is
  because the disturbance cancels out the effect of the controller and drives
  the system in the opposite direction. On the other hand, when \(\theta>7/3\),
  the cost difference curve is only positive definite near the origin, but
  negative elsewhere, so the trajectories remain bounded for all time, although
  they do not converge to the origin. In this case, high parameter values push
  the system in the same direction as the input, and input saturation moderates
  the effect of overshoot at high parameter values. %
}{%
  For \(-1<\theta<7/3\), we have strong stability as \(\Delta
  V_2^0(\cdot,\theta)\) is negative definite. For \(\theta\geq 7/3\), the sign
  of \(\Delta V_2^0(\cdot,\theta)\) is ambiguous, and we have robust stability.
  For \(\theta<-1\), the trajectories are unbounded as \(\Delta
  V_2^0(\cdot,\theta)\) is positive definite. %
}%
We point out the existing literature on inherent robustness is not sufficient to
predict strong stability whenever \(-1<\theta<7/3\). %

\section{Robust and strong stability}\label{sec:robust:strong}
Consider the closed-loop system
\begin{align}\label{eq:plant:cl}
  x^+ &= f_c(x,\theta) \defas f(x,\kappa_N(x),\theta), & \theta&\in\Theta
\end{align}
where \(\Theta\subseteq\real^{n_\theta}\). Let
\(\phi_c(k;x,\boldsymbol{\uptheta})\) denote solutions to \cref{eq:plant:cl} at
time \(k\), given an initial state \(x\in\mathcal{X}_N\) and a sufficiently long
parameter sequence \(\theta\in\Theta\). If
\(\Theta\defas\set{\theta\in\real^{n_\theta} | |\theta|\leq\delta}\), it is
convenient to write \cref{eq:plant:cl} as
\(x^+=f_c(x,\theta),|\theta|\leq\delta\). We define robustly positive invariant
(RPI) sets for \cref{eq:plant:cl} as follows.
\begin{definition}[Robust positive invariance]\label{defn:rpi}
  A set \(X\subseteq\real^n\) is \emph{robustly positive invariant} for the
  system \(x^+=f_c(x,\theta),\theta\in\Theta\) if \(f_c(x,\theta)\in X\) for all
  \(x\in X\) and \(\theta\in\Theta\).
\end{definition}

In this section, we present stability definitions and results for
\cref{eq:plant:cl}. For brevity, asymptotic and exponential definitions and
results are consolidated into the same statement. The main difference between
our definitions and results and existing ones in the literature is the
restriction of the state to the RPI set \(X\) and the disturbance
to the arbitrary set \(\Theta\). %
\ifthenelse{\boolean{LongVersion}}{%
  Additionally, while in this section we consider the closed-loop system with
  exogenous disturbances \cref{eq:plant:cl}, the system could be equivalently
  interpreted as an open-loop system with arbitrary inputs, as is typically
  considered in the stability literature. %
}{}

\subsection{Robust stability}
We define robust asymptotic stability (RAS) similarly to input-to-state
stability (ISS) from~\cite{jiang:wang:2001}. Likewise, we define robust
exponential stability (RES) similarly to input-to-state exponential stability
(ISES) from~\cite{grune:sontag:wirth:1999}.
\begin{definition}[Robust stability]\label{defn:ras}
  A system \(x^+=f_c(x,\theta),\theta\in\Theta\) is \emph{robustly
    asymptotically stable} (in a RPI set \(X\subseteq\real^n\)) if there exists
  \(\beta\in\calKL\) and \(\gamma\in\calK\) such that
  \begin{equation}\label{eq:ras}
    |\phi_c(k;x,\boldsymbol{\uptheta})| \leq \beta(|x|,k) +
    \gamma(\|\boldsymbol{\uptheta}\|_{0:k-1})
  \end{equation}
  for all \(k\in\nnegint\), \(x\in X\), and \(\boldsymbol{\uptheta}\in\Theta^k\).
  If, additionally, \(\beta(s,k)=cs\lambda^k\) for some \(c>0\) and
  \(\lambda\in(0,1)\), we say \(x^+=f_c(x,\theta),\theta\in\Theta\) is
  \emph{robustly exponentially stable} (in \(X\)).
\end{definition}

\begin{definition}[ISS/ISES Lyapunov function]\label{defn:lyap:iss}
  A function \(V:X\rightarrow\nnegreal\) is an \emph{ISS Lyapunov function} (in
  an RPI set \(X\subseteq\real^n\), for the system
  \(x^+=f_c(x,\theta),\theta\in\Theta\)) if there exist functions
  \(\alpha_1,\alpha_2,\alpha_3\in\calKinf\) and \(\sigma\in\calK\) such that
  \begin{subequations}\label{eq:lyap:iss}
    \begin{align}
      \alpha_1(|x|) &\leq V(x) \leq \alpha_2(|x|) \label{eq:lyap:iss:a} \\
      V(f_c(x,\theta)) &\leq V(x) - \alpha_3(|x|) + \sigma(|\theta|).
                         \label{eq:lyap:iss:b}
    \end{align}
  \end{subequations}
  for all \(x\in X\) and \(\theta\in\Theta\). If, additionally,
  \(\alpha_i(\cdot)\defas a_i{(\cdot)}^b\) for some \(a_i,b>0\) and each
  \(i\in\intinterval{1}{3}\), we say \(V\) is an \emph{ISES Lyapunov function}
  (in \(X\), for \(x^+=f_c(x,\theta),\theta\in\Theta\)).
\end{definition}

The result below is a generalization
of~\cite[Prop.~19]{allan:bates:risbeck:rawlings:2017} to include arbitrary
disturbance sets and the exponential case\ifthenelse{\boolean{LongVersion}}{
  (see \Cref{app:lyap} for proof)}{}. %
\begin{theorem}[ISS/ISES Lyapunov theorem]\label{thm:lyap:robust}
  The system \(x^+=f_c(x,\theta),\theta\in\Theta\) is RAS (RES) in an RPI set
  \(X\subseteq\real^n\) if it admits an ISS (ISES) Lyapunov function in \(X\).
\end{theorem}
\ifthenelse{\boolean{LongVersion}}{}{%
  \begin{proof}
    While a self-contained proof can be found in \cite{kuntz:rawlings:2024d}, we
    provide a sketch of the proof as follows.

    As in \cite{allan:bates:risbeck:rawlings:2017}, the asymptotic case follows
    identically to the proof \cite[Lem.~3.5]{jiang:wang:2001}, noting that it is
    unnecessary to invoke continuity of \(f_c\) and \(V\), and, without loss of
    generality, we can restrict the state and disturbance to \(X\) and
    \(\Theta\), respectively.

    For the exponential case, we immediately have
    \begin{align}
      a_1|x(k)|^b \leq V(x(k))
      &\leq \lambda_0V(x(k-1)) + \sigma(|\theta(k-1)|) \nonumber \\
      &\leq \lambda_0^k V(x) + \sum_{i=1}^k \lambda_0^{i-1}\sigma(\theta(k-i)) \nonumber \\
      &\leq a_2\lambda_0^k|x|^b + \frac{\sigma(\|\boldsymbol{\uptheta}\|_{0:k-1})}{1-\lambda_0} \label{eq:ses:1}
    \end{align}
    for all \(k\in\nnegint\), \(x\in X\), and
    \(\boldsymbol{\uptheta}\in\Theta^k\), where \(\lambda_0\defas
    1-\frac{a_2}{a_3} \in (0,1)\) and \(x(j) \defas
    \phi_c(k;x,\boldsymbol{\uptheta})\) for all \(j\in\intinterval{0}{k}\). The
    desired bound follows by applying the function \(((\cdot)/a_1)^{1/b}\) to both
    sides and using the triangle inequality when \(b\leq 1\) and the definition
    of convexity when \(b>1\), appropriately defining \(c>0\) and
    \(\lambda\in(0,1)\) for each case.
  \end{proof}
}

\subsection{Strong stability}
We take strong asymptotic stability (SAS) as a time-invariant version of the
conclusion of~\cite[Prop.~2.2]{jiang:wang:2002}. Strong exponential stability
(SES) is defined similarly.
\begin{definition}[Strong stability]\label{defn:stable:strong}
  A system \(x^+=f_c(x,\theta),\theta\in\Theta\) is \emph{strongly
    asymptotically stable} (in a RPI set \(X\subseteq\real^n\)) if there exists
  \(\beta\in\calKL\) such that
  \begin{equation}\label{eq:sas}
    |\phi_c(k;x,\boldsymbol{\uptheta})| \leq \beta(|x|,k)
  \end{equation}
  for all \(k\in\nnegint\), \(x\in X\), and \(\boldsymbol{\uptheta}\in\Theta^k\).
  If, additionally, \(\beta(s,k) \defas cs\lambda^k\) for all \(s\geq 0\) and
  \(k\in\nnegint\), and some \(c>0\) and \(\lambda\in(0,1)\), we say
  \(x^+=f_c(x,\theta),\theta\in\Theta\) is \emph{strongly exponentially stable}
  (in \(X\)).
\end{definition}

\begin{definition}[Lyapunov function]\label{defn:lyap:strong}
  A function \(V:X\rightarrow\nnegreal\) is a \emph{Lyapunov function} (in a RPI
  set \(X\subseteq\real^n\), for the system
  \(x^+=f(x,\theta),\theta\in\Theta\)), if there exist functions
  \(\alpha_1,\alpha_2\in\calKinf\) and a continuous function
  \(\alpha_3\in\calPD\) such that
  \begin{subequations}\label{eq:lyap:strong}
    \begin{align}
      \alpha_1(|x|) &\leq V(x) \leq \alpha_2(|x|) \label{eq:lyap:strong:a} \\
      V(f_c(x,\theta)) &\leq V(x) - \alpha_3(|x|) \label{eq:lyap:strong:b}
    \end{align}
  \end{subequations}
  for all \(x\in X\) and \(\theta\in\Theta\). If, additionally,
  \(\alpha_i(\cdot)\defas a_i{(\cdot)}^b\) for some \(a_i,b>0\) and each
  \(i\in\intinterval{1}{3}\), we say \(V\) is an \emph{exponential Lyapunov
    function} (in \(X\), for \(x^+=f_c(x,\theta),\theta\in\Theta\)).
\end{definition}

The following theorem
generalizes~\cite[Prop.~13]{allan:bates:risbeck:rawlings:2017}
and~\cite[Lem.~15]{pannocchia:rawlings:wright:2011} to include arbitrary
disturbance sets\ifthenelse{\boolean{LongVersion}}{ (see \Cref{app:lyap} for
  proof)}{}. %
\begin{theorem}\label{thm:lyap:strong}
  The system \(x^+=f_c(x,\theta),\theta\in\Theta\) is SAS (SES) in a RPI set
  \(X\subseteq\real^n\) if it admits a Lyapunov function (an exponential
  Lyapunov function) in \(X\).
\end{theorem}
\ifthenelse{\boolean{LongVersion}}{}{%
  \begin{proof}
    See \cite{kuntz:rawlings:2024d} for a self-contained version of the
    following proof sketch.

    For the asymptotic case, since the proof of \cite[Lem.~2.8]{jiang:wang:2002}
    still holds when parts relating to continuity of \(f_c\) and \(V\) are
    dropped (cf.~\cite[Prop.~12]{kuntz:rawlings:2024d}), there exist
    functions \(\alpha,\rho\in\calKinf\) such that \(W(f_c(x,\theta)) \leq W(x)
    - \alpha(|x|)\) for all \(x\in X\) and \(\theta\in\Theta\), where
    \(W\defas\rho\circ V\). Moreover \(\hat{\alpha}_1(|x|)\leq
    W(x)\leq\hat{\alpha}_2(|x|)\) for all \(x\in X\), where
    \(\hat{\alpha}_i\defas\rho\circ\alpha_i\in\calKinf,i\in\intinterval{1}{2}\).
    In other words, \(W\) is a Lyapunov function on \(X\) for the system
    \cref{eq:plant:cl}, but with a \(\calKinf\)-function cost decrease. The rest
    of the proof of the asymptotic part follows identically to the proof of the
    relevant part of \cite[Thm.~1(1)]{jiang:wang:2002}.

    Following the proof of \Cref{thm:lyap:robust}, the exponential case is
    established by setting \(\sigma\equiv 0\) in~\cref{eq:ses:1} and applying
    the function \(((\cdot)/a_1)^{1/b}\) to both sides.
  \end{proof}
}

\section{Inherent robustness of MPC}\label{sec:mpc:robust}
In this section, we review results on the inherent robustness of nominal MPC.\@
\ifthenelse{\boolean{LongVersion}}{}{%
  See \cite{kuntz:rawlings:2024d} for direct proofs of the results in this
  section. %
}%
\Cref{thm:mpc:robust} follows as a special case of the suboptimal MPC robustness
result~\cite[Thm.~21]{allan:bates:risbeck:rawlings:2017}%
\ifthenelse{\boolean{LongVersion}}{ %
  (see \Cref{app:mpc:robust} for proof). %
}{.}
\begin{theorem}\label{thm:mpc:robust}
  Suppose \Cref{assum:cont,assum:cons,assum:stabilizability,assum:posdef} hold.
  Let \(\rho>0\) and \(\mathcal{S}\defas\textnormal{lev}_\rho V_N^0\). Then
  there exist \(\delta>0\), \(\alpha_2\in\calKinf\), and \(\sigma\in\calK\) such
  that
  \begin{subequations}\label{eq:mpc:robust:lyap}
    \begin{align}
      \alpha_1(|x|) &\leq V_N^0(x) \leq \alpha_2(|x|) \label{eq:mpc:robust:lyap:a} \\
      V_N^0(f_c(x,\theta)) &\leq V_N^0(x) - \alpha_1(|x|) +
                             \sigma(|\theta|) \label{eq:mpc:robust:lyap:b}
    \end{align}
  \end{subequations}
  for all \(x\in\mathcal{S}\) and \(|\theta|\leq\delta\), and the system
  \(x^+=f_c(x,\theta),|\theta|\leq\delta\) is RAS in the RPI set
  \(\mathcal{S}\).
\end{theorem}

A key step of the proof of \Cref{thm:mpc:robust} and the main results is to
establish the following robust descent property: %
\ifthenelse{\boolean{OneColumn}}{%
  \begin{equation}\label{eq:mpc:descent:robust}
    V_N^0(f_c(x,\theta)) \leq V_N^0(x) - \ell(x,\kappa_N(x)) +
    V_N(f_c(x,\theta),\tilde{\mathbf{u}}(x)) - V_N(\hat
    f_c(x),\tilde{\mathbf{u}}(x)).
  \end{equation}
}{%
  \begin{multline}\label{eq:mpc:descent:robust}
    V_N^0(f_c(x,\theta)) \leq V_N^0(x) - \ell(x,\kappa_N(x)) \\
    + V_N(f_c(x,\theta),\tilde{\mathbf{u}}(x)) - V_N(\hat
    f_c(x),\tilde{\mathbf{u}}(x)).
  \end{multline}
}%
In fact, it is shown~\cref{eq:mpc:descent:robust} can be achieved on any
sublevel set of \(V_N^0\) and a sufficiently small neighborhood
\(|\theta|\leq\delta\). We restate this intermediate result in the following
proposition%
\ifthenelse{\boolean{LongVersion}}{ %
  (see \Cref{app:mpc:rpi} for proof). %
}{.}
\begin{proposition}\label{prop:mpc:rpi}
  Suppose \Cref{assum:cont,assum:cons,assum:stabilizability,assum:posdef} hold.
  Let \(\rho>0\) and \(\mathcal{S}\defas\textnormal{lev}_\rho V_N^0\). There
  exists \(\delta>0\) such that \cref{eq:mpc:descent:robust} holds for all
  \(x\in\mathcal{S}\) and \(|\theta|\leq\delta\) and \(\mathcal{S}\) is RPI for
  \(x^+=f_c(x,\theta),|\theta|\leq\delta\).
\end{proposition}

With quadratic costs (\Cref{assum:quad}),
\Cref{assum:cont,assum:cons,assum:stabilizability} also imply inherent
\emph{exponential} robustness of MPC.\@ \Cref{thm:mpc:robust:exp} follows as a
special case of the suboptimal MPC robustness result
\cite[Thm.~18]{pannocchia:rawlings:wright:2011}%
\ifthenelse{\boolean{LongVersion}}{ %
  (see \Cref{app:mpc:robust:exp} for proof). %
}{.}
\begin{theorem}
  \label{thm:mpc:robust:exp}
  Suppose \Cref{assum:cont,assum:cons,assum:stabilizability,assum:quad} hold.
  Let \(\rho>0\) and \(\mathcal{S}\defas\textnormal{lev}_\rho V_N^0\). There
  exist \(\delta,c_2>0\) and \(\sigma\in\calK\) such that
  \begin{subequations}\label{eq:mpc:robust:lyap:exp}
    \begin{align}
      c_1|x|^2 &\leq V_N^0(x) \leq c_2|x|^2 \label{eq:mpc:robust:lyap:exp:a} \\
      V_N^0(f_c(x,\theta)) &\leq V_N^0(x) - c_1|x|^2 +
                             \sigma(|\theta|) \label{eq:mpc:robust:lyap:exp:b}
    \end{align}
  \end{subequations}
  for all \(x\in\mathcal{S}\) and \(|\theta|\leq\delta\), where
  \(c_1\defas\underline{\sigma}(Q)\), and the system
  \(x^+=f_c(x,\theta),|\theta|\leq\delta\) is RES in the RPI set
  \(\mathcal{S}\).
\end{theorem}

\section{Stability of MPC despite mismatch}\label{sec:mpc:strong}
In this section, we investigate two approaches to guarantee strong stability of
the closed-loop system \cref{eq:plant:cl}. First, we take a direct approach and
assume the existence of an ISS Lyapunov function that achieves a certain maximum
increase due to mismatch. In general, an additional scaling condition is
required for the mismatch term, although it is automatically satisfied for
quadratic cost MPC.\@ While these new Lyapunov assumptions are difficult to
check, we can easily construct error bounds that imply the maximum Lyapunov
increase for \(V_N^0\) via the standard MPC assumptions
(\Cref{assum:cont,assum:cons,assum:stabilizability,assum:posdef,%
  assum:quad}) and one or both of \Cref{assum:steady-state,assum:diff}.

\subsection{Maximum Lyapunov increase}\label{ssec:mpc:strong:lyap}
We begin with the direct approach. The goal here is not (necessarily) to provide
the means to check if a given MPC is strongly stabilizing, but to (i) identify a
set of conditions for which an ISS Lyapunov function also guarantees strong
stability (not only for the closed-loop MPC but for a general class of systems)
and (ii) provide a path towards proving certain classes of nominal MPCs are
strongly stabilizing.

\subsubsection{Asymptotic case}
For inherent robustness, a maximum increase of the form
\cref{eq:mpc:robust:lyap:b} is proven for the optimal value function \(V_N^0\).
However, since the perturbation term \(\sigma(|\theta|)\) is uniform in \(|x|\),
strong stability is not demonstrated for nonzero \(\theta\). Under
\Cref{assum:steady-state}, we might assume the perturbation vanishes in either
of the limits \(|x|\rightarrow 0\) or \(|\theta|\rightarrow 0\). In this sense,
the perturbation should be class-\(\calK\) in \(|x|\) whenever \(|\theta|\) is
fixed, and vice versa. We call these functions \emph{joint \(\calK\)-functions}
or \emph{\(\calK^2\)-functions} and define them as follows.
\begin{definition}[Class \(\calK^2\)]\label{defn:kfunc:joint}
  The class of \emph{joint \(\calK\)-functions}, denoted
  \(\calK^2\) 
  is the class of continuous functions
  \(\gamma:\nnegreal^2\rightarrow\nnegreal\) such that
  \(\gamma(s,\cdot),\gamma(\cdot,s)\in\calK\) 
  for all \(s>0\).
\end{definition}

To achieve strong stability, we assume the existence of an ISS Lyapunov function
with a \(\calK^2\)-function perturbation term, rather than the standard
\(\calK\)-function perturbation term. Moreover, we require the perturbation to
decay faster than the nominal cost decrease in the limit \(|x|\rightarrow 0\) so
that the descent property of \Cref{defn:lyap:strong} is achieved for
sufficiently small \(\theta\).
\begin{assumption}
  \label{assum:lyap}%
  There exists a l.s.c.~function \(V:\real^n\rightarrow\nnegereal\) such that,
  for each \(\rho>0\), there exist \(\delta_0>0\),
  \(\alpha_1,\alpha_2,\alpha_3\in\calKinf\), and \(\gamma_V\in\calK^2\)
  satisfying the following:
  \begin{enumerate}[(a)]
  \item \(\mathcal{S}\defas \textnormal{lev}_\rho V \subseteq \mathcal{X}_N\),
  \item for each \(x\in\mathcal{S}\) and \(|\theta|\leq\delta_0\), we have
    \begin{subequations}\label{eq:lyap}
      \begin{align}
        \alpha_1(|x|) &\leq V(x) \leq \alpha_2(|x|) \label{eq:lyap:a} \\
        V(f_c(x,\theta)) &\leq V(x) - \alpha_3(|x|) + \gamma_V(|x|,|\theta|);
                           \label{eq:lyap:b}
      \end{align}
    \end{subequations}
  \item and there exists \(\tau>0\) such that
    \begin{equation}\label{eq:mpc:mismatch:posdef}
      \limsup_{s\rightarrow 0^+} \frac{\gamma_V(s,\tau)}{\alpha_3(s)} < 1.
    \end{equation}
  \end{enumerate}
\end{assumption}

\begin{remark}
  We assume \(V\) is l.s.c.~to ensure \(\mathcal{S}\) is closed and can be used
  as a domain of attraction. Note l.s.c.~of \(V_N^0\) is compatible with the
  jump to \(V_N^0(x)=\infty\) when \(x\not\in\mathcal{X}_N\).
\end{remark}

With \Cref{assum:lyap}, we have our first main result.
\begin{theorem}\label{thm:mpc:mismatch}
  Suppose \Cref{assum:lyap} holds with \(V:\real^n\rightarrow\nnegereal\). For
  each \(\rho>0\), there exists \(\delta>0\) for which
  \(x^+=f_c(x,\theta),|\theta|\leq\delta\) is SAS in the RPI set
  \(\mathcal{S}\defas \textnormal{lev}_\rho V\).
\end{theorem}

To prove \Cref{thm:mpc:mismatch}, we require a preliminary result related to the
ability of a given \(\calK^2\)-function to lower bound another given
\(\calK\)-function (see \Cref{app:kfunc:joint} for proof).
\begin{proposition}\label{prop:kfunc:joint}
  Let \(\alpha\in\calKinf\) and \(\gamma\in\calK^2\). If there exists \(\tau>0\)
  such that %
  \ifthenelse{\boolean{OneColumn}}{%
    \[
      \limsup_{s\rightarrow 0^+} \frac{\gamma(s,\tau)}{\alpha(s)} < 1
    \]
  }{%
    \(\limsup_{s\rightarrow 0^+} \gamma(s,\tau)/\alpha(s)<1\),
  }%
  then, for each \(\rho>0\), there exists \(\delta>0\) such that
  \(\gamma(s,t)<\alpha(s)\) for all \(s\in(0,\rho]\) and \(t\in[0,\delta]\).
\end{proposition}

Finally, we prove \Cref{thm:mpc:mismatch}.

\begin{proof}[Proof of \Cref{thm:mpc:mismatch}]%
  By \Cref{assum:lyap}(a,b) there exists \(\delta_0>0\),
  \(\alpha_1,\alpha_2,\alpha_3\in\calKinf\), and \(\gamma_V\in\calK^2\) such
  that \(\mathcal{S}\subseteq\mathcal{X}_N\) and \cref{eq:lyap} holds for each
  \(x\in\mathcal{S}\) and \(|\theta|\leq\delta_0\). Let
  \(\varepsilon_0\defas\sup_{x\in\mathcal{S}} |x|>0\).\footnote{If
    \(\mathcal{S}=\set{0}\), the conclusion would hold trivially, so we can
    assume \(\mathcal{S}\neq\set{0}\) without loss of generality.} By
  \Cref{assum:lyap}(c) and \Cref{prop:kfunc:joint}, there exists \(\delta_1>0\)
  such that \(\alpha_3(s) > \gamma_V(s,t)\) for all \(s\in(0,\varepsilon_0]\)
  and \(t\in[0,\delta_1]\). With \(\delta\defas\min\set{\delta_0,\delta_1}\),
  the function
  \[
    \sigma(s) \defas \begin{cases} \alpha_3(s) - \gamma_V(s,\delta), & 0\leq
      s\leq\varepsilon_0 \\ \alpha_3(\varepsilon_0) -
      \gamma_V(\varepsilon_0,\delta), & s>\varepsilon_0 \end{cases}
  \]
  is both class-\(\calPD\) and continuous. By \cref{eq:lyap:b}, we have
  \[
    V(f_c(x,\theta)) - V(x) \leq -\alpha_3(|x|) + \gamma_V(|x|,\delta) = -\sigma(|x|)
  \]
  for all \(x\in\mathcal{S}\) and \(|\theta|\leq\delta\). Moreover,
  \(V(x)\leq\rho\) implies
  \[
    V(f_c(x,\theta)) \leq V(x) - \sigma(|x|) \leq \rho
  \]
  so \(\mathcal{S}=\textnormal{lev}_\rho V\) must be RPI.\@
  Finally, \(x^+=f_c(x,\theta),|\theta|\leq\delta\) is SAS in \(\mathcal{S}\) by
  \Cref{thm:lyap:strong}.
\end{proof}

\begin{remark}
  One might na{\"i}vely assume that the closed-loop system \cref{eq:plant:cl} is
  SAS under only \Cref{assum:lyap}(a,b). However, if the scaling condition
  \Cref{assum:lyap}(c) does not hold, then it may be the case that we cannot
  shrink \(t\) small enough to make \(\alpha_3(\cdot) - \gamma_V(\cdot,t)\)
  positive definite in a sufficiently large neighborhood of the origin, let
  alone any neighborhood at all. Thus \Cref{assum:lyap}(a,b) alone are
  insufficient to show \(V\) is a Lyapunov function for the closed-loop system
  \cref{eq:plant:cl}. This is illustrated in the example of
  \Cref{ssec:example:sqrt}\ifthenelse{\boolean{LongVersion}}{ and in the
    following examples}{}.
\end{remark}

\ifthenelse{\boolean{LongVersion}}{%
  \begin{example}\label{ex:kfunc:1}
    Let \(\alpha_3(s)\defas s^2\), \(\gamma_V(s,t)\defas st\), and
    \(L\defas\limsup_{s\rightarrow 0^+} \frac{\gamma_V(s,t)}{\alpha_3(s)}\).
    Then \(\alpha_3\in\calKinf\) and \(\gamma_V\in\calK^2\), but \(L =
    \lim_{s\rightarrow 0^+} t/s = \infty\) for each \(t>0\). In fact, since
    \(\sigma_t(s) \defas \alpha_3(s) - \gamma_V(s,t) = s^2-st\), \(\sigma_t\) is
    negative definite near the origin for each \(t>0\).
  \end{example}

  \begin{example}\label{ex:kfunc:2}
    Let \(\alpha_3(s)\defas s\), \(\gamma_V(s,t)\defas\frac{2st}{s+t}\), and
    \(L\defas\limsup_{s\rightarrow 0^+} \frac{\gamma_V(s,t)}{\alpha_3(s)}\).
    Then \(\alpha_3\in\calKinf\) and \(\gamma_V\in\calK^2\), but \(L =
    \lim_{s\rightarrow 0^+} \frac{2t}{s+t} = 2\) for each \(t>0\). Moreover,
    since \(\sigma_t(s) \defas \alpha_3(s) - \gamma_V(s,t) = s-\frac{2st}{s+t} =
    \frac{s^2-st}{s+t}\), \(\sigma_t\) is negative definite near the origin for
    each \(t>0\).
  \end{example}

  \begin{remark}
    While \Cref{assum:posdef} implies \cref{eq:lyap} can be satisfied with
    \(\alpha_3\defas\alpha_1\), it may be the case that
    \cref{eq:mpc:mismatch:posdef} is not satisfied. For example, suppose in some
    neighborhood of the origin, that \(\ell(x,u) \defas |x|^2+|u|\),
    \(\kappa_N(x)\defas -x\), and \((f,\ell,V_f)\) are Lipschitz on compact
    sets. Then \(\gamma_V(s,t) \defas Lst\), \(\alpha_1(s) \defas s^2\), and
    \(\alpha_3(s) \defas s^2 + s\) satisfy
    \cref{eq:mismatch,eq:bound:posdef,eq:lyap:b} for some \(L>0\). While
    \(\limsup_{s\rightarrow 0^+} \gamma_V(s,t)/\alpha_1(s)=\infty\)
    for each \(t>0\), we have \(\limsup_{s\rightarrow 0^+}
    \gamma_V(s,t)/\alpha_3(s) = Lt\) and therefore \cref{eq:mpc:mismatch:posdef}
    holds for any \(\tau\in[0,1/L)\).
  \end{remark}
}{}

\begin{remark}\label{rem:lyap}
  To achieve \Cref{assum:lyap}(a), it is necessary to have \(V(x)=\infty\) for
  all \(x\not\in\mathcal{X}_N\). Under
  \Cref{assum:cont,assum:cons,assum:stabilizability,assum:posdef}, this is
  automatically achieved by the optimal value function \(V_N^0\), since,
  according to the convention of~\cite{rockafellar:wets:1998}, we have
  \(V_N^0(x)=\infty\) for infeasible problems.
\end{remark}

\ifthenelse{\boolean{LongVersion}}{%
  \begin{remark}
    A restricted version of \Cref{assum:steady-state} is automatically satisfied
    under \Cref{assum:lyap}(b). To see this, we set \(x=0\) in \cref{eq:lyap} to
    give \(f_c(0,\theta) = f(0,\kappa_N(0),\theta)=0\) for all
    \(|\theta|\leq\delta\) and some \(\delta>0\). If, additionally,
    \Cref{assum:cont,assum:cons,assum:posdef} are satisfied, we have
    \[
      \tilde\alpha_1(|(x,\kappa_N(x))|) \leq \tilde\alpha_1(|(x,\kappa_N(x))|)
      \leq V_N^0(x) \leq \tilde\alpha_2(|x|)
    \]
    for some \(\tilde\alpha_1,\tilde\alpha_2\in\calKinf\), which implies
    \(\kappa_N(0)=0\), so \(f(0,0,\theta)=0\) for all \(|\theta|\leq\delta\).
  \end{remark}
}{}

\subsubsection{Exponential case}
To achieve strong \emph{exponential} stability, \Cref{assum:lyap} is
strengthened to require power law versions of the bounds in \cref{eq:lyap}.
Since identical exponents are required, the scaling condition
\Cref{assum:lyap}(c) is automatically satisfied.
\begin{assumption}
  \label{assum:lyap:exp}%
  There exists a l.s.c.~function \(V:\real^n\rightarrow\nnegereal\) such that,
  for each \(\rho>0\), there exist \(\delta_0,a_1,a_2,a_3,b>0\) and
  \(\sigma_V\in\calKinf\) satisfying the following:
  \begin{enumerate}[(a)]
  \item \(\mathcal{S}\defas \textnormal{lev}_\rho V \subseteq \mathcal{X}_N\);
    and
  \item for each \(x\in\mathcal{S}\) and \(|\theta|\leq\delta_0\), we have
  \begin{subequations}\label{eq:lyap:exp}
    \begin{align}
      a_1|x|^b &\leq V(x) \leq a_2|x|^b \label{eq:lyap:exp:a} \\
      V(f_c(x,\theta)) &\leq V(x) - a_3|x|^b + \sigma_V(|\theta|)|x|^b.
                         \label{eq:lyap:exp:b}
    \end{align}
  \end{subequations}
  \end{enumerate}
\end{assumption}

With \Cref{assum:lyap:exp}, we have our second main result.
\begin{theorem}\label{thm:mpc:mismatch:exp}
  Suppose \Cref{assum:lyap:exp} holds with \(V:\real^n\rightarrow\nnegereal\).
  For each \(\rho>0\), there exists \(\delta>0\) for which
  \(x^+=f_c(x,\theta),|\theta|\leq\delta\) is SES in the RPI set
  \(\mathcal{S}\defas \textnormal{lev}_\rho V\).
\end{theorem}
\begin{proof}
  \Cref{assum:lyap:exp} gives \(\delta_0,a_1,a_2,a_3,b>0\) such that
  \(\mathcal{S}\subseteq\mathcal{X}_N\) and \cref{eq:lyap:exp} holds for each
  \(x\in\mathcal{S}\) and \(|\theta|\leq\delta_0\). Let
  \(\delta_1\in(0,\sigma_V^{-1}(a_3))\) and
  \(\delta\defas\min\set{\delta_0,\delta_1}>0\). Then, by \cref{eq:lyap:exp:b},
  \[
    V(f_c(x,\theta)) - V(x) \leq -[a_3 - \sigma_V(\delta)]|x|^b = -a_4|x|^b
  \]
  for all \(x\in\mathcal{S}\) and \(|\theta|\leq\delta\), where \(a_4\defas a_3
  - \sigma_V(\delta) \geq a_3 - \sigma_V(\delta_1) > 0\). But this means that
  \(V(x)\leq\rho\) implies
  \[
    V(f_c(x,\theta)) \leq V(x) - a_4|x|^b \leq \rho
  \]
  so \(\mathcal{S}=\textnormal{lev}_\rho V\) must be RPI.\@ Finally,
  \(x^+=f_c(x,\theta),|\theta|\leq\delta\) is SES in \(\mathcal{S}\) by
  \Cref{thm:lyap:strong}.
\end{proof}

\ifthenelse{\boolean{LongVersion}}{%
  \begin{remark}
    \Cref{rem:lyap} also applies to \Cref{assum:lyap:exp}(a): we require
    \(V(x)=\infty\) for all \(x\not\in\mathcal{X}_N\).
  \end{remark}

  \begin{remark}
    A restricted version of \Cref{assum:steady-state} is automatically satisfied
    under \Cref{assum:lyap:exp}(b). Setting \(x=0\) in \cref{eq:lyap:exp} gives
    \(f_c(0,\theta) = f(0,\kappa_N(0),\theta)=0\) for all \(|\theta|\leq\delta\)
    and some \(\delta>0\). If, additionally,
    \Cref{assum:cont,assum:cons,assum:quad} are satisfied, we have
    \[
      c_1|(x,\kappa_N(x))|^2 \leq c_1|(x,\kappa_N(x))|^2 \leq V_N^0(x) \leq
      c_2|x|^2
    \]
    for some \(c_1,c_2>0\), which implies \(\kappa_N(0)=0\), so
    \(f(0,0,\theta)=0\) for all \(|\theta|\leq\delta\).
  \end{remark}
}{}


\subsection{Error bounds}\label{ssec:mpc:strong:err}
While the maximum Lyapunov increases \cref{eq:lyap:b,eq:lyap:exp:b} are
difficult to verify directly, they are in fact satisfied for the optimal value
function (i.e., \(V\defas V_N^0\)) under fairly general conditions, as we show
in \Cref{ssec:mpc:strong}. First, however, we require bounds on the error due to
mismatch.

\subsubsection{Model error bounds}
Stability of MPC under mismatch was first investigated
by~\cite{santos:biegler:1999,santos:biegler:castro:2008}, who considered, for a
fixed parameter \(\theta\in\real^{n_\theta}\), the following power law bound:
\begin{equation}\label{eq:mismatch:santos}
  |f(x,u,\theta) - \hat f(x,u)| \leq c|x|
\end{equation}
for some \(c>0\) and all \((x,u)\in\real^n\times\real^m\). However, the bound
\cref{eq:mismatch:santos} does not account for changing or unknown
\(\theta\in\real^{n_\theta}\) and is uniform in \(u\in\real^m\), thus ruling out
the motivating example from \Cref{ssec:example:scalar}. To handle the former
issue, we can take \(c=\sigma_f(|\theta|)\) for some \(\sigma_f\in\calKinf\).
For the latter issue, it suffices to either replace \(|x|\) with \(|(x,u)|\),
i.e.,
\begin{equation}\label{eq:mismatch:exp:f}
  |f(x,u,\theta) - \hat f(x,u)| \leq \sigma_f(|\theta|)|(x,u)|
\end{equation}
or consider a bound on the closed-loop error, i.e.,
\begin{equation}\label{eq:mismatch:exp:fc}
  |f_c(x,\theta) - \hat f_c(x)| \leq \tilde\sigma_f(|\theta|)|x|
\end{equation}
for all \(x\in\mathcal{S}\), \(u\in\mathbb{U}\), and
\(\theta\in\real^{n_\theta}\), where \(\sigma_f,\tilde\sigma_f\in\calKinf\) and
\(\mathcal{S}\subseteq\real^n\) is an appropriately chosen compact set.

In the following propositions, we derive the bounds
\cref{eq:mismatch:exp:f,eq:mismatch:exp:fc} using Taylor's theorem and
\Cref{assum:cont,assum:cons,assum:stabilizability,assum:quad,%
  assum:steady-state,assum:diff} (see
\Cref{app:mismatch:exp:f,app:mismatch:exp:fc} for proofs).
\begin{proposition}\label{prop:mismatch:exp:f}
  Suppose \Cref{assum:cont,assum:cons,assum:steady-state,assum:diff} hold. For
  each compact set \(\mathcal{S}\subseteq\real^n\), there exists
  \(\sigma_f\in\calKinf\) such that \cref{eq:mismatch:exp:f} holds for all
  \(x\in\mathcal{S}\), \(u\in\mathbb{U}\), and \(\theta\in\real^{n_\theta}\).
\end{proposition}
\begin{proposition}\label{prop:mismatch:exp:fc}
  Suppose \Cref{assum:cont,assum:cons,assum:stabilizability,assum:quad,%
    assum:steady-state,assum:diff} hold. For each compact set
  \(\mathcal{S}\subseteq\mathcal{X}_N\), there exists
  \(\tilde\sigma_f\in\calKinf\) such that \cref{eq:mismatch:exp:fc} holds for
  all \(x\in\mathcal{S}\) and \(\theta\in\real^{n_\theta}\).
\end{proposition}

More generally, we could consider \(\calK^2\)-function bounds,
\begin{align}
  |f(x,u,\theta)-\hat f(x,u)|
  &\leq \gamma_f(|(x,u)|,|\theta|) \label{eq:mismatch:f} \\
  |f_c(x,\theta)-\hat f_c(x)|
  &\leq \tilde\gamma_f(|x|,|\theta|) \label{eq:mismatch:fc}
\end{align}
for all \(x\in\mathcal{S}\) and \(\theta\in\Theta\), where
\(\gamma_f,\tilde\gamma_f\in\calK^2\), and \(\mathcal{S}\subseteq\real^n\) and
\(\Theta\subseteq\real^{n_\theta}\) are appropriately chosen compact sets. In
the following propositions, we derive the bounds
\cref{eq:mismatch:f,eq:mismatch:fc} using
\Cref{assum:cont,assum:cons,assum:stabilizability,assum:quad,%
  assum:steady-state} (see \Cref{app:mismatch:f,app:mismatch:fc} for proofs).
\begin{proposition}\label{prop:mismatch:f}
  Suppose \Cref{assum:cont,assum:cons,assum:steady-state} hold. For any compact
  sets \(\mathcal{S}\subseteq\real^n\) and \(\Theta\subseteq\real^{n_\theta}\),
  there exists \(\gamma_f\in\calK^2\) satisfying \cref{eq:mismatch:f} for all
  \(x\in\mathcal{S}\), \(u\in\mathbb{U}\), and \(\theta\in\Theta\).
\end{proposition}
\begin{proposition}\label{prop:mismatch:fc}
  Suppose \Cref{assum:cont,assum:cons,assum:stabilizability,assum:posdef,%
    assum:steady-state} hold. For any compact sets
  \(\mathcal{S}\subseteq\mathcal{X}_N\) and \(\Theta\subseteq\real^{n_\theta}\),
  there exists \(\tilde\gamma_f\in\calK^2\) satisfying \cref{eq:mismatch:fc} for
  all \(x\in\mathcal{S}\) and \(\theta\in\Theta\).
\end{proposition}

\subsubsection{Suboptimal cost error bounds}
Ultimately, we require a maximum Lyapunov increase of the form \cref{eq:lyap:b}
or \cref{eq:lyap:exp:b}. The robust descent property
\cref{eq:mpc:descent:robust} suggests a path through imposing an error bound on
the suboptimal cost function \(V_N(f_c(x,\theta),\tilde{\mathbf{u}}(x))\), i.e.,
\begin{equation}\label{eq:mismatch:exp}
  |V_N(f_c(x,\theta),\tilde{\mathbf{u}}(x)) -
  V_N(\hat f_c(x),\tilde{\mathbf{u}}(x))| \leq \sigma_V(|\theta|)|x|^2
\end{equation}
where \(\sigma_V\in\calKinf\). In \Cref{prop:mismatch:exp}, we establish
\cref{eq:mismatch:exp} under \Cref{assum:cont,assum:cons,assum:stabilizability,%
  assum:quad,assum:steady-state,assum:diff} (see \Cref{app:mismatch:exp} for
proof).
\begin{proposition}\label{prop:mismatch:exp}
  Suppose \Cref{assum:cont,assum:cons,assum:stabilizability,assum:quad,%
    assum:steady-state,assum:diff} hold and let
  \(\mathcal{S}\subseteq\mathcal{X}_N\) be compact. Then there exists
  \(\sigma_V\in\calKinf\) such that \cref{eq:mismatch:exp} holds for all
  \(x\in\mathcal{S}\) and \(\theta\in\real^{n_\theta}\).
\end{proposition}

Similarly, we can derive a \(\calK^2\)-function version of
\cref{eq:mismatch:exp} under
\Cref{assum:cont,assum:cons,assum:stabilizability,assum:posdef,%
  assum:steady-state} (see \Cref{app:mismatch:VN} for proof).
\begin{proposition}\label{prop:mismatch}
  Suppose \Cref{assum:cont,assum:cons,assum:stabilizability,assum:posdef,%
    assum:steady-state} hold. Let \(\mathcal{S}\subseteq\mathcal{X}_N\) and
  \(\Theta\subseteq\real^{n_\theta}\) be compact. Then there exists
  \(\gamma_V\in\calK^2\) such that, for each \(x\in\mathcal{S}\) and
  \(\theta\in\Theta\),
  \begin{equation}\label{eq:mismatch}
    |V_N(f_c(x,\theta),\tilde{\mathbf{u}}(x)) -
    V_N(\hat f_c(x),\tilde{\mathbf{u}}(x))| \leq \gamma_V(|x|,|\theta|).
  \end{equation}
\end{proposition}

\subsection{Stability despite mismatch}\label{ssec:mpc:strong}
\subsubsection{General costs}
Finally, we are in a position to construct a maximum Lyapunov increase
\cref{eq:lyap:b} or \cref{eq:lyap:exp:b}. For general costs, this is
accomplished in the following proposition.

\begin{proposition}\label{prop:lyap}
  Suppose \Cref{assum:cont,assum:cons,assum:stabilizability,assum:posdef,%
    assum:steady-state} hold. Then \Cref{assum:lyap}(a,b) hold with \(V\defas
  V_N^0\).
\end{proposition}
\begin{proof}
  Let \(\rho>0\), \(\mathcal{S}\defas\textnormal{lev}_\rho V_N^0\), and
  \(V\defas V_N^0\). Then \(\mathcal{S}\subseteq\mathcal{X}_N\) trivially.
  Since \(V_N^0\) is l.s.c.~\cite[Lem.~7.18]{bertsekas:shreve:1978},
  \(\mathcal{S}\) is closed. By \Cref{thm:mpc:robust}, there exists
  \(\alpha_2\in\calKinf\) satisfying \cref{eq:lyap:a} for all
  \(x\in\mathcal{S}\). Then
  \(|x|\leq\alpha_1^{-1}(V(x))\leq\alpha_1^{-1}(\rho)\) for all
  \(x\in\mathcal{S}\), so \(\mathcal{S}\) is compact.

  By \Cref{prop:mpc:rpi}, there exists \(\delta_0>0\) such that \(\mathcal{S}\)
  is RPI for \(x^+=f_c(x,\theta),|\theta|\leq\delta_0\) and
  \cref{eq:mpc:descent:robust} holds for all \(x\in\mathcal{S}\) and
  \(|\theta|\leq\delta_0\). Moreover, for each \(x\in\mathcal{S}\) and
  \(|\theta|\leq\delta_0\), \cref{eq:mismatch} holds for some
  \(\gamma_V\in\calK^2\) by \Cref{prop:mismatch}. Finally, combining
  \cref{eq:bound:posdef,eq:mpc:descent:robust,eq:mismatch} gives
  \cref{eq:lyap:b} with \(\alpha_3\defas\alpha_1\).
\end{proof}

\Cref{assum:lyap}(a,b) alone do not guarantee strong stability. However, we can
strengthen the hypothesis of \Cref{prop:lyap} with a scaling requirement to
guarantee strong stability.

\begin{theorem}\label{cor:mpc:mismatch}
  Suppose \Cref{assum:cont,assum:cons,assum:stabilizability,assum:posdef,%
    assum:steady-state} hold. Let \(\rho>0\) and
  \(\mathcal{S}\defas\textnormal{lev}_\rho V_N^0\). Then \cref{eq:lyap} holds
  for all \(x\in\mathcal{S}\) and \(|\theta|\leq\delta_0\) with \(V\defas
  V_N^0\) and some \(\delta_0>0\), \(\alpha_1,\alpha_2,\alpha_3\in\calKinf\),
  and \(\gamma_V\in\calK^2\). If, additionally, there exists \(\tau>0\)
  satisfying \cref{eq:mpc:mismatch:posdef}, then there exists \(\delta>0\) such
  that \(x^+=f_c(x,\theta),|\theta|\leq\delta\) is SAS in the RPI set
  \(\mathcal{S}\).
\end{theorem}
\begin{proof}
  The first part follows from \Cref{prop:lyap}, and the second part follows from
  \Cref{thm:mpc:mismatch}.
\end{proof}

\subsubsection{Quadratic costs}
For quadratic costs, we construct \cref{eq:lyap:exp:b} in the following
proposition.

\begin{proposition}\label{prop:lyap:exp}
  Suppose \Cref{assum:cont,assum:cons,assum:stabilizability,assum:quad,%
    assum:steady-state,assum:diff} hold. Then \Cref{assum:lyap:exp} holds with
  \(b\defas 2\) and \(V\defas V_N^0\).
\end{proposition}
\begin{proof}
  Let \(\rho>0\), \(V\defas V_N^0\), and
  \(\mathcal{S}\defas\textnormal{lev}_\rho V\). Since \Cref{assum:quad} implies
  \Cref{assum:posdef}, we have from the first paragraph of the proof of
  \Cref{prop:lyap} that \(\mathcal{S}\) is compact.

  \Cref{thm:mpc:robust:exp} also implies \cref{eq:lyap:exp:a} holds for all
  \(x\in\mathcal{S}\), with \(a_1,a_2>0\) and \(b\defas 2\). By
  \Cref{prop:mpc:rpi}, there exists \(\delta_0>0\) such that \(\mathcal{S}\) is
  RPI for \(x^+=f_c(x,\theta),|\theta|\leq\delta_0\) and
  \cref{eq:mpc:descent:robust} holds for all \(x\in\mathcal{S}\) and
  \(|\theta|\leq\delta_0\). Moreover, for each \(x\in\mathcal{S}\) and
  \(|\theta|\leq\delta_0\), \cref{eq:mismatch:exp} holds for some
  \(\sigma_V\in\calKinf\) by \Cref{prop:mismatch}, and combining
  \cref{eq:mpc:descent:robust,eq:mismatch} gives \cref{eq:lyap:exp:b}.
\end{proof}

Our third and final main result follows immediately from
\Cref{thm:mpc:mismatch,prop:lyap:exp}.
\begin{theorem}\label{cor:mpc:mismatch:exp}
  Suppose \Cref{assum:cont,assum:cons,assum:stabilizability,assum:quad,%
    assum:steady-state,assum:diff} holds. For each \(\rho>0\), there exists
  \(\delta>0\) for which \(x^+=f_c(x,\theta),|\theta|\leq\delta\) is SES in the
  RPI set \(\mathcal{S}\defas \textnormal{lev}_\rho V_N^0\).
\end{theorem}
  \begin{proof}
    By \Cref{prop:lyap:exp}, \Cref{assum:lyap:exp} holds with \(V\defas V_N^0\),
    and by \Cref{thm:mpc:mismatch:exp}, there exists \(\delta>0\) for which
    \(\mathcal{S}\) is RPI and \(x^+=f_c(x,\theta),|\theta|\leq\delta\) is SES
    in \(\mathcal{S}\).
  \end{proof}

\section{Examples}\label{sec:examples}
In this section, we illustrate the results through several examples. First, we
consider a non-differentiable system that satisfies \Cref{assum:lyap}(a,b) but
not \Cref{assum:lyap}(c), and is not SAS.\@ %
\ifthenelse{\boolean{LongVersion}}{%
  Second, we consider a non-differentiable example that nonetheless satisfies
  \Cref{assum:lyap:exp} and is therefore SES.\@ %
}{}%
Finally, we consider the inverted pendulum system to showcase how the nominal
MPC handles different types of mismatch. Notably, we consider (i) discretization
errors, (ii) unmodeled dynamics, and (iii) incorrectly estimated input gains. %
\ifthenelse{\boolean{LongVersion}}{}{%
  See~\cite{kuntz:rawlings:2024d} for an example of a non-differentiable system
  that nonetheless satisfies \Cref{assum:lyap:exp} and is therefore SES.\@ %
}%

\subsection{Strong asymptotic stability counterexample}\label{ssec:example:sqrt}
\begin{figure}
  \centering
  \ifthenelse{\boolean{OneColumn}}{%
    \includegraphics[width=0.6\linewidth,page=2]{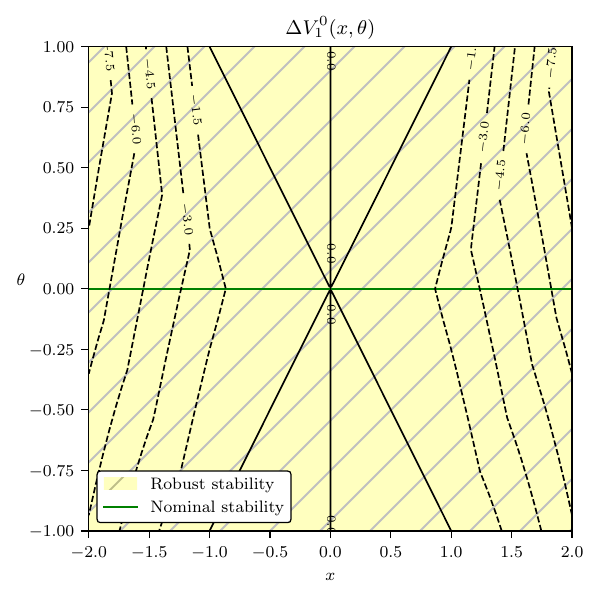}
  }{%
    \includegraphics[width=\linewidth,page=2]{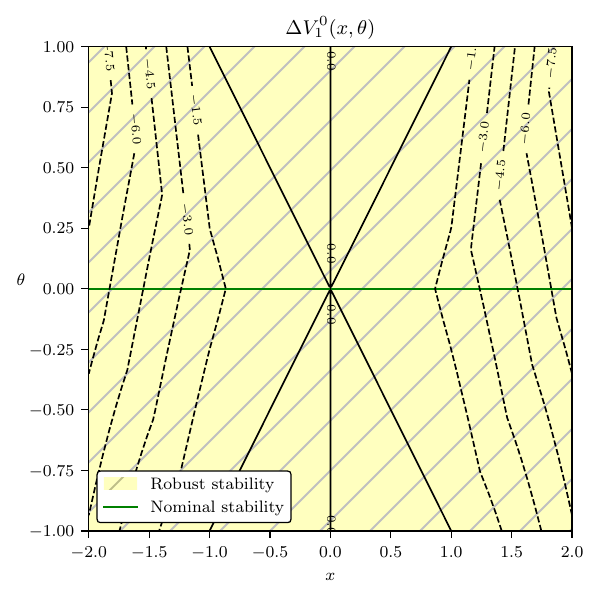}
  }%
  \caption{Contours of the cost difference for the MPC of \cref{eq:plant:sqrt}.}%
  \label{fig:sqrt:contour}
\end{figure}

\begin{figure}
  \centering
  \ifthenelse{\boolean{OneColumn}}{%
    \includegraphics[width=0.6\linewidth,page=3]{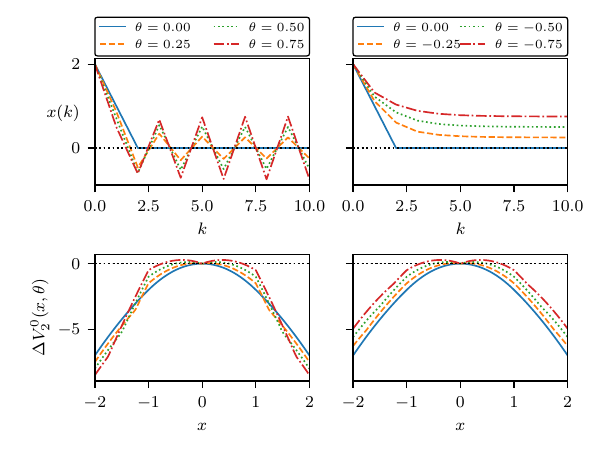}
  }{%
    \includegraphics[width=\linewidth,page=3]{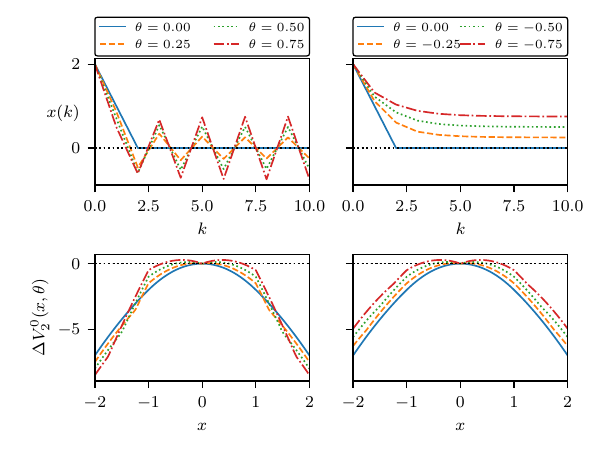}
  }%
  \caption{For (left) nonnegative and (right) nonpositive values of \(\theta\),
    the (top) closed-loop trajectories for the MPC of \cref{eq:plant:sqrt} with
    initial state \(x=2\), and (bottom) cost differences of the same MPC as a
    function of \(x\).}%
  \label{fig:sqrt:traj}
\end{figure}

Consider the scalar system
\begin{equation}\label{eq:plant:sqrt}
  x^+ = f(x,u,\theta) \defas \sigma(x + (1+\theta)u)
\end{equation}
where \(\sigma\) is the \emph{signed square root} defined as
\(\sigma(y)\defas\textnormal{sgn}(y)\sqrt{|y|}\) for each \(y\in\real\). We
define a nominal MPC with \(\mathbb{U}\defas[-1,1]\), \(\ell(x,u)\defas
x^2+u^2\), \(V_f(x)\defas 4x^2\), \(\mathbb{X}_f\defas[-1,1]\), and
\(N\defas 1\).


In \Cref{app:example:sqrt}, it is shown the closed-loop system
\(x^+=f(x,\kappa_1(x),\theta),|\theta|\leq 3\) is RES on
\(\mathcal{X}_1=[-2,2]\) with the nominal control law
\(\kappa_1(x)\defas-\textnormal{sat}(x)\). Additionally, it is shown
\Cref{assum:lyap}(a,b) is satisfied with \(V\defas V_1^0\), and \cref{eq:lyap:b}
holds for all \(x\in\mathcal{S} \defas \textnormal{lev}_2 V_1^0 = [-1,1]\) and
\(|\theta|\leq \delta_0\defas 3\) with \(\alpha_3(s)\defas 2s^2\), and
\(\gamma_V(s,t)\defas st+4\sqrt{st}\). But this implies \(\lim_{s\rightarrow
  0^+} \gamma_V(s,t)/\alpha_3(s) = \infty\) for each \(t>0\), so
\Cref{assum:lyap}(c) is not satisfied.

However, \Cref{assum:lyap} is only sufficient, not necessary, for establishing
strong stability. But we have \(V_1^0(x)=2x^2\) and
\begin{align*}
  \Delta V_1^0(x,\theta)
  &\defas V_1^0(f(x,\kappa_1(x),\theta)) - V_1^0(x) \\
  &= 2[\sigma(\theta x)]^2 - 2x^2 = 2(|\theta|-|x|)|x| > 0.
\end{align*}
for each \(0<|x|<|\theta|\leq 1\), so the state always gets pushed out of
\((-|\theta|,|\theta|)\) unless it starts at the origin or \(\theta=0\). In
other words, the MPC only provides inherent robustness, not strong stability,
even though \Cref{assum:lyap}(a,b) is satisfied.

In \Cref{fig:sqrt:contour}, we plot contours of the cost difference \(\Delta
V_1^0(x,\theta)\), and in \Cref{fig:sqrt:traj} we plot closed-loop trajectories
and the cost difference curve \(\Delta V_1^0(\cdot,\theta)\) for several values
of \(\theta\). Only with \(\theta=0\) does the trajectory converge to the origin
and the cost difference curve remain negative definite. For each \(\theta\neq
0\), the cost difference is positive definite near the origin, and the
trajectory does not converge to the origin.

\ifthenelse{\boolean{LongVersion}}{%
\subsection{Non-differentiable yet strongly exponential
  stable}\label{ssec:example:sin}
\begin{figure}
  \centering
  \ifthenelse{\boolean{OneColumn}}{%
    \includegraphics[width=0.6\linewidth,page=2]{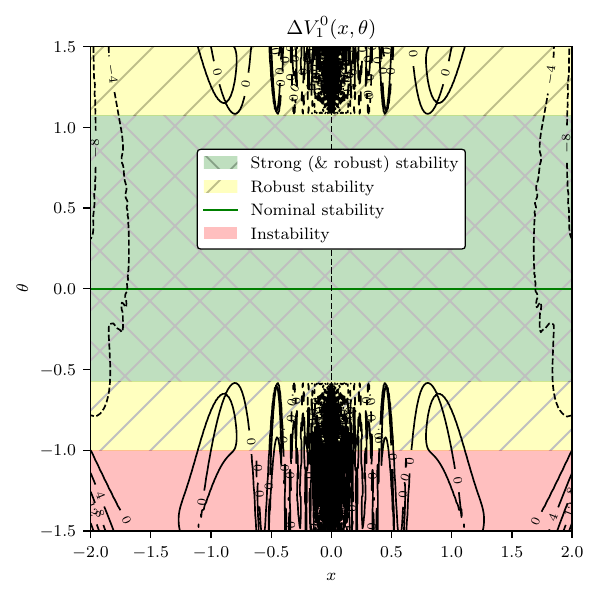}
  }{%
    \includegraphics[width=\linewidth,page=2]{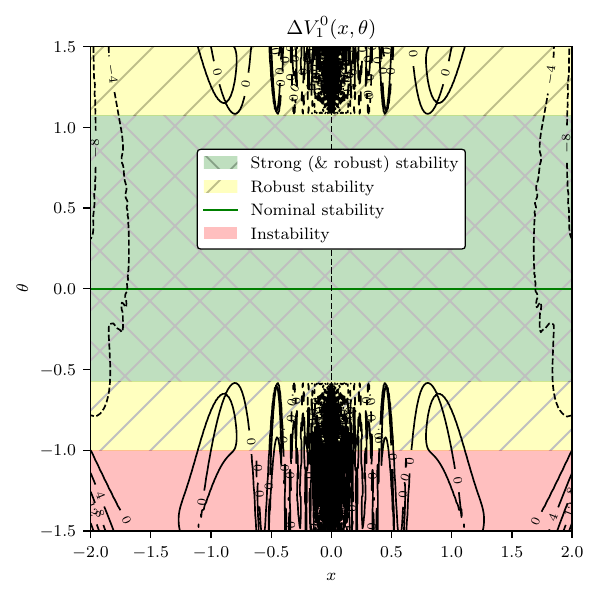}
  }%
  \caption{Contours of the cost difference for the MPC of \cref{eq:plant:sin}.}%
  \label{fig:sin:contour}
\end{figure}

\begin{figure}
  \centering
  \ifthenelse{\boolean{OneColumn}}{%
    \includegraphics[width=0.6\linewidth,page=3]{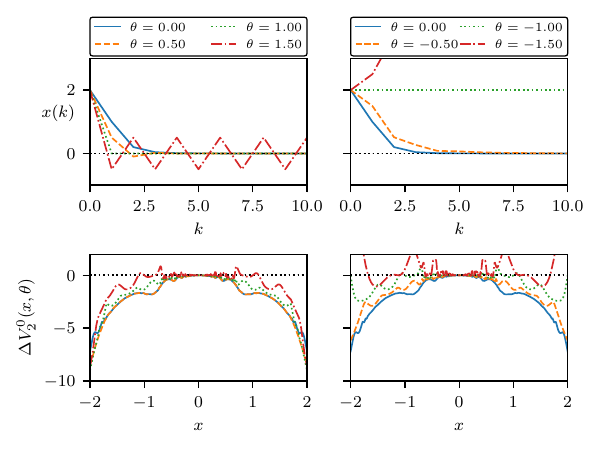}
  }{%
    \includegraphics[width=\linewidth,page=3]{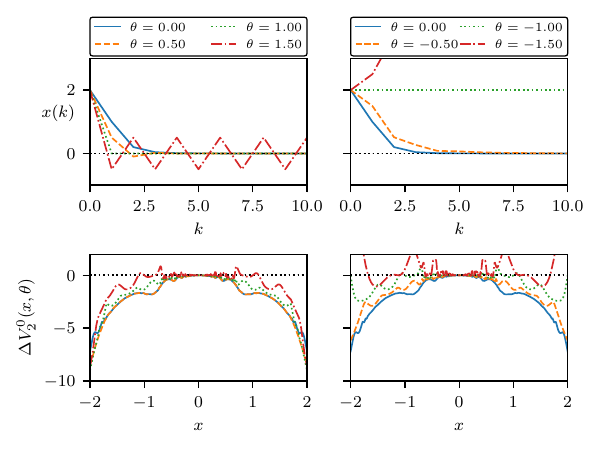}
  }%
  \caption{For (left) nonnegative and (right) nonpositive values of \(\theta\),
    the (top) closed-loop trajectories for the MPC of \cref{eq:plant:sin} with
    initial state \(x=2\), and (bottom) cost differences of the same MPC as a
    function of \(x\).}%
  \label{fig:sin:traj}
\end{figure}

Consider the scalar system
\begin{equation}\label{eq:plant:sin}
  x^+ = f(x,u,\theta) \defas x + (1/2)\gamma(x) + (1+\theta)u
\end{equation}
where \(\gamma:\real\rightarrow\real\) is defined as
\[
  \gamma(x)\defas \begin{cases} 0, & x=0, \\ |x|\sin(2\pi/x), & x\neq
    0. \end{cases}
\]
While the function \(\gamma\) is continuous, it is not differentiable at the
origin. We define a nominal MPC with \(\mathbb{U}\defas[-1,1]\),
\(\ell(x,u)\defas x^2+u^2\), \(V_f(x)\defas 4x^2\),
\(\mathbb{X}_f\defas[-1,1]\), and \(N\defas 1\).

In \Cref{app:example:sin}, we show the closed-loop system
\(x^+=f(x,\kappa_1(x),\theta),|\theta|\leq 1\) is RES on
\(\mathcal{X}_1=[-2,2]\) with the nominal control law \(\kappa_1(x)\defas
-\textnormal{sat}((4/5)x + (2/5)\gamma(x))\). Moreover, it is shown that
\Cref{assum:lyap:exp} is satisfied, and by \Cref{thm:mpc:mismatch:exp} (and its
proof), the closed-loop system
\(x^+=f(x,\kappa_1(x),\theta),|\theta|\leq\delta\defas 0.5\) is SES on
\(\mathcal{X}_1=[-2,2]\).

To establish a clearer picture of robust and strong stability for the
closed-loop, we plot in \Cref{fig:sin:contour} contours of the cost difference
\(\Delta V_1^0(x,\theta) \defas V_1^0(f(x,\kappa_1(x),\theta)) - V_1^0(x)\), and
in \Cref{fig:sin:traj} closed-loop trajectories and the cost difference curve
\(\Delta V_1^0(\cdot,\theta)\) for several values of \(\theta\). For \(\theta\)
between \(\theta_0\approx 0.57\) and \(\theta_1\approx 1.08\), the closed-loop
system is strongly stable, with trajectories converging to the origin, and a
negative definite cost difference curve. Outside of this range but with
\(\theta\in[-1,1.5]\), the closed-loop system is still robustly stable, with a
cost difference curve of ambiguous sign but trajectories converging to a
neighborhood of the origin. Finally, for \(\theta<-1\), trajectories are
unbounded because \(\mathcal{X}_1\) is not RPI.\@
}{}

\subsection{Upright pendulum}\label{ssec:example:pendulum}

\begin{figure*}
  \centering
  \ifthenelse{\boolean{LongVersion}}{%
    \subfloat[Unmodeled dynamics\label{fig:pendulum:traj:a}]{%
      \includegraphics[width=0.49\linewidth,page=1]{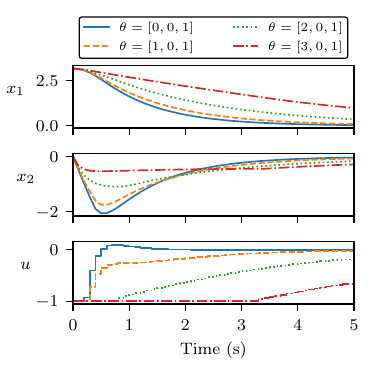}%
    }\hfill
    \subfloat[Underestimated gain\label{fig:pendulum:traj:b}]{%
      \includegraphics[width=0.49\linewidth,page=2]{pendulum_example.pdf}%
    }

    \subfloat[Overestimated gain\label{fig:pendulum:traj:c}]{%
      \includegraphics[width=0.49\linewidth,page=3]{pendulum_example.pdf}%
    }\hfill
    \subfloat[Breakdown cases\label{fig:pendulum:traj:d}]{%
      \includegraphics[width=0.49\linewidth,page=4]{pendulum_example.pdf}%
    } %
  }{%
    \subfloat[Unmodeled dynamics\label{fig:pendulum:traj:a}]{%
      \includegraphics[width=0.33\linewidth,page=1]{pendulum_example.pdf}%
    }\hfill
    \subfloat[Underestimated gain\label{fig:pendulum:traj:b}]{%
      \includegraphics[width=0.33\linewidth,page=2]{pendulum_example.pdf}%
    }\hfill
    \subfloat[Overestimated gain\label{fig:pendulum:traj:c}]{%
      \includegraphics[width=0.33\linewidth,page=3]{pendulum_example.pdf}%
    } %
  }%
  \caption{Simulated closed-loop trajectories for the MPC of \cref{eq:pendulum}
    from the resting position \(x(0)=(\pi,0)\) to the upright position
    \(x_s=(0,0)\) for various values of \((\theta_1,\theta_2)\in\real^2\).}%
  \label{fig:pendulum:traj}
\end{figure*}

Consider the nondimensionalized pendulum system
\begin{equation}\label{eq:pendulum}
  \dot x = F(x,u,\theta) \defas
  \begin{bmatrix} x_2 \\ \sin x_1 - \theta_1^2x_2 + (\hat k+\theta_2)u \end{bmatrix}
\end{equation}
where \(x_1,x_2\in\real\) are the angle and angular velocity, \(u\in[-1,1]\) is
the (signed and normalized) motor voltage, \(\theta_1\in\real\) is an air
resistance factor, \(\hat k>0\) is the estimated gain of the motor, and
\(\theta_2\in\real\) is the error in the motor gain estimate. Let
\(\psi(t;x,u,\theta)\) denote the solution to the differential equation
\cref{eq:pendulum} at time \(t\geq 0\) given an initial condition \(x(0)=x\),
constant input signal \(u(t)=u\), and parameters \(\theta\). We model the
continuous-time system \cref{eq:pendulum} as
\begin{equation}\label{eq:plant:pendulum}
  x^+ = f(x,u,\theta) \defas x + \Delta F(x,u,\theta) + \theta_3 r(x,u,\theta)
\end{equation}
where \(r\) is a residual function given by
\[
  r(x,u,\theta) \defas \int_0^\Delta [F(\psi(t;x,u,\theta),u,\theta) -
  F(x,u,\theta)]dt.
\]
Assuming a zero-order hold on the input \(u\), the system \cref{eq:pendulum} is
discretized (exactly) as \cref{eq:plant:pendulum} with \(\theta_3=1\). Since we
model the system with \(\theta=0\) as
\begin{equation}\label{eq:model:pendulum}
  x^+ = \hat f(x,u) \defas f(x,u,0) = x +
  \Delta\begin{bmatrix} x_2 \\ \sin x_1 + \hat ku \end{bmatrix}
\end{equation}
we do not need access to \(r\) to design the nominal MPC.\@

For the following simulations, let the model gain be \(\hat k=5\textnormal{ rad}
/ \textnormal{s}^2\), the sample time be \(\Delta = 0.1 \textnormal{ s}\), and
define a nominal MPC with \(N\defas 20\), \(\mathbb{U}\defas[-1,1]\),
\(\ell(x,u)\defas |x|^2+u^2\), \(V_f(x)\defas |x|_{P_f}^2\),
\(\mathbb{X}_f\defas\textnormal{lev}_{c_f} V_f\), and
\(c_f\defas\underline{\sigma}(P_f)/8\), where \(P_f = %
\begin{bsmallmatrix} 31.133\ldots & 10.196\ldots \\
  10.196\ldots & 10.311\ldots \end{bsmallmatrix}\) is shown, in %
\Cref{app:example:pendulum}, to satisfy \Cref{assum:stabilizability} with the
terminal law \(\kappa_f(x)\defas -2x_1-2x_2\).
\Cref{assum:cont,assum:cons,assum:quad,assum:steady-state} are satisfied
trivially, and \Cref{assum:diff} is satisfied since continuous differentiability
of \(F\) implies continuous differentiability of \(\psi\) (and therefore also
\(r\) and \(f\))~\cite[Thm.~3.3]{hale:1980}. Thus, the conclusion of
\Cref{cor:mpc:mismatch:exp} holds for some \(\delta>0\), and if we can take
\(\delta>1\), the nominal MPC is inherently strongly stabilizing with
\((\theta_1,\theta_2)\in\real^2\) sufficiently small.

In \Cref{fig:pendulum:traj}, we simulate the closed-loop system
\(x^+=f(x,\kappa_{20}(x),\theta)\) for some fixed
\((\theta_1,\theta_2,1)\in\real^3\). Note that all of these simulations include
discretization errors. \Cref{fig:pendulum:traj:a} showcases the ability of MPC
to handle unmodeled dynamics (i.e., a missing air resistance term). In
\Cref{fig:pendulum:traj:b}, the gain of the motor is increased until the nominal
controller is severely underdamped. In \Cref{fig:pendulum:traj:c}, the gain of
the motor is decreased until the motor cannot overcome the force of gravity and
strong stability is not achieved. \ifthenelse{\boolean{LongVersion}}{%
  In \Cref{fig:pendulum:traj:d}, we plot cases where the errors as made so
  extreme as to prevent stability. %
}{}

\section{Conclusion}\label{sec:conclusion}
We establish conditions under which MPC is strongly stabilizing despite
plant-model mismatch in the form of parameter errors. Namely, it suffices to
assume the existence of a Lyapunov function with a maximum increase, suitably
bounded level sets, and a scaling condition (\Cref{assum:lyap,assum:lyap:exp}).
While we are not able to show the assumptions hold in general, when the MPC has
quadratic costs it is possible to show that continuous differentiability of the
dynamics implies strong stability (\Cref{thm:mpc:mismatch:exp}). When the
\(\calK^2\)-function bound is not properly scaled, the MPC may not be
stabilizing, as illustrated in the examples. In this sense, while MPC is not
inherently stabilizing under mismatch \emph{in general}, there is a common class
of cost functions (quadratic costs) for which nominal MPC is inherently
stabilizing under mismatch.

Several questions about the strong stability of MPC remain unanswered.
While quadratic costs are used in many control problems, it may be possible to
generalize \Cref{cor:mpc:mismatch:exp} to other useful classes of stage costs,
such as \(q\)-norm costs,
or costs with exact penalty functions for soft state constraints
(cf.~\cite{santos:biegler:castro:2008,kuntz:rawlings:2024f}). We propose the
direct approach to strong exponential stability
(\Cref{assum:lyap:exp,thm:mpc:mismatch:exp}) provides a path to generalizing
\Cref{cor:mpc:mismatch:exp} to other classes of stage costs, output feedback, or
semidefinite costs.
Nonlinear MPC is computationally difficult to implement online. Therefore it
would be worth extending this work to include the suboptimal MPC algorithm from
\cite{allan:bates:risbeck:rawlings:2017} using the approach therein.
Finally, while our analysis only considers discrete-time MPC and an effective
zero-order hold on the parameter variations, the continuous-time extensions
(both continuous-time MPC are discrete-time MPC with continuously varying
\(\theta\)) are worthy pursuits.

While systems with fixed and known setpoints are a useful and interesting class
of problems, many systems have parameter-dependent, time-varying setpoints. To
track setpoints under mismatch, offset-free MPC can be used. Theory on nonlinear
offset-free MPC is limited, typically relying on stability of the closed-loop
system to guarantee offset-free
performance~\citep{pannocchia:gabiccini:artoni:2015}. In
\cite{kuntz:rawlings:2024f}, we extend the theory in this
\ifthenelse{\boolean{LongVersion}}{report}{article} to consider offset-free MPC
with quadratic costs, and we establish sufficient conditions for closed-loop
stability and guaranteed offset-free performance for a general class of
differentiable systems subject to time-varying setpoints, persistent
disturbances, and plant-model mismatch. However, many issues in offset-free MPC
theory, including necessary conditions for offset-free performance, nonquadratic
costs, and nondifferentiable systems, are not yet well understood.

\appendix
\ifthenelse{\boolean{LongVersion}}{%
\section{Lyapunov proofs}\label[appendix]{app:lyap}
In this appendix, we prove some of the Lyapunov results of
\Cref{sec:robust:strong}.

\subsection{Proof of \cref{thm:lyap:robust}}\label[appendix]{app:lyap:robust}
Suppose \(X\) is RPI for \cref{eq:plant:cl}. Let the functions
\(V:X\rightarrow\nnegreal\), \(\alpha_i\in\calKinf,i\in\intinterval{1}{3}\), and
\(\sigma\in\calK\) satisfy \cref{eq:lyap:iss} for all \(x\in X\) and
\(\theta\in\Theta\).

\paragraph{Asymptotic case}
The proof of the asymptotic case follows similarly to Lemma~3.5 of
\cite{jiang:wang:2001}. The main difference is the restriction of the state to
an RPI set \(X\) for all \(k\in\nnegint\). First, we rewrite
\cref{eq:lyap:iss:b} as
\begin{equation}\label{eq:ras:1}
  V(f_c(x,\theta)) \leq (\id - \alpha_4)(V(x)) + \sigma(|\theta|)
\end{equation}
for all \(x\in X\) and \(\theta\in\Theta\), where
\(\alpha_4\defas\alpha_3\circ\alpha_2^{-1}\in\calKinf\). By Lemma B.1 of
\cite{jiang:wang:2001}, we can assume \(\id-\alpha_4\in\calK\) without loss of
generality. Let \(\rho\in\calKinf\) such that \(\id-\rho\in\calKinf\).

For the rest of this part of the proof, fix \(x\in X\) and
\(\boldsymbol{\uptheta}\in\Theta^\infty\), and, for simplicity, let \(x(k)
\defas \phi_c(k;x,\boldsymbol{\uptheta})\) for all \(k\in\nnegint\). Note that
\(x(k) \in X\) for all \(k\in\nnegint\) since \(X\) is RPI.\@ Let \(b \defas
\alpha_4^{-1}(\rho^{-1}(\sigma(\|\boldsymbol{\uptheta}\|)))\) and \(D\defas
\set{x\in X | V(x) \leq b}\). The following intermediate result is required.
\begin{proposition}
  If there exists \(k_0\in\nnegint\) such that \(x(k_0)\in D\), then \(x(k)\in
  D\) for all \(k\geq k_0\).
\end{proposition}
\begin{proof}
  Suppose \(k\geq k_0\) and \(x(k)\in D\). Then \(V(x(k))\leq b\) and
  by \cref{eq:ras:1},
  \begin{align*}
    V(x(k+1))
    &\leq (\id - \alpha_4)(V(x(k))) + \sigma(\|\boldsymbol{\uptheta}\|) \\
    &\leq (\id - \alpha_4)(b) + \sigma(\|\boldsymbol{\uptheta}\|) \\
    &= \underbrace{-(\id - \rho)(\alpha_4(b))}_{\leq 0} + b
      \underbrace{- \rho(\alpha_4(b)) + \sigma(\|\boldsymbol{\uptheta}\|)}_{=0} \leq b.
  \end{align*}
  The result follows by induction.
\end{proof}

Next, let \(j_0 \defas \min\set{ k\in\nnegint | x(k)\in D}\). The above lemma
gives \(V(x(k)) \leq \hat\gamma(\|\boldsymbol{\uptheta}\|)\) for all \(k\geq
j_0\), where \(\hat\gamma\defas\alpha_4^{-1}\circ\rho^{-1}\circ\sigma\). On the
other hand, if \(k<j_0\), then we have \(\rho(\alpha_4(V(x(k)))) >
\sigma(\|\boldsymbol{\uptheta}\|)\) and therefore
\begin{align*}
  V(x(k+1)) - V(x(k))
  &\leq -\alpha_4(V(x(k))) + \sigma(\|\boldsymbol{\uptheta}\|) \\
  &= -\alpha_4(V(x(k))) + \rho(\alpha_4(V(x(k))))
    -\rho(\alpha_4(V(x(k)))) + \sigma(\|\boldsymbol{\uptheta}\|) \\
  &\leq -\alpha_4(V(x(k))) + \rho(\alpha_4(V(x(k)))) \\
  &= -((\id - \rho)\circ\alpha_4)(V(x(k)))
\end{align*}
for all \(k\in\nnegint\). Since \((\id - \rho)\circ\alpha_4\in\calKinf\), by
Lemma 4.3 of \cite{jiang:wang:2001}, there exists \(\hat\beta\in\calKL\) such
that
\[
  \alpha_1(|x(k)|) \leq V(x(k)) \leq \hat\beta(V(x(0)),k) = \hat{\beta}(V(x),k)
  \leq \hat\beta(\alpha_2(|x|),k)
\]
for all \(k\in\nnegint\). Combining the above inequalities gives
\[
  |x(k)| \leq \max\{\beta(|x|,k), \; \gamma(\|\boldsymbol{\uptheta}\|)\} \leq
  \beta(|x|,k) + \gamma(\|\boldsymbol{\uptheta}\|)
\]
for all \(k\in\nnegint\), where \(\beta(s,k) \defas
\alpha_1^{-1}(\hat\beta(\alpha_2(s),k))\) is class-\(\calKL\), and \(\gamma
\defas \alpha_1^{-1}\circ\hat\gamma\). Finally, causality lets us drop future
terms of \(\theta\) from the signal norm in the above inequality and simply
write~\cref{eq:ras}. %

\paragraph{Exponential case}
Suppose, additionally, that \(\alpha_i \defas
a_i\id^b,i\in\intinterval{1}{3}\). %
Without loss of generality, we have \(a_3<a_2\). Then \cref{eq:lyap:iss} can be
rewritten
\begin{align}
  V(f_c(x,\theta))
  &\leq V(x) - a_3|x|^b + \sigma(|\theta|) \nonumber \\
  &\leq V(x) - \frac{a_3}{a_2}V(x) + \sigma(|\theta|) \nonumber \\
  &= \lambda_0 V(x) + \sigma(|\theta|) \label{eq:res:1}
\end{align}
for all \(x\in X\) and \(\theta\in\Theta\), where \(\lambda_0\defas
1-\frac{a_3}{a_2} \in(0,1)\). Since \(X\) is RPI, we can recursively apply the
inequality \cref{eq:res:1} to give
\begin{align*}
  V(\phi_c(k;x,\boldsymbol{\uptheta}))
  &\leq \lambda_0^k V(x) + \sum_{i=1}^k \lambda_0^{i-1}\sigma(|\theta(k-i)|) \\
  &\leq \lambda_0^k V(x) + \left( \sum_{i=1}^k \lambda_0^{i-1} \right)
    \max_{i\in\intinterval{0}{k-1}} \sigma(|\theta(i)|) \\
  &\leq \lambda_0^k V(x) + \frac{\max_{i\in\intinterval{0}{k-1}}
    \sigma(|\theta(i)|)}{1-\lambda_0} \\
  &= \lambda_0^k V(x) +
    \frac{\sigma(\|\boldsymbol{\uptheta}\|_{0:k-1})}{1-\lambda_0} \\
  &= a_2|x|^b\lambda_0^k +
    \frac{\sigma(\|\boldsymbol{\uptheta}\|_{0:k-1})}{1-\lambda_0}
\end{align*}
for all \(k\in\nnegint\), \(x\in X\), and \(\boldsymbol{\uptheta}\in\Theta^k\).
If \(b\geq 1\), then, by the triangle inequality for the \(b\)-norm, we have
\begin{align*}
  |\phi_c(k;x,\boldsymbol{\uptheta})|
  &\leq \left( \frac{V(\phi_c(k;x,\boldsymbol{\uptheta}))}{a_1} \right)^{1/b} \\
  &\leq \frac{1}{a_1^{1/b}} \left( a_2|x|^b\lambda_0^k +
    \frac{\sigma(\|\boldsymbol{\uptheta}\|_{0:k-1})}{1-\lambda_0} \right)^{1/b} \\
  &\leq \left( \frac{a_2}{a_1}  \right)^{1/b}|x|(\lambda_0^b)^k +
    \left( \frac{\sigma(\|\boldsymbol{\uptheta}\|_{0:k-1})}{a_1(1-\lambda_0)} \right)^{1/b} \\
  &\leq c|x|\lambda^k + \gamma(\|\boldsymbol{\uptheta}\|_{0:k-1})
\end{align*}
for all \(k\in\nnegint\), \(x\in X\), and \(\boldsymbol{\uptheta}\in\Theta^k\),
where \(\lambda\defas \lambda_0^{1/b} \in (0,1)\), \(c\defas\left(
  \frac{a_2}{a_1} \right)^{1/b} > 0\), and \(\gamma(\cdot)\defas\left(
  \frac{\sigma(\cdot)}{a_1(1-\lambda_0)} \right)^{1/b} \in \calK\). On the other
hand, if \(b\in(0,1)\), then \(1/b\geq 1\), so by convexity of
\((\cdot)^{1/b}\), we have
\begin{align*}
  |\phi_c(k;x,\boldsymbol{\uptheta})|
  &\leq \left( \frac{2}{a_1} \right)^{1/b} \left(
    \frac{1}{2}a_2|x|^b\lambda_0^k + \frac{1}{2}\frac{
    \sigma(\|\boldsymbol{\uptheta}\|_{0:k-1})}{1-\lambda_0} \right)^{1/b} \\
  &\leq \frac{1}{2}\left( \frac{2a_2}{a_1} \right)^{1/b}|x|(\lambda_0^b)^k +
    \frac{1}{2}\left( \frac{2\sigma(\|\boldsymbol{\uptheta}\|_{0:k-1})}{
    a_1(1-\lambda_0)} \right)^{1/b} \\
  &\leq c|x|\lambda^k + \gamma(\|\boldsymbol{\uptheta}\|_{0:k-1})
\end{align*}
for all \(k\in\nnegint\), \(x\in X\), and \(\boldsymbol{\uptheta}\in\Theta^k\),
where \(\lambda\defas \lambda_0^{1/b} \in (0,1)\), \(c\defas \frac{1}{2}\left(
  \frac{2a_2}{a_1} \right)^{1/b} > 0\), and \(\gamma(\cdot) \defas
\frac{1}{2}\left( \frac{2\sigma(\cdot)}{a_1(1-\lambda_0)} \right)^{1/b}
\in\calK\). In either case, \cref{eq:ras} is satisfied with \(\beta(s,k)\defas
cs\lambda^k\) and some \(\gamma\in\calK\) for some (appropriately chosen)
\(c>0\) and \(\lambda\in(0,1)\). %
\hspace*{\fill}~\QED %

\subsection{Proof of~\Cref{thm:lyap:strong}}\label[appendix]{app:lyap:strong}
As in the proof of \Cref{thm:lyap:robust}, we suppose \(X\) is RPI for
\cref{eq:plant:cl}, and let the functions \(V:X\rightarrow\nnegreal\) and
\(\alpha_i\in\calKinf,i\in\intinterval{1}{3}\) satisfy \cref{eq:lyap:strong} for
all \(x\in X\) and \(\theta\in\Theta\).

\paragraph{Asymptotic case}
For the asymptotic case, we first require the following preliminary result.

\begin{proposition}\label{prop:lyap:posdef}
  If \cref{eq:plant:cl} admits a Lyapunov function \(V\) on an RPI set \(X\),
  then there exist functions \(\alpha,\rho\in\calKinf\) such that
  \begin{equation}\label{eq:lyap:posdef}
    W(f_c(x,\theta)) \leq W(x) - \alpha(|x|)
  \end{equation}
  for all \(x\in X\) and \(\theta\in\Theta\), where \(W\defas \rho\circ V\).
\end{proposition}
\begin{proof}
  The proof follows closely to that of Lemma 2.8 in \cite{jiang:wang:2002}, with
  the main difference being we do not assume continuity of either \(f_c\) or
  \(V\), and we consider a time-invariant system and Lyapunov function. We begin
  by rewriting \cref{eq:lyap:strong:b} as
  \[
    V(f_c(x,\theta)) - V(x) \leq -\hat{\alpha}_3(V(x))
  \]
  where \(\hat{\alpha}_3(s) \defas \min\set{ \alpha_3(r) | \alpha_2^{-1}(s) \leq
    r \leq \alpha_1^{-1}(s) }\). The minimum exists because the objective
  \(\alpha_3\) is continuous and the constraint set
  \([\alpha_2^{-1}(s),\alpha_1^{-1}(s)]\) is compact. Moreover, \(\alpha_3\) is
  continuous and \(\alpha_3\in\calPD\).

  Let \(\rho_0\in\calKinf\) be any smooth function such that
  \(\rho_0(s/2)\hat{\alpha}_3(s) \geq s\) for all \(s\geq 1\). Let \(\rho(s)
  \defas s + \int_0^s \rho_0(r)dr\), which is smooth and class-\(\calKinf\) by
  construction. Moreover, we have the derivative \(\rho'(s) \defas (d\rho/ds)(s)
  = 1 + \rho_0(s)\). Finally, let \(W \defas \rho \circ V\). Then we have
  \[
    \hat{\alpha}_1(|x|) \leq W(x) \leq \hat{\alpha}_2(|x|)
  \]
  for all \(x\in X\), where \(\hat{\alpha}_i \defas
  \rho\circ\alpha_i,i\in\intinterval{1}{2}\).

  For the remainder of the proof, fix \(x\in X\) and \(\theta\in\Theta\), and
  let \(V \defas V(x)\) and \(V^+ \defas V(f_c(x,\theta))\). Notice that \(V^+
  \leq V\) by \cref{eq:lyap:iss:b}. Suppose \(V \geq 1\). From the mean value
  theorem, there exists \(\tau\in[0,1]\) such that
  \[
    \rho(V^+) - \rho(V) = \rho'(V^+ + \tau(V-V^+))(V^+-V).
  \]
  If \(V^+ \leq V/2\), then, since \(\rho'(s) \geq 1\) for all \(s\geq 0\), we
  have
  \[
    \rho(V^+) - \rho(V) \leq -V/2.
  \]
  On the other hand, if \(V^+ \geq V/2\), then
  \[
    \rho'(V^+ + \tau(V - V^+)) \geq \rho'(V^+) > \rho_0(V/2)
  \]
  and therefore
  \begin{align*}
    \rho(V^+) - \rho(V)
    &\leq \rho_0(V/2)(V^+-V) \\
    &\leq -\rho_0(V/2)\hat{\alpha}_3(V) \\
    &\leq -V
  \end{align*}
  where the last inequality follows by construction of \(\rho_0\) and because
  \(V\geq 1\). Combining these inequalities gives
  \[
    \rho(V^+) - \rho(V) \leq -V/2.
  \]
  On the other hand, if \(V \leq 1\), then we have
  \begin{align*}
    \rho(V^+) - \rho(V)
    &= V^+ + \int_0^{V^+} \rho_0(s)ds - \left( V + \int_0^V \rho_0(s)ds \right) \\
    &\leq V^+ - V \leq -\hat{\alpha}_3(V).
  \end{align*}

  Let \(\alpha_4\) be any \(\calKinf\)-function satisfying \(\alpha_4(s) \leq
  \hat{\alpha}_3\) for all \(s\in[0,1]\) and \(\alpha_4(s) \leq s/2\) for all
  \(s>1\). Define \(\alpha\defas\alpha_4\circ\alpha_1\in\calKinf\). Then
  \begin{align*}
    W(f_c(x,\theta)) - W(x)
    &\leq \begin{cases} -V(x)/2, & V(x) > 1 \\
            -\hat{\alpha}_3(V(x)), & V(x) \leq 1 \end{cases} \\
    &\leq -\alpha_4(V(x)) \\
    &\leq -\alpha_4(\alpha_1(|x|)) \\
    &= -\alpha(|x|)
  \end{align*}
  for all \(x\in X\) and \(\theta\in\Theta\).
\end{proof}

Returning to the proof of the asymptotic case of \Cref{thm:lyap:strong},
\Cref{prop:lyap:posdef} implies the existence of functions
\(\alpha,\rho\in\calKinf\) satisfying \cref{eq:lyap:posdef} for all \(x\in X\)
and \(\theta\in\Theta\), where \(W\defas\rho\circ V\). We can rewrite
\cref{eq:lyap:posdef} as
\begin{equation}\label{eq:sas:1}
  W(f_c(x,\theta)) \leq (\id - \alpha_4)(W(x))
\end{equation}
for all \(x\in X\) and \(\theta\in\Theta\), where \(\alpha_4 \defas
\alpha\circ\hat{\alpha}_2^{-1}\in\calKinf\). As in the proof of
\Cref{thm:lyap:robust}, we can assume \(\id-\alpha_4\in\calK\) without loss of
generality (Lemma B.1 of \cite{jiang:wang:2001}). Moreover, there exists
\(\hat\beta\in\calKL\) such that, for all \(x\in X\), \(k\in\nnegint\), and
\(\boldsymbol{\uptheta}\in\Theta^k\), \(W(\phi_c(k;x,\boldsymbol{\uptheta}))
\leq \hat\beta(V(x),k)\) (Lemma 4.3 of \cite{jiang:wang:2001}), and therefore
\[
  \hat{\alpha}_1(|\phi_c(k;x,\boldsymbol{\uptheta})|) \leq
  W(\phi_c(k;x,\boldsymbol{\uptheta})) \leq \hat\beta(W(x),k) \leq
  \hat\beta(\hat{\alpha}_2(|x|),k).
\]
Finally, with \(\beta(s,k) \defas
\hat\alpha_1^{-1}(\hat\beta(\hat\alpha_2(s),k))\), we have the desired
inequality \cref{eq:sas} for all \(x\in X\), \(k\in\nnegint\), and
\(\boldsymbol{\uptheta}\in\Theta^k\).

\paragraph{Exponential case}
Suppose, additionally, that \(\alpha_i \defas
a_i\id^b,i\in\intinterval{1}{3}\). %
Without loss of generality, we have \(a_3<a_2\). Then \cref{eq:lyap:strong} can
be rewritten
\begin{equation}\label{eq:ses:1}
  V(f_c(x,\theta)) \leq V(x) - a_3|x|^b \leq V(x) - \frac{a_3}{a_2}V(x) =
  \lambda_0 V(x)
\end{equation}
for all \(x\in X\) and \(\theta\in\Theta\), where \(\lambda_0\defas
1-\frac{a_3}{a_2} \in(0,1)\). Since \(X\) is RPI, we can recursively apply the
inequality \cref{eq:ses:1} to give
\[
  V(\phi_c(k;x,\boldsymbol{\uptheta})) \leq \lambda_0^k V(x) \leq
  a_2|x|^b\lambda_0^k
\]
for all \(k\in\nnegint\), \(x\in X\), and \(\boldsymbol{\uptheta}\in\Theta^k\).
Finally, using the lower bound of \cref{eq:lyap:strong:a}, we have \cref{eq:sas}
for all \(k\in\nnegint\), \(x\in X\), and \(\boldsymbol{\uptheta}\in\Theta^k\),
where \(\beta(s,k)\defas cs\lambda^k\), \(\lambda\defas \lambda_0^{1/b} \in
(0,1)\), and \(c\defas\left( \frac{a_2}{a_1} \right)^{1/b} > 0\). %
\hspace*{\fill}~\QED %
}{}

\ifthenelse{\boolean{LongVersion}}{
\section{Nominal MPC stability}\label[appendix]{app:mpc:stable}
In this appendix, we provide sketches of the proofs of the nominal MPC stability
results \Cref{thm:mpc:stable,thm:mpc:stable:exp}. First, the lower bound
\(V_N^0(x)\geq\alpha_1(|x|)\) follows immediately from \Cref{assum:posdef}.
Next, consider the following proposition from
\cite{allan:bates:risbeck:rawlings:2017}.
\begin{proposition}[Prop.~20~of~\cite{allan:bates:risbeck:rawlings:2017}]\label{prop:cont:2}%
  Let \(C\subseteq D\subseteq\real^n\), with \(C\) compact, \(D\) closed, and
  \(f : D \rightarrow \real^m\) continuous. Then there exists
  \(\alpha\in\calKinf\) such that \(|f(x)-f(y)|\leq\alpha(|x-y|)\) for all
  \(x\in C\) and \(y\in D\).
\end{proposition}
Under \Cref{assum:cont,assum:cons,assum:stabilizability,assum:posdef}, we can
establish the following bounds via \Cref{prop:cont:2},\footnote{\Cref{eq:bound:Vf}
  follows immediately from \Cref{prop:cont:2} and \Cref{assum:cont,assum:cons}.
  For \cref{eq:bound:VN0}, see \cite[Prop.~2.16]{rawlings:mayne:diehl:2020}.}
\begin{align}
  V_f(x) &\leq \alpha_f(|x|), & \forall x&\in\mathbb{X}_f \label{eq:bound:Vf} \\
  V_N^0(x) &\leq \alpha_2(|x|), & \forall x&\in\mathcal{X}_N \label{eq:bound:VN0}
\end{align}
for some \(\alpha_f,\alpha_2\in\calKinf\). To establish the cost difference
bound, first note that, under
\Cref{assum:cons,assum:stabilizability}, we have
\[
  V_f(\hat f(x,\kappa_f(x))) \leq V_f(x) - \ell(x,\kappa_f(x)) \leq c_f
\]
for all \(x\in\mathbb{X}_f\). Therefore \(\mathbb{X}_f\) is positive invariant
for \(x^+=\hat f(x,\kappa_f(x))\). But this means \(\mathcal{X}_N\) is
positively invariant because, for each \(x\in\mathcal{X}_N\),
\(\tilde{\mathbf{u}}(x)\) steers the system into \(\mathbb{X}_f\) in \(N-1\)
moves and keeps it there, meaning \(\hat f_c(x)\in\mathcal{X}_N\). Finally,
\Cref{assum:stabilizability} implies
\begin{equation}\label{eq:mpc:descent}
  V_N^0(\hat f_c(x)) \leq V_N(\hat f_c(x),\tilde{\mathbf{u}}(x)) \leq V_N^0(x) -
  \ell(x,\kappa_N(x))
\end{equation}
for all \(x\in\mathcal{X}_N\)~\cite[pp.~116--117]{rawlings:mayne:diehl:2020}.
Therefore \(V_N^0(\hat f_c(x))\leq V_N^0(x) - \alpha_1(|x|)\) by
\Cref{assum:posdef}.


Let \(\rho>0\) and \(\mathcal{S}\defas\textnormal{lev}_\rho V_N^0\). As noted in
the main text, we have \(\mathcal{S}\subseteq\mathcal{X}_N\) by definition of the
sublevel set.
%
\Cref{assum:cons,assum:quad} implies
\(\underline{\sigma}(P_f)|x|^2\leq V_f(x)\leq c_f\) for all
\(x\in\mathbb{X}_f\), so we have \(|x|\leq\varepsilon \defas
\sqrt{c_f/\underline{\sigma}(P_f)}\) for all \(x\in\mathbb{X}_f\). Then with
\(c_2\defas\max\set{\overline{\sigma}(P_f),\rho/\varepsilon^2}\), we can write
\[
  V_N^0(x) \leq
  \begin{cases}
    V_f(x) \leq \overline{\sigma}(P_f)|x|^2 \leq c_2|x|^2,
    & |x| \leq \varepsilon, \\
    \rho \leq c_2\varepsilon^2 \leq c_2|x|^2,
    & |x| \geq \varepsilon.
  \end{cases}
\]
for each \(x\in\mathcal{S}\). Finally, \(V_N^0\) is an exponential Lyapunov
function in \(\mathcal{S}\) for \(x^+=\hat f_c(x)\).

\section{Inherent robustness of MPC}
This appendix contains proofs of the inherent robustness results from
\Cref{sec:mpc:robust}. %
Throughout, we require the following proposition, which follows from
\Cref{prop:cont:2}.
\begin{proposition}\label{prop:mpc:bounds}
  Suppose \Cref{assum:cont,assum:cons} holds and let \(\tilde V_f(\cdot,\cdot)
  \defas V_f(\hat\phi(N;\cdot,\cdot))\). Then, for any compact set
  \(\mathcal{S}\subseteq\mathcal{X}_N\), there exist
  \(\alpha_a,\alpha_b,\alpha_\theta\in\calKinf\) such that, for each
  \(x\in\mathcal{S}\) and \(\theta\in\real^{n_\theta}\),
  \begin{align}
    |\tilde V_f(x^+,\tilde{\mathbf{u}}(x)) - \tilde V_f(\hat x^+,\tilde{\mathbf{u}}(x))|
    &\leq \alpha_a(|x^+-\hat x^+|) \label{eq:mpc:robust:1} \\
    |V_N(x^+,\tilde{\mathbf{u}}(x)) - V_N(\hat x^+,\tilde{\mathbf{u}}(x))|
    &\leq \alpha_b(|x^+-\hat x^+|) \label{eq:mpc:robust:2} \\
    |f_c(x,\theta) - \hat f_c(x)| &\leq \alpha_\theta(|\theta|)
                                    \label{eq:mpc:robust:3}
  \end{align}
  where \(x^+\defas f_c(x,\theta)\) and \(\hat x^+\defas\hat f_c(x)\).
\end{proposition}
\begin{proof}
  \Cref{assum:cont,assum:cons} guarantee \(\tilde{\mathbf{u}}(x)\) is
  well-defined for all \(x\in\mathcal{X}_N\)
  \cite[Prop.~2.4]{rawlings:mayne:diehl:2020}.
  Define \(C_0\defas\mathcal{S}\times\mathbb{U}\times\set{0}\) and
  \(C_1\defas\mathcal{S}\times\mathbb{U}^N\). Then \(C_0\) and \(C_1\) are
  compact, \(f\) is continuous, and \(\tilde V_f\) and \(V_N\) are continuous as
  they are finite compositions of continuous functions. By
  \Cref{prop:cont:2}, there exist
  \(\alpha_a,\alpha_b,\alpha_\theta\in\calKinf\) such that
  \begin{align*}
    |\tilde V_f(x^+,\mathbf{u}) - \tilde V_f(\hat x^+,\hat{\mathbf{u}})|
    &\leq \alpha_a(|(x^+-\hat x^+,\mathbf{u}-\hat{\mathbf{u}})|) \\
    |V_N(x^+,\mathbf{u}) - V_N(\hat x^+,\hat{\mathbf{u}})|
    &\leq \alpha_b(|(x^+-\hat x^+,\mathbf{u}-\hat{\mathbf{u}})|) \\
    |f(x,u,\theta) - \hat f(\hat x,\hat u)|
    &\leq \alpha_\theta(|(x-\hat x,u-\hat u,\theta)|)
  \end{align*}
  for all \((\hat x,\hat u,0)\in C_0\), \((\hat x^+,\hat{\mathbf{u}})\in C_1\),
  \((x,u,\theta)\in\real^{n+m+n_\theta}\), and
  \((x^+,\mathbf{u})\in\real^{n+Nm}\). Specializing the above inequalities to
  \(x=\hat x\), \(\hat x^+=\hat f_c(x)\), \(x^+=f_c(x,\theta)\), \(u=\hat
  u=\kappa_N(x)\), and \(\mathbf{u} = \hat{\mathbf{u}} = \tilde{\mathbf{u}}(x)\)
  gives \cref{eq:mpc:robust:1,eq:mpc:robust:2,eq:mpc:robust:3} for all
  \(x\in\mathcal{S}\) and \(\theta\in\real^{n_\theta}\).
\end{proof}

\subsection{Proof of \Cref{prop:mpc:rpi}}\label[appendix]{app:mpc:rpi}
Suppose \Cref{assum:cont,assum:cons,assum:stabilizability,assum:posdef} hold.
Let \(\rho>0\) and \(\mathcal{S}\defas\textnormal{lev}_\rho V_N^0\). First, we
have \(\alpha_1,\alpha_2,\alpha_f\in\calKinf\) satisfying the bounds
\cref{eq:bound:posdef,eq:bound:Vf,eq:bound:VN0,eq:mpc:descent,eq:mpc:lyap} from
the assumptions and \Cref{thm:mpc:stable} (and its proof). Next, we let
\(x\in\mathcal{S}\) and \(\theta\in\real^{n_\theta}\), and define \(\tilde
V_f(\cdot,\cdot) \defas V_f(\hat\phi(N;\cdot,\cdot))\), \(x^+\defas
f_c(x,\theta)\), and \(\hat x^+\defas \hat f_c(x)\), throughout. By
\Cref{prop:mpc:bounds}, there exist
\(\alpha_a,\alpha_b,\alpha_\theta\in\calKinf\) satisfying the bounds
\cref{eq:mpc:robust:1,eq:mpc:robust:2,eq:mpc:robust:3}.

(a)---\textit{Robust feasibility}: By nominal feasibility, we have \(\hat
x^0(N;\hat x^+)\in\mathbb{X}_f\) and therefore \(V_f(\hat x^0(N;x))\leq c_f\).
By construction of the warm start, we have
\(\hat\phi(N;x^+,\tilde{\mathbf{u}}(x))=\hat f(\hat x^0(N;x),\kappa_f(\hat
x^0(N;x)))\) and therefore
\begin{align*}
  \tilde V_f(\hat x^+,\tilde{\mathbf{u}}(x))
  &= V_f(\hat\phi(N;\hat x^+,\tilde{\mathbf{u}}(x))) \\
  &= V_f(\hat f(\hat x^0(N;x),\kappa_f(\hat x^0(N;x)))) \\
  &\leq V_f(\hat x^0(N;x)) - \alpha_1(|\hat x^0(N;x)|)
\end{align*}
where the inequality follows from \Cref{assum:stabilizability,assum:posdef}. If
\(V_f(\hat x^0(N;x))\geq c_f/2\), then \(|\hat
x^0(N;x)|\geq\alpha_f^{-1}(c_f/2)\) and \(\tilde V_f(\hat
x^+,\tilde{\mathbf{u}}(x)) \leq c_f - \alpha_1(\alpha_f^{-1}(c_f/2))\). On the
other hand, if \(V_f(\hat x^0(N;x)) < c_f/2\), then \(\tilde V_f(\hat
x^+,\tilde{\mathbf{u}}(x)) < c_f/2\). In summary,
\[
  \tilde V_f(\hat x^+,\tilde{\mathbf{u}}(x)) \leq c_f - \gamma_1
\]
where \(\gamma_1\defas\min\set{c_f/2, \alpha_1(\alpha_f^{-1}(c_f/2)}>0\).
Combining the above inequality with \cref{eq:mpc:robust:1,eq:mpc:robust:3}
gives
\[
  \tilde V_f(x^+,\tilde{\mathbf{u}}(x)) \leq c_f - \gamma_1 +
  \alpha_a(\alpha_\theta(|\theta|)).
\]
Therefore, so long as \(|\theta| \leq \delta_1 \defas
\alpha_\theta^{-1}(\alpha_a^{-1}(\gamma_1))\), we have
\(V_f(\hat\phi(N;x^+,\tilde{\mathbf{u}}(x))) = \tilde
V_f(x^+,\tilde{\mathbf{u}}(x)) \leq c_f\), which implies
\(\hat\phi(N;x^+,\tilde{\mathbf{u}}(x))\in\mathbb{X}_f\), and therefore
\((x^+,\tilde{\mathbf{u}}(x))\in\mathcal{Z}_N\).

(b)---\emph{Descent property}: Suppose \(|\theta|\leq\delta_1\). Then
\((x^+,\tilde{\mathbf{u}}(x))\in\mathcal{Z}_N\) by part (a), so the inequality
\(V_N^0(x^+) \leq V_N(x^+,\tilde{\mathbf{u}}(x))\) follows by optimality.
Combining this inequality with the nominal descent property
\cref{eq:mpc:descent} gives the robust descent property
\cref{eq:mpc:descent:robust}.

(c)---\textit{Positive invariance of \(\mathcal{S}\)}: Suppose again that
\(|\theta|\leq\delta_1\). Then the inequality \cref{eq:mpc:robust:3} holds
from part (b), and combining it with
\cref{eq:mpc:descent:robust,eq:mpc:robust:2} gives
\begin{equation}\label{eq:mpc:robust:4}
  V_N^0(x^+) \leq V_N^0(x) - \alpha_1(|x|) + \alpha_b(\alpha_\theta(|\theta|)).
\end{equation}
If \(V_N^0(x)\geq \rho/2\), then \(|x|\geq\alpha_2^{-1}(\rho/2)\) and
\(V_N^0(x^+) \leq \rho - \alpha_1(\alpha_2^{-1}(\rho/2)) +
\alpha_b(\alpha_\theta(|\theta|))\). On the other hand, if \(V_N^0(x) < \rho/2\),
then \(V_N^0(x^+) < \rho/2 + \alpha_b(\alpha_\theta(|\theta|))\). Then
\[
  V_N^0(x^+) \leq \rho - \gamma_2 + \alpha_b(\alpha_\theta(|\theta|))
\]
where \(\gamma_2\defas\min\set{\rho/2, \alpha_1(\alpha_2^{-1}(\rho/2)}>0\).
Therefore \(V_N^0(x^+)\leq\rho\) and \(x^+\in\mathcal{S}\) so long as
\(|\theta|\leq\delta\defas\min\set{\delta_1,\delta_2}\) where \(\delta_2
\defas \alpha_\theta^{-1}(\alpha_b^{-1}(\gamma_2))\). %
\hspace*{\fill}~\QED %

\subsection{Proof of \Cref{thm:mpc:robust}}\label[appendix]{app:mpc:robust}
Suppose \Cref{assum:cont,assum:cons,assum:stabilizability,assum:posdef} hold.
Let \(\rho>0\) and \(\mathcal{S}\defas\textnormal{lev}_\rho V_N^0\). From
\Cref{thm:mpc:stable}, there exists \(\alpha_2\in\calKinf\) such that
\cref{eq:mpc:lyap:a} holds for all \(x\in\mathcal{S}\subseteq\mathcal{X}_N\),
where \(\alpha_1\in\calKinf\) is from \Cref{assum:posdef}. By
\Cref{prop:mpc:bounds}, there exist \(\alpha_b,\alpha_\theta\in\calK\) such that
\cref{eq:mpc:robust:2,eq:mpc:robust:3} hold for all \(x\in\mathcal{S}\) and
\(\theta\in\real^{n_\theta}\). By \Cref{prop:mpc:rpi}, there exists \(\delta>0\)
such that \cref{eq:mpc:descent:robust} holds for all \(x\in\mathcal{S}\) and
\(|\theta|\leq\delta\), and \(\mathcal{S}\) is RPI for
\(x^+=f_c(x,\theta),|\theta|\leq\delta\). As in the proof of
\Cref{prop:mpc:rpi}, we can combine
\cref{eq:mpc:descent:robust,eq:mpc:robust:2,eq:mpc:robust:3} to give
\cref{eq:mpc:robust:4} for all \(x\in\mathcal{S}\) and \(|\theta|\leq\delta\),
which is the desired cost decrease bound with
\(\sigma\defas\alpha_b\circ\alpha_\theta\in\calK\). Thus, part (a) is
established, and part (b) follows by \Cref{thm:lyap:robust}. %
\hspace*{\fill}~\QED %

\subsection{Proof of \Cref{thm:mpc:robust:exp}}\label[appendix]{app:mpc:robust:exp}
Suppose \Cref{assum:cont,assum:cons,assum:stabilizability,assum:quad} hold. Let
\(\rho>0\) and \(\mathcal{S}\defas\textnormal{lev}_\rho V_N^0\). All the
conditions of \Cref{thm:mpc:stable:exp,thm:mpc:robust} are satisfied. Thus,
there exists \(c_2>0\) such that \cref{eq:mpc:lyap:exp} holds for all
\(x\in\mathcal{S}\) with \(c_1\defas\underline{\sigma}(Q)>0\). Moreover, we can
substitute \(\alpha_1(\cdot)\defas c_1(\cdot)^2\) and \(\alpha_2(\cdot)\defas
c_2(\cdot)^2\) into the proof of \Cref{thm:mpc:robust} to construct \(\delta>0\)
and \(\sigma\in\calK\) such that \cref{eq:mpc:robust:lyap:exp} holds for all
\(x\in\mathcal{S}\) and \(|\theta|\leq\delta\). Therefore, by
\Cref{thm:lyap:robust}, \(x^+=f_c(x,\theta),|\theta|\leq\delta\) is ISES in
\(\mathcal{S}\). %
\hspace*{\fill}~\QED %
}{}

\ifthenelse{\boolean{LongVersion}}{%
\section{Proofs of strong stability results}\label[appendix]{app:mpc:mismatch}
In this appendix we prove strong stability results from
\Cref{sec:mpc:strong}. %
}{}%
The following preliminary results are required throughout.

\ifthenelse{\boolean{LongVersion}}{}{%
  \begin{proposition}[Prop.~20~of~\cite{allan:bates:risbeck:rawlings:2017}]\label{prop:cont:2}%
    Let \(C\subseteq D\subseteq\real^n\), with \(C\) compact, \(D\) closed, and
    \(f : D \rightarrow \real^m\) continuous. Then there exists
    \(\alpha\in\calKinf\) such that \(|f(x)-f(y)|\leq\alpha(|x-y|)\) for all
    \(x\in C\) and \(y\in D\).
  \end{proposition}
}

\begin{proposition}\label{prop:mismatch:exp:xu}
  Suppose \Cref{assum:cont,assum:cons,assum:stabilizability,assum:quad} hold.
  Let \(\rho>0\) and \(\mathcal{S}\defas\textnormal{lev}_\rho V_N^0\). There
  exist \(c_x,c_u>0\) such that
  \begin{align}
    |\hat x^0(k;x)| &\leq c_x|x|,
    & \forall\, x&\in\mathcal{S},\; k\in\intinterval{0}{N}.
                   \label{eq:mismatch:exp:x} \\
    |u^0(k;x)| &\leq c_u|x|,
    & \forall\, x&\in\mathcal{S},\; k\in\intinterval{0}{N-1}.
                   \label{eq:mismatch:exp:u}
  \end{align}
\end{proposition}
\begin{proof}
  By \Cref{thm:mpc:robust:exp}, we have the upper bound
  \cref{eq:mpc:robust:lyap:exp:a} for all \(x\in\mathcal{S}\) and some
  \(c_2>0\). Moreover, since \(Q,R,P_f\) are positive definite, we can write,
  for each \(x\in\mathcal{S}\) and \(k\in\intinterval{0}{N-1}\),
  \begin{align*}
    \underline{\sigma}(Q)|\hat x^0(k;x)|^2
    &\leq |\hat x^0(k;x)|_Q^2 \leq V_N^0(x) \leq c_2|x|^2 \\
    \underline{\sigma}(P_f)|\hat x^0(N;x)|^2
    &\leq |\hat x^0(N;x)|_{P_f}^2 \leq V_N^0(x) \leq c_2|x|^2 \\
    \underline{\sigma}(R)|u^0(k;x)|^2
    &\leq |u^0(k;x)|_R^2 \leq V_N^0(x) \leq c_2|x|^2.
  \end{align*}
  Thus, with \(c_x\defas\max\set{\sqrt{c_2/\underline{\sigma}(Q)},
    \sqrt{c_2/\underline{\sigma}(P_f)}}\) and
  \(c_u\defas\sqrt{c_2/\underline{\sigma}(R)}\), we have
  \cref{eq:mismatch:exp:x,eq:mismatch:exp:u}.
\end{proof}

\begin{proposition}\label{prop:kfunc:comp}
  For each \(\alpha\in\calK\) and \(\gamma\in\calK^2\), let
  \(\gamma_1(s,t)\defas\alpha(\gamma(s,t))\),
  \(\gamma_2(s,t)\defas\gamma(\alpha(s),t)\), and
  \(\gamma_3(s,t)\defas\gamma(s,\alpha(t))\) for each \(s,t\geq 0\). Then
  \(\gamma_1,\gamma_2,\gamma_3\in\calK^2\).
\end{proposition}
\begin{proof}
  These facts follow directly from the closure of \(\calK\) under
  composition~\citep{kellett:2014}. For example, for each \(s\geq 0\), we have
  \(\gamma_2(\cdot,s) = \gamma(\alpha(\cdot),s)\in\calK\) by closure under
  composition, \(\gamma_2(s,\cdot)=\gamma(\alpha(s),\cdot)\in\calK\) trivially,
  and \(\gamma_2\) is continuous as it is a composition of continuous functions.
\end{proof}

\subsection{Proof of \Cref{prop:kfunc:joint}}\label[appendix]{app:kfunc:joint}
Let \(\alpha\in\calKinf\), \(\gamma\in\calK^2\), and \(\tau>0\). Define %
\ifthenelse{\boolean{LongVersion}}{%
  \[
    \tilde\gamma(s,t) \defas \sup_{\tilde s\in(0,s)} \frac{\gamma(\tilde
      s,t)}{\alpha(\tilde s)}
  \]
}{%
  \(\tilde\gamma(s,t) \defas \sup_{\tilde s\in(0,s)} \gamma(\tilde s,t) /
  \alpha(\tilde s)\) %
}%
for each \(s,t>0\), so that %
\ifthenelse{\boolean{LongVersion}}{%
  \[
    L \defas \limsup_{s\rightarrow 0^+} \frac{\gamma(s,\tau)}{\alpha(s)} =
    \lim_{s\rightarrow 0^+} \tilde\gamma(s,\tau).
  \]
}{%
  \(L \defas \limsup_{s\rightarrow 0^+} \gamma(s,\tau)/\alpha(s) =
  \lim_{s\rightarrow 0^+} \tilde\gamma(s,\tau)\). %
} %
Suppose the hypothesis holds, i.e., \(L<1\). Then there exists \(\rho_0>0\) such
that \(|\tilde\gamma(s,\tau) - L| < 1 - L\) for all \(s\in(0,\rho_0]\). But
\(\tilde\gamma(s,t)\geq 0\) and \(L\geq 0\) for all \(s,t>0\), so
\(\tilde\gamma(s,\tau) < 1\) for all \(s\in(0,\rho_0]\) by the reverse triangle
inequality. Therefore
\[
  \frac{\gamma(s,t)}{\alpha(s)} \leq \frac{\gamma(s,\tau)}{\alpha(s)} \leq
  \tilde\gamma(s,\tau) < 1
\]
and \(\gamma(s,t)<\alpha(s)\) for all \(s\in(0,\rho_0]\) and \(t\in[0,\tau]\).

Fix \(\rho>0\). If \(\rho\leq\rho_0\), the proof is complete with
\(\delta\defas\tau\). Otherwise, we must enlarge the interval in \(s\) by
shrinking the interval in \(t\). For each \(t\in(0,\tau]\), let
\[
  \gamma_0(t) \defas \inf\set{s>0 | \gamma(s,t)\geq\alpha(s)}.
\]
Since \(\gamma(s,t) < \alpha(s)\) for each \(s\in(0,\rho_0]\) and
\(t\in[0,\tau]\), we have \(\gamma_0(t) > 0\) for all \(t\in(0,\tau]\). Then, by
continuity of \(\alpha\) and \(\gamma\), the first nonzero point at which
\(\alpha\) and \(\gamma\) intersect, if it exists, must be equal to
\(\gamma_0(t)\). Otherwise, \(\gamma_0(t)\) is infinite. Note that \(\gamma_0\)
is a strictly decreasing function on \((0,\tau]\) since, for any
\(t\in(0,\tau]\), we have \(\gamma(\gamma_0(t),t') < \gamma(\gamma_0(t),t) =
\alpha(\gamma_0(t))\) for all \(t'\in(0,t)\). Moreover, \(\lim_{t\rightarrow
  0^+} \gamma_0(t) = \infty\) since, if \(\gamma_0\) was upper bounded by some
\(\overline\gamma>0\), we could take
\(\gamma(\overline\gamma,t)\geq\alpha(\overline\gamma)>0\) for all
\(t\in(0,\tau]\), a contradiction of the fact that \(\gamma(s,\cdot)\in\calK\)
for all \(s>0\). Then there must exist \(\delta\in(0,\tau]\) such that
\(\gamma_0(\delta)>\rho\), which implies \(\gamma(s,t)<\alpha(s)\) for all
\(s\in(0,\rho]\) and \(t\in[0,\delta]\). %
\hspace*{\fill}~\QED %

\subsection{Proof of
  \Cref{prop:mismatch:exp:f}}\label[appendix]{app:mismatch:exp:f}
Suppose \Cref{assum:cont,assum:cons,assum:steady-state,assum:diff} hold, let
\(\mathcal{S}\subseteq\real^n\) be compact, and define \(z\defas(x,u)\). %
By \Cref{prop:cont:2}, for each \(i\in\intinterval{1}{n}\), there exists
\(\sigma_i\in\calKinf\) such that
\begin{equation}\label{eq:mismatch:exp:f:1}
  | \partial_z f_i(z,\theta) - \partial_z \hat{f}_i(\tilde{z}) | \leq
  \sigma_i(|(z-\tilde{z},\theta)|)
\end{equation}
for all \(z,\tilde{z}\in\mathcal{S}\times\mathbb{U}\) and
\(\theta\in\real^{n_\theta}\). Next, let \(\mathcal{Z}\) denote the convex
hull of \(\mathcal{S}\times\mathbb{U}\). Then \(tz\in\mathcal{Z}\) for all
\(t\in[0,1]\) and \(z\in\mathcal{Z}\). By Taylor's
theorem~\cite[Thm.~12.14]{apostol:1974}, for each \(i\in\intinterval{1}{n}\)
and \((z,\theta) \in \mathcal{Z}\times\Theta\), there exists
\(t_i(z,\theta)\in(0,1)\) such that
\begin{equation}\label{eq:mismatch:exp:f:2}
  f_i(z,\theta) - \hat f_i(z) = [\partial_z f_i(t_i(z,\theta)z,\theta) -
  \partial_z \hat{f}_i(t_i(z,\theta)z)]z.
\end{equation}
Combining \cref{eq:mismatch:exp:f:1,eq:mismatch:exp:f:2} gives, for each
\((z,\theta)\in\mathcal{S}\times\mathbb{U}\times\real^{n_\theta}\),
\[
  |f(z,\theta) - \hat f(z)| \leq \sum_{i=1}^n |f_i(z,\theta) - \hat f_i(z)|
  \leq \sum_{i=1}^n \sigma_i(|\theta|)|z|
\]
and therefore \cref{eq:mismatch:exp:f} holds with
\(\sigma_f\defas\sum_{i=1}^n\sigma_i\). %
\hspace*{\fill}~\QED %

\subsection{Proof of
  \Cref{prop:mismatch:exp:fc}}\label[appendix]{app:mismatch:exp:fc}
Suppose \Cref{assum:cont,assum:cons,assum:stabilizability,assum:quad,%
  assum:steady-state,assum:diff} hold. Let \(\mathcal{S}\subseteq\mathcal{X}_N\)
be compact. %
By \Cref{prop:mismatch:exp:xu}, there exists \(c_u>0\) such that \(|\kappa_N(x)|
= |u^0(0;x)| \leq c_u|x|\), and therefore \(|(x,\kappa_N(x))| \leq |x| +
|\kappa_N(x)| \leq (1+c_u)|x|\), for all \(x\in\mathcal{S}\). Moreover, by
\Cref{prop:mismatch:exp:f}, there exists \(\sigma_f\in\calKinf\) such that
\[
  |f_c(x,\theta) - \hat f_c(x)| \leq \sigma_f(|\theta|)|(x,\kappa_N(x))| \leq
  \tilde\sigma_f(|\theta|) |x|
\]
for all \(x\in\mathcal{S}\) and \(\theta\in\real^{n_\theta}\), where
\(\tilde\sigma_f \defas \sigma_f(1+c_u)
\ifthenelse{\boolean{LongVersion}}{\in\calKinf}{}\). %
\hspace*{\fill}~\QED %

\subsection{Proof of \Cref{prop:mismatch:f}}\label[appendix]{app:mismatch:f}
Suppose \Cref{assum:cont,assum:cons,assum:steady-state} hold. Let
\(\mathcal{S}\subseteq\real^n\) and \(\Theta\subseteq\real^{n_\theta}\) be
compact. %
Without loss of generality, assume \(\mathcal{S}\) and \(\Theta\) contain the
origin. Then \(C\defas\mathcal{S}\times\mathbb{U}\times\Theta\) is compact, and
by \Cref{prop:cont:2}, there exists \(\sigma_f\in\calKinf\) such that
\begin{equation}\label{eq:mismatch:f:1}
  |f(x,u,\theta)-f(\tilde x,\tilde u,\tilde\theta)| \leq
  \sigma_f(|(x,u,\theta)-(\tilde x,\tilde u,\tilde \theta)|)
\end{equation}
for all \((x,u,\theta),(\tilde x,\tilde u,\tilde\theta)\in C\). Specializing
\cref{eq:mismatch:f:1} to \((\tilde x,\tilde u,\tilde\theta)=(x,u,0)\in C\)
gives
\begin{equation}\label{eq:mismatch:f:2}
  |f(x,u,\theta)-\hat f(x,u)| \leq \sigma_f(|\theta|)
\end{equation}
for all \((x,u,\theta)\in C\). On the other hand, specializing
\cref{eq:mismatch:f:1} to \((\tilde x,\tilde u,\tilde\theta)=(0,0,\theta)\in C\)
gives
\[
  |f(x,u,\theta)| = |f(x,u,\theta)-f(0,0,\theta)| \leq \sigma_f(|(x,u)|)
\]
and therefore
\ifthenelse{\boolean{OneColumn}}{%
  \begin{equation}\label{eq:mismatch:f:3}
    |f(x,u,\theta)-\hat f(x,u)| \leq |f(x,u,\theta)| + |\hat f(x,u)| \leq
    2\sigma_f(|(x,u)|)
  \end{equation}
}{%
  \begin{multline}\label{eq:mismatch:f:3}
    |f(x,u,\theta)-\hat f(x,u)| \leq |f(x,u,\theta)| + |\hat f(x,u)| \\
    \leq 2\sigma_f(|(x,u)|)
  \end{multline}
}%
for all \((x,u,\theta)\in C\). Combining
\cref{eq:mismatch:f:2,eq:mismatch:f:3} gives
\[
  |f(x,u,\theta) - \hat f(x,u)| \leq \min\{ 2\sigma_f(|(x,u)|),
  \sigma_f(|\theta|) \}
\]
for all \((x,u,\theta)\in C\), which is an upper bound that is clearly
continuous, nondecreasing in each \(|x|\) and \(|\theta|\), and zero if either
\(|x|\) or \(|\theta|\) is zero. To make the upper bound strictly increasing,
pick any \(\sigma_1,\sigma_2\in\calK\) and let \(\gamma_f(s,t) \defas
\min\set{2\sigma_f(s),\sigma_f(t)} + \sigma_1(s)\sigma_2(t)\) for each \(s,t\geq
0\). Then \(\gamma_f\in\calK^2\) satisfies \cref{eq:mismatch:f} for all
\((x,u,\theta)\in C\). %
\hspace*{\fill}~\QED %

\subsection{Proof of \Cref{prop:mismatch:fc}}\label[appendix]{app:mismatch:fc}
Suppose \Cref{assum:cont,assum:cons,assum:stabilizability,assum:posdef,%
  assum:steady-state} hold. Let \(\mathcal{S}\subseteq\mathcal{X}_N\) and
\(\Theta\subseteq\real^{n_\theta}\) be compact. Using the bounds
\cref{eq:mpc:robust:lyap:a,eq:bound:posdef} with \(u=\kappa_N(x)\), we have, for
each \(x\in\mathcal{X}_N\),
\[
  \alpha_1(|\kappa_N(x)|) \leq \ell(x,\kappa_N(x)) \leq V_N^0(x) \leq
  \alpha_2(|x|).
\]
Thus, \(|(x,\kappa_N(x))| \leq |x| + |\kappa_N(x)| \leq \alpha(|x|)\) for all
\(x\in\mathcal{X}_N\), where \(\alpha(\cdot)\defas (\cdot) +
\alpha_1^{-1}(\alpha_2(\cdot))\in\calKinf\). By \Cref{prop:mismatch:f}, there
exists \(\gamma_f\in\calK^2\) such that, for all \(x\in\mathcal{S}\) and
\(\theta\in\Theta\),
\begin{align*}
  |f_c(x,\theta) - \hat f_c(x)|
  &\leq \gamma_f(|(x,\kappa_N(x))|,|\theta|) \\
  &\leq \gamma_f(\alpha(|x|),|\theta|) \asdef \tilde\gamma_f(|x|,|\theta|)
\end{align*}
where \(\tilde\gamma_f\in\calK^2\) by \Cref{prop:kfunc:comp}. %
\hspace*{\fill}~\QED %

\subsection{Proof of
  \Cref{prop:mismatch:exp}}\label[appendix]{app:mismatch:exp}
Suppose \Cref{assum:cont,assum:cons,assum:stabilizability,assum:quad,%
  assum:steady-state,assum:diff} hold and let
\(\mathcal{S}\subseteq\mathcal{X}_N\) be compact. Throughout, we fix
\(x\in\mathcal{S}\) and \(\theta\in\Theta\), making any constructions
independently of \(x\) and \(\theta\). For brevity, let \(\hat x^+ \defas \hat
f_c(x)\), \(x^+ \defas f_c(x,\theta)\), \(\hat{x}^+(k) \defas
\hat{\phi}(k;\hat{x}^+,\tilde{\mathbf{u}}(x))\), and \(x^+(k) \defas
\hat{\phi}(k;x^+,\tilde{\mathbf{u}}(x))\). First, we can write
\begin{align}
  e_{V_N}^+
  &\defas V_N(x^+,\tilde{\mathbf{u}}(x)) -
    V_N(\hat{x}^+,\tilde{\mathbf{u}}(x)) \nonumber \\
  &= e_{V_f}^+ + \sum_{k=0}^{N-1} 2[e_x^+(k)]^\top Q\hat{x}^+(k) +
  |e_x^+(k)|_Q^2 \label{eq:mismatch:exp:5} \\
  e_{V_f}^+
  &\defas V_f(x^+(N)) - V_f(\hat{x}^+(N)) \nonumber \\
  &= 2[e_x^+(N)]^\top P_f\hat{x}^+(N) + |e_x^+(N)|_{P_f}^2 \label{eq:mismatch:exp:1}
\end{align}
where \(e_x^+(k) \defas x^+(k) - \hat{x}^+(k)\).

Next, we establish bounds on the individual terms in
\cref{eq:mismatch:exp:1,eq:mismatch:exp:5}.
By \Cref{prop:mismatch:exp:xu}, there exists \(c_x>0\) such that, for each
\(k\in\intinterval{0}{N-1}\), we have
\begin{equation}\label{eq:mismatch:exp:2}
  |\hat{x}^+(k)| = |\hat x^0(k+1;x)| \leq c_x|x|.
\end{equation}
By \Cref{assum:stabilizability,assum:quad}, whenever \(x\in\mathbb{X}_f\) we
have %
\ifthenelse{\boolean{OneColumn}}{%
  \begin{equation*}
    \underline{\sigma}(P_f)|\hat f(x,\kappa_f(x))|^2 \leq
    V_f(\hat f(x,\kappa_f(x))) \leq V_f(x) - \underline{\sigma}(Q)|x|^2 \leq
    [\overline{\sigma}(P_f) - \underline{\sigma}(Q)]|x|^2
  \end{equation*}
}{%
  \begin{multline*}
    \underline{\sigma}(P_f)|\hat f(x,\kappa_f(x))|^2
    \leq V_f(\hat f(x,\kappa_f(x))) \\
    \leq V_f(x) - \underline{\sigma}(Q)|x|^2
    \leq [\overline{\sigma}(P_f) - \underline{\sigma}(Q)]|x|^2
  \end{multline*}
}%
and therefore %
\ifthenelse{\boolean{LongVersion}}{%
  \[
    |\hat f(x,\kappa_f(x))| \leq \gamma_f|x|
  \]
}{%
  \(|\hat f(x,\kappa_f(x))| \leq \gamma_f|x|\) %
}%
where \(\gamma_f\defas\sqrt{[\overline{\sigma}(P_f) -
  \underline{\sigma}(Q)]/\underline{\sigma}(P_f)}\). Then \(\hat
x^0(N;x)\in\mathbb{X}_f\) gives %
\ifthenelse{\boolean{OneColumn}}{%
  \begin{equation}\label{eq:mismatch:exp:3}
    |\hat{x}^+(N)| = |\hat f(\hat{x}^0(N;x),\kappa_f(\hat{x}^0(N;x)))|
    \leq \gamma_f|\hat x^0(N;x)| \leq \gamma_fc_x|x|.
  \end{equation}
}{%
  \begin{multline}\label{eq:mismatch:exp:3}
    |\hat{x}^+(N)| = |\hat f(\hat{x}^0(N;x),\kappa_f(\hat{x}^0(N;x)))| \\
    \leq \gamma_f|\hat x^0(N;x)| \leq \gamma_fc_x|x|.
  \end{multline}
}%
Since \((\mathcal{S},\mathbb{U},\Theta)\) are each compact and \(f\) is
continuous, \(\mathcal{S}_0\defas f(\mathcal{S},\mathbb{U},\Theta)\) and
\(\mathcal{S}_{k+1}\defas\hat f(\mathcal{S}_k,\mathbb{U})\) are compact for all
\(k\in\nnegint\) (by induction). Then
\(\overline{\mathcal{S}}\defas\bigcup_{k=0}^N \mathcal{S}_k\) is compact, and,
since \(\hat f\) is Lipschitz continuous on bounded sets, there exists \(L_f>0\)
such that \(|\hat f(x_1,u_1)-\hat f(x_2,u_2)|\leq L_f|(x_1,u_1)-(x_2,u_2)|\) for
all \(x_1,x_2\in\overline{\mathcal{S}}\) and \(u_1,u_2\in\mathbb{U}\). Then %
\ifthenelse{\boolean{OneColumn}}{%
  \begin{equation*}
    |e_x^+(k+1)| = |\hat f(x^+(k),u^0(k+1;x)) -
    \hat f(\hat{x}^+(k),u^0(k+1;x))| \leq L_f|e_x^+(k)|
  \end{equation*}
}{%
  \(|e_x^+(k+1)| \leq L_f|e_x^+(k)|\) %
}%
for each \(k\in\intinterval{0}{N-1}\), and, for each \(k\in\intinterval{0}{N}\),
we have
\begin{equation}\label{eq:mismatch:exp:4}
  |e_x^+(k)| \leq L_f^k|x^+-\hat x^+|.
\end{equation}

Combining \cref{eq:mismatch:exp:1,eq:mismatch:exp:3,eq:mismatch:exp:4} gives
\begin{equation}\label{eq:mismatch:exp:Vf}
  |e_{V_f}^+| \leq c_{a,1}|x||x^+-\hat x^+| + c_{a,2}|x^+-\hat x^+|^2.
\end{equation}
with \(c_{a,1}\defas 2L_f^N\gamma_fc_x\overline{\sigma}(P_f)\) and
\(c_{a,2}\defas L_f^{2N}\overline{\sigma}(P_f)\).
Combining \cref{eq:mismatch:exp:Vf,eq:mismatch:exp:2,%
  eq:mismatch:exp:4,eq:mismatch:exp:5} gives
\begin{equation}\label{eq:mismatch:exp:VN}
  |e_{V_N}^+| \leq c_{b,1}|x||x^+-\hat x^+| + c_{b,2}|x^+-\hat x^+|^2
\end{equation}
with \(c_{b,1}\defas c_{a,1} + 2\overline{\sigma}(Q)\sum_{k=0}^{N-1} L_f^kc_x\)
and \(c_{b,2}\defas c_{a,2} + \overline{\sigma}(Q)\sum_{k=0}^{N-1} L_f^{2k}\).
By \Cref{prop:mismatch:exp:f}, there exists \(\tilde\sigma_f\in\calKinf\) such
that \(|x^+-\hat{x}^+|\leq\tilde{\sigma}_f(|\theta|)|x|\), and combining this
inequality with \cref{eq:mismatch:exp:VN}, we have \cref{eq:mismatch:exp} with
\(\sigma_V\defas c_{b,1}\tilde{\sigma}_f + c_{b,2}\tilde{\sigma}_f^2
\ifthenelse{\boolean{LongVersion}}{\in\calKinf}{}\). %
\hspace*{\fill}~\QED %

\subsection{Proof of \Cref{prop:mismatch}}\label[appendix]{app:mismatch:VN}
Suppose \Cref{assum:cont,assum:cons,assum:stabilizability,assum:posdef,%
  assum:steady-state} hold. Let \(\mathcal{S}\subseteq\mathcal{X}_N\) and
\(\Theta\subseteq\real^{n_\theta}\) be compact. By \Cref{prop:cont:2}, there
exists \(\alpha_b\in\calKinf\) such that
\begin{equation}\label{eq:mismatch:1}
  V_N(x_1,\mathbf{u}_1) - V_N(x_2,\mathbf{u}_2) \leq
  \alpha_b(|(x_1,\mathbf{u}_1)-(x_2,\mathbf{u}_2)|)
\end{equation}
for all \((x_1,\mathbf{u}_1),(x_2,\mathbf{u}_2)\in
f(\mathcal{S},\mathbb{U},\Theta)\times\mathbb{U}^N\). Specializing
\cref{eq:mismatch:1} to \(x_1=x^+\defas f_c(x,\theta)\), \(x_2=\hat x^+\defas
f_c(x)\), and \(\mathbf{u}_1=\mathbf{u}_2=\tilde{\mathbf{u}}(x)\) gives
\begin{equation}\label{eq:mismatch:2}
  |V_N(x^+,\tilde{\mathbf{u}}(x)) - V_N(\hat x^+,\tilde{\mathbf{u}}(x))| \leq
  \alpha_b(|x^+-\hat x^+|)
\end{equation}
for each \(x\in\mathcal{S}\) and \(\theta\in\Theta\). By \Cref{prop:mismatch:fc}
there exists \(\tilde\gamma_f\in\calK^2\) satisfying \cref{eq:mismatch:fc} for
all \(x\in\mathcal{S}\) and \(\theta\in\Theta\). Finally, combining
\cref{eq:mismatch:2,eq:mismatch:fc} gives \cref{eq:mismatch} with
\(\gamma_V(s,t) \defas \alpha_b(\tilde\gamma_f(s,t))\) for all \(s,t\geq 0\),
where \(\gamma_V\in\calK^2\) by \Cref{prop:kfunc:comp}. %
\hspace*{\fill}~\QED %

\ifthenelse{\boolean{LongVersion}}{%
\section{Examples}\label[appendix]{app:examples}

}{}%
\subsection{Strong asymptotic stability counterexample}\label[appendix]{app:example:sqrt}
Consider the plant \cref{eq:plant:sqrt} and MPC defined in
\Cref{ssec:example:sqrt}. We aim to show the closed-loop system
\(x^+=f(x,\kappa_1(x),\theta),|\theta|\leq\delta\) is RES with \(\delta=3\), but
not inherently strongly stabilizing for any \(\delta>0\).
By Lipschitz continuity of \(x^2\) on bounded sets and \(1/2\)-H{\"o}lder
continuity of \(\sqrt{|x|}\),
\begin{align}
  |x^2-y^2| &\leq 4|x-y|, & \forall\; &x,y\in[-2,2], \label{eq:lipschitz:pow2}\\
  |\sigma(x)-\sigma(y)| &\leq 2\sqrt{|x-y|}, & \forall\; &x,y\in\real. \label{eq:holder:sqrt}
\end{align}
\ifthenelse{\boolean{LongVersion}}{%
  To show \cref{eq:lipschitz:pow2}, note that, for each \(\delta>0\), we have
  \begin{align*}
    |x^2-y^2| = |x+y||x-y| \leq 2\delta|x-y|
  \end{align*}
  for all \(x,y\in[-\delta,\delta]\), and take \(\delta=2\) to give
  \cref{eq:lipschitz:pow2}. For \cref{eq:holder:sqrt}, we first show
  \(\sqrt{(\cdot)}\) is \(1/2\)-H{\"o}lder continuous on \(\nnegreal\): %
  \ifthenelse{\boolean{OneColumn}}{%
    \[
      |\sqrt{x}-\sqrt{y}| = \frac{|x-y|}{\sqrt{x}+\sqrt{y}} \leq
      \frac{|x-y|}{\sqrt{x}+\sqrt{y}} = \sqrt{|x-y|}
      \frac{\sqrt{|x-y|}}{\sqrt{x}+\sqrt{y}} \leq \sqrt{|x-y|}
    \]
  }{%
    \begin{multline*}
      |\sqrt{x}-\sqrt{y}| = \frac{||x|-|y||}{|\sqrt{x}+\sqrt{y}|} \leq
      \frac{|x-y|}{|\sqrt{x}+\sqrt{y}|} \\
      = \sqrt{|x-y|}\frac{\sqrt{|x-y|}}{\sqrt{x}+\sqrt{y}} \leq \sqrt{|x-y|}
    \end{multline*}
  }%
  for all \(x,y\geq 0\), where the last inequality follows by the triangle
  inequality. Then we automatically get
  \(|\sigma(x)-\sigma(y)|\leq\sqrt{|x-y|}\) if \(x,y\geq 0\). On the other hand,
  if \(x\geq 0\) and \(y\leq 0\), we have
  \[
    |\sigma(x)-\sigma(y)| = |\sqrt{x}+\sqrt{y}|\leq \sqrt{x} + \sqrt{-y} \leq
    2\sqrt{x-y}.
  \]
  Finally, flipping the signs of the prior arguments gives
  \cref{eq:holder:sqrt}. %
}{}

First, we derive the control law. The terminal set can be reached in a single
move if and only if \(|x|\leq 2\), so we have the steerable set
\(\mathcal{X}_1=[-2,2]\). Consider the problem \emph{without} the terminal
constraint.
The objective is
\[
  V_1(x,u) = x^2 + u^2 + 4|x+u|
\]
which is increasing in \(u\) if \(x>1\) and \(|u|\leq 1\), and decreasing in
\(u\) if \(x<-1\) and \(|u|\leq 1\). Thus \(V_1(x,\cdot)\) is minimized (over
\(|u|\leq 1\)) by \(\mathbf{u}^0(x) = -\textnormal{sgn}(x)\) for all
\(x\not\in[-1,1]\). On the other hand, if \(|x|\leq 1\), then \(V_1(x,\cdot)\)
is decreasing on \([-1,-x)\) and increasing on \((-x,1]\). Thus \(V_1(x,\cdot)\)
is minimized (over \(|u|\leq 1\)) by \(\mathbf{u}^0(x) = -x\) so long as
\(|x|\leq 1\). In summary, we have the control law \(\kappa_1(x)\defas
-\textnormal{sat}(x)\).
But
\[
  |\hat f(x,\kappa_1(x))| =
  \begin{cases} 0, & |x|\leq 1 \\
    |x-\textnormal{sgn}(x)| = |x| - 1, & 0<|x|\leq 2 \end{cases}
\]
so \(u=\kappa_1(x)\) drives each state in \(\mathcal{X}_1=[-2,2]\) to the
terminal constraint \(\mathbb{X}_f=[-1,1]\). Therefore \(\kappa_1\) is also the
control law of the problem \emph{with} the terminal constraint. %
\ifthenelse{\boolean{LongVersion}}{%
  The control law \(\kappa_1\) is plotted, along with the unforced dynamics
  \(\hat f(\cdot,0)\), against \(x\in\mathcal{X}_1\) in \Cref{fig:sqrt:nominal}
  (top left). %
}{}

\ifthenelse{\boolean{LongVersion}}{%
  \begin{figure}
    \centering
    \ifthenelse{\boolean{OneColumn}}{%
      \includegraphics[width=0.75\linewidth,page=1]{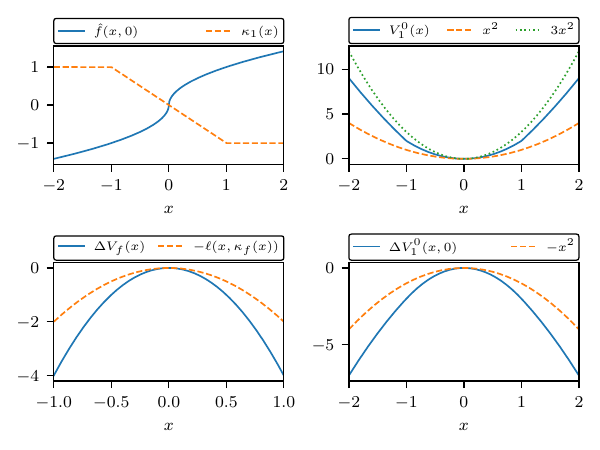}
    }{%
      \includegraphics[width=\linewidth,page=1]{sqrt_example.pdf}
    }
    \caption{For the MPC of \cref{eq:plant:sqrt}, plots of (top left) the open-loop
      dynamics and control law, (bottom left) the terminal cost difference, (top
      right) the optimal value function, and (bottom right) the cost difference,
      each with the relevant (nominal) bounds from
      \Cref{assum:stabilizability,eq:mpc:robust:lyap:exp}.}%
    \label{fig:sqrt:nominal}
  \end{figure}
}{}

\ifthenelse{\boolean{LongVersion}}{%
  \Cref{assum:cont,assum:posdef} are satisfied by definition, \Cref{assum:cons} is
  satisfied with \(c_f\defas 8\), and \Cref{assum:stabilizability} is
  satisfied with \(\kappa_f(x)\defas -x\) since \(\hat f(x,\kappa_f(x))=0\) and %
  \ifthenelse{\boolean{OneColumn}}{%
    \[
      \Delta V_f(x) \defas V_f(\hat f(x,\kappa_f(x))) - V_f(x) = -4x^2 \leq
      -2x^2 = -\ell(x,\kappa_f(x))
    \]
  }{%
    \begin{multline*}
      \Delta V_f(x) \defas V_f(\hat f(x,\kappa_f(x))) - V_f(x) = -4x^2 \\
      \leq -2x^2 = -\ell(x,\kappa_f(x))
    \end{multline*}
  }%
  for all \(x\in\mathbb{X}_f\). %
  \ifthenelse{\boolean{LongVersion}}{%
    See \Cref{fig:sqrt:nominal} (bottom left) for plots of \(\Delta V_f\) and
    \(-\ell(\cdot,\kappa_f(\cdot))\). 
  }{}%
  Therefore, by \Cref{thm:mpc:robust}, the closed-loop system \(x^+ =
  f(x,\kappa_1(x),\theta),|\theta|\leq\delta\) is RAS on
  \(\mathcal{X}_1=[-2,2]\) with ISS Lyapunov function \(V_1^0\) for some
  \(\delta>0\). Our next goal is to find such a \(\delta>0\).

}{%
  It is easy to check that
  \Cref{assum:cont,assum:cons,assum:stabilizability,assum:posdef} hold, so by
  \Cref{thm:mpc:robust}, the closed-loop system \(x^+ =
  f(x,\kappa_1(x),\theta),|\theta|\leq\delta\) is RAS on \(\mathcal{X}_1=[-2,2]\) with
  ISS Lyapunov function \(V_1^0\) for some \(\delta>0\). Our next goal is to
  find such a \(\delta>0\). %
}%

\ifthenelse{\boolean{LongVersion}}{%
  First, however, let us establish that \(V_1^0\) is a Lyapunov function for the
  modeled closed-loop \(x^+=\hat f(x,\kappa_1(x))\) in \(\mathcal{X}_1=[-2,2]\).
  We already have \(V_1^0(x)\geq x^2\) for all \(|x|\leq 2\). For the upper
  bound, we have
  \[
    V_1^0(x) = V_1(x,\kappa_1(x)) = \begin{cases} 2x^2, & |x|\leq 1, \\ x^2 +
      4|x| - 3, & 1<|x|\leq 2 \end{cases}
  \]
  for each \(|x|\leq 2\). But the polynomials \(-2x^2\pm 4x-3\) have no real
  roots, so \(4|x|-3 < 2x^2\), and the above inequality gives \(V_1^0(x) \leq
  3x^2\) for all \(|x|\leq 2\). Moreover, by \cref{eq:mpc:descent}, we have
  \(\Delta V_1^0(x,\theta) \leq -x^2\), so \(x^+=\hat f(x,\kappa_1(x))\) is in
  fact \emph{exponentially} stable on \(\mathcal{X}_1=[-2,2]\). We plot
  \(V_1^0\) and \(\Delta V_1^0(\cdot,0)\defas V_1^0(\hat
  f(\cdot,\kappa_1(\cdot)))-V_1^0(\cdot)\), along with their exponential
  Lyapunov bounds, in \Cref{fig:sqrt:nominal} (right). %
}{} %

For robust positive invariance, let \(|x|\leq 2\), \(\theta\in\real\),
\(x^+\defas f(x,\kappa_1(x),\theta)\), \(\hat x^+\defas\hat f(x,\kappa_1(x))\)
and note that %
\ifthenelse{\boolean{LongVersion}}{%
  \[
    x^+ = \sigma(\sigma^{-1}(\hat x^+) - \theta\textnormal{sat}(x))
  \]
  where \(\sigma^{-1}(x)=\textnormal{sgn}(x)|x|^2\), and therefore
}{}%
\[
  |x^+| \leq \sqrt{|\hat x^+|^2 + |\theta||\textnormal{sat}(x)|} \leq \sqrt{1 +
    |\theta|}.
\]
Then \(|x^+|\leq 2\) so long as \(|\delta|\leq 3\), so \(\mathcal{X}_1=[-2,2]\)
is RPI for \(x^+=f(x,\kappa_1(x),\theta),|\theta|\leq 3\).

By continuity of \(f\), \(V_1^0\), and \(\kappa_1\) and \Cref{prop:cont:2},
there exists \(\sigma\in\calKinf\) such that \(|V_1^0(x^+) - V_1^0(\hat x^+)|
\leq \sigma(|\theta|)\) and therefore \(V_1^0(x^+) \leq V_1^0(\hat x^+) +
|V_1^0(x^+) - V_1^0(\hat x^+)| \leq V_1^0(x) - x^2 + \sigma(|\theta|)\) for all
\(|x|\leq 2\) and \(|\theta|\leq 3\), where \(x^+\defas
f(x,\kappa_1(x),\theta)\) and \(\hat x^+\defas\hat f(x,\kappa_1(x))\). Therefore
\(x^+=f(x,\kappa_1(x),\theta),|\theta|\leq 3\) is not only RAS, but \emph{RES}
on \(\mathcal{X}_1\) by \Cref{thm:lyap:robust}.

We now aim to show strong stability is \emph{not} achieved. For simplicity, we
consider \(\mathcal{S} \defas \textnormal{lev}_2 V_1^0 = [-1,1] = \mathbb{X}_f\)
as the candidate basin of attraction. Let \(|x|\leq 1\), \(|\theta|\leq 3\),
\(x^+\defas f(x,\kappa_1(x),\theta)\), and \(\hat x^+\defas \hat
f(x,\kappa_1(x))\). %
Moreover,
\(\ell(x,\kappa_1(x)) \geq 2|x|^2 \asdef \alpha_3(|x|)\). Next, we have
\(\kappa_1(x)=-x\), \(x^+ = \sigma(x\theta)\), and \(\hat x^+ = 0\). Therefore %
\ifthenelse{\boolean{OneColumn}}{%
  \begin{align*}
    |V_1(x^+,\tilde{\mathbf{u}}(x)) - V_1(\hat x^+,\tilde{\mathbf{u}}(x))|
    &= |(x^+)^2 + 4|x^+|| \leq |x^+|^2 + 4|x^+| \\
    &\leq |x||\theta| + 4\sqrt{|x||\theta|} \asdef \gamma_V(|x|,|\theta|)
  \end{align*}
}{%
  \begin{align*}
    |V_1&(x^+,\tilde{\mathbf{u}}(x)) - V_1(\hat x^+,\tilde{\mathbf{u}}(x))| \\
        &= |(x^+)^2 + 4|x^+|| \leq |x^+|^2 + 4|x^+| \\
        &\leq |x||\theta| + 4\sqrt{|x||\theta|} \asdef \gamma_V(|x|,|\theta|)
  \end{align*}
}%
where \(\gamma_V\in\calK^2\). For each \(t>0\), we have
\(\gamma_V(s,t)/\alpha_3(s) \ifthenelse{\boolean{LongVersion}}{=
  (st+4\sqrt{st})/(2s^2)}{} = t/(2s) + 2\sqrt{t}/s^{3/2}\), so
\(\lim_{s\rightarrow 0^+} \gamma_V(s,t)/\alpha_3(s) = \infty\) for all \(t>0\),
and \cref{eq:mpc:mismatch:posdef} is not satisfied.

As mentioned in the main text, \cref{eq:mpc:mismatch:posdef} is sufficient but
not necessary. The cost difference curve is positive definite, as
\[
  \Delta V_1^0(x,\theta) = 2[\sigma(\theta x)]^2 - 2x^2 = 2(|\theta|-|x|)|x| > 0
\]
for any \(0<|x|<|\theta|\leq 1\). In other words, \(\theta\) can be arbitrarily
small but nonzero, and the cost difference curve will remain positive definite
near the origin.
\ifthenelse{\boolean{LongVersion}}{
\subsection{Nonlinearizable yet inherently strongly stabilizing}\label[appendix]{app:example:sin}
Consider the plant \cref{eq:plant:sin} and MPC defined in
\Cref{ssec:example:sin}. We aim to show the closed-loop system
\(x^+=f(x,\kappa_1(x),\theta),|\theta|\leq\delta\) is RES in \(\mathcal{X}_1\)
with \(\delta=1\), and SES with \(\delta=1/2\).

To derive the control law, we first consider the problem \emph{without} the
terminal constraint (i.e., \(\mathbb{X}_f=\real\)). We have the objective
\[
  V_1(x,u) = x^2 + u^2 + 4\left( x + (1/2)\gamma(x) + u \right)^2.
\]
Taking the partial derivative in \(u\),
\[
  \frac{\partial V_1}{\partial u}(x,u) = 8x + 4\gamma(x) + 10u
\]
and setting that to zero gives the optimal input
\[
  \mathbf{u}^0(x) = -g(x) \defas -(4/5)x - (2/5)\gamma(x)
\]
whenever \(|g(x)|\leq 1\). Otherwise the solution saturates at \(\mathbf{u}^0(x)
= -\textnormal{sgn}(g(x))\), so we have \(\mathbf{u}^0(x) = \kappa_1(x)\defas
-\textnormal{sat}(g(x))\) for all \(|x|\leq 2\).

To see where the control law \(\kappa_1(x)\) saturates, first note
\[
  \frac{d^2g}{dx^2}(x) = \frac{2}{5}\frac{d^2\gamma}{dx^2}(x) =
  -\frac{8\pi^2\sin(2\pi/x)}{5|x|^3}
\]
for all \(x\neq 0\), so \(g(x)\) is strictly concave on \(x\in[1/(n-1/2),1/n]\)
and strictly convex on \(x\in[1/n,1/(n+1/2)]\) for each \(n\in\allint\).
Therefore \(g(x)\) achieves a local maximum on each \(x\in[1/(n-1/2),1/n]\), and
the maximum is strictly decreasing with \(n\). The last, and greatest, of these
local maxima on \(|x|\leq 2\) is achieved on \(2/3\leq x\leq 1\). Through
numerical optimization, we find \(\max_{0\leq x\leq 1} g(x) = \max_{2/3\leq
  x\leq 1} g(x) \approx 0.9849\). By strict convexity of \(g(x)\) on
\(x\in[1,2]\), \(g(1)=4/5\), and \(g(2)=8/5\), we have \(\max_{1\leq x\leq 2}
g(x) = g(2) = 8/5\). Therefore \(g(x)\) intersects the horizontal line at
\(u=1\) exactly once over \(x\in[-2,2]\), and it does so at some
\(x^*\in[1,2]\), which we can numerically verify is \(x^*\approx 1.6989\). By
symmetry, \(g(x)\) intersects \(u=-1\) at \(-x^*\). Finally, because \(g(x)\) is
strictly convex (concave) on \([1,2]\) (\([-2,-1]\)), it saturates on
\((x^*,2]\) (and \([-2,-x^*)\)) and we have
\[
  \kappa_1(x) =
  \begin{cases} -(4/5)x - (2/5)\gamma(x), & |x|\leq x^*, \\
    -\textnormal{sgn}(x), & x^*<|x|\leq 2. \end{cases}
\]

For the problem \emph{with} the terminal constraint, we have %
\ifthenelse{\boolean{OneColumn}}{%
  \[
    |\hat f(x,\kappa_1(x))| = |(1/5)x+(3/5)\gamma(x)| \leq (1/5)x +
    (3/5)|\gamma(x)| \leq 4/5
  \]
}{%
  \begin{multline*}
    |\hat f(x,\kappa_1(x))| = |(1/5)x+(3/5)\gamma(x)| \\
    \leq (1/5)x + (3/5)|\gamma(x)| \leq 4/5
  \end{multline*}
}%
for each \(x\in[0,1]\),  %
\ifthenelse{\boolean{OneColumn}}{%
  \[
    |\hat f(x,\kappa_1(x))| = |(1/5)x+(3/5)\gamma(x)| \leq |(1/5)x -
    (3/5)|\gamma(x)|| \leq (2/5)|x| \leq 4/5
  \]
}{%
  \begin{multline*}
    |\hat f(x,\kappa_1(x))| = |(1/5)x+(3/5)\gamma(x)| \\
    \leq |(1/5)x - (3/5)|\gamma(x)|| \leq (2/5)|x| \leq 4/5
  \end{multline*}
}%
for each \(x\in[1,x^*]\), and
\ifthenelse{\boolean{OneColumn}}{%
  \[
    |\hat f(x,\kappa_1(x))| = |x + (1/2)\gamma(x) - 1|
    \leq |x| - 1 - (1/2)|\gamma(x)| \leq |x| - 1 \leq 1
  \]
}{%
  \begin{multline*}
    |\hat f(x,\kappa_1(x))| = |x + (1/2)\gamma(x) - 1| \\
    \leq |x| - 1 - (1/2)|\gamma(x)| \leq |x| - 1 \leq 1
  \end{multline*}
}%
for each \(x\in[x^*,2]\), where we have used the fact that \(\gamma(x)\leq 0\)
for all \(x\in[1,2]\). Therefore \(|\hat f(x,\kappa_1(x))|\leq 1\) for all
\(x\in[0,2]\), and the same holds for all \(x\in[-2,0]\) by symmetry. Therefore
the terminal constraint \(\mathbb{X}_f=[-1,1]\) is automatically satisfied by
the unconstrained control law, so \(\kappa_1(x)\) is also the control law for
the MPC \emph{with} the terminal constraint. %
\ifthenelse{\boolean{LongVersion}}{%
  In \Cref{fig:sin:nominal} (top left), we plot \(\kappa_1\) and \(\hat
  f(\cdot,0)\) on \(\mathcal{X}_1\). %
}{}

\ifthenelse{\boolean{LongVersion}}{%
  \begin{figure}
    \centering
    \ifthenelse{\boolean{OneColumn}}{%
      \includegraphics[width=0.75\linewidth,page=1]{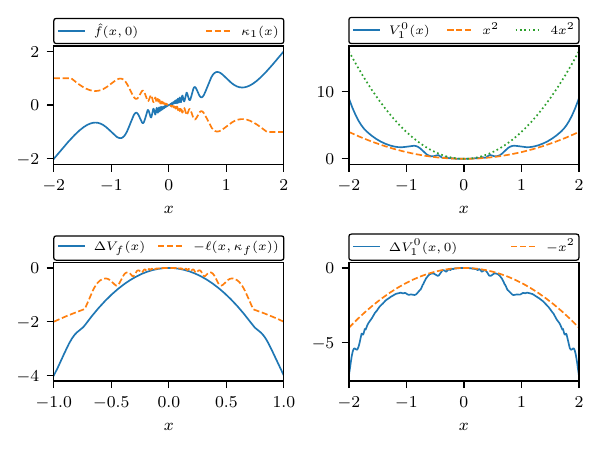}
    }{%
      \includegraphics[width=\linewidth,page=1]{sin_example.pdf}
    }%
    \caption{For the MPC of \cref{eq:plant:sin}, we plot as a function of \(x\)
      (top left) the open-loop dynamics and control law, (bottom left) the
      terminal cost difference, (top right) the optimal value function, and
      (bottom right) the cost difference, each with the relevant (nominal) bounds
      from \Cref{assum:stabilizability,eq:mpc:robust:lyap:exp}.}%
    \label{fig:sin:nominal}
  \end{figure}
}{}

\ifthenelse{\boolean{LongVersion}}{%
  \Cref{assum:cont,assum:quad} are satisfied by definition, and
  \Cref{assum:cons} is satisfied with \(c_f\defas 4\). Let \(\kappa_f(x)\defas
  -(1/2)(x+\gamma(x))\) for all \(|x|\leq 1\). Then
  \[
    |\kappa_f(x)| \leq (1/2)(|x| + |\gamma(x)|) \leq |x| \leq 1
  \]
  for all \(|x|\leq 1\), so \(u=\kappa_f(x)\) is feasible in the terminal
  constraint. Moreover, \(\hat f(x,\kappa_f(x)) = (1/2)x\), so %
  \ifthenelse{\boolean{OneColumn}}{%
    \[
      \Delta V_f(x) \defas V_f(\hat f(x,\kappa_f(x))) - V_f(x) +
      \ell(x,\kappa_f(x)) = -2x^2 + |\kappa_f(x)|^2 \leq -x^2 \leq 0
    \]
  }{%
    \begin{multline*}
      \Delta V_f(x) \defas V_f(\hat f(x,\kappa_f(x))) - V_f(x) + \ell(x,\kappa_f(x)) \\
      = -2x^2 + |\kappa_f(x)|^2 \leq -x^2 \leq 0
    \end{multline*}
  }%
  and \Cref{assum:stabilizability} is satisfied. See \Cref{fig:sin:nominal}
  (bottom left) for plots of \(\Delta V_f\) and
  \(-\ell(\cdot,\kappa_f(\cdot))\). By \Cref{thm:mpc:robust:exp}, the
  closed-loop system \(x^+=f(x,\kappa_1(x),\theta),|\theta|\leq\delta\) is RES
  on \(\mathcal{X}_1\) with the ISS Lyapunov function \(V_1^0\) for some
  \(\delta>0\). Our next aim is to find such a \(\delta>0\). %

}{%
  \Cref{assum:cont,assum:cons,assum:stabilizability,assum:quad} are easily
  checked with \(c_f\defas 16\), \(\kappa_f(\cdot)\defas
  -(1/2)((\cdot)+\gamma(\cdot))\), and \(\alpha_1(\cdot)\defas
  (\cdot)^2\in\calKinf\), so there exists \(\delta>0\) such that the closed-loop
  system \(x^+=f(x,\kappa_1(x),\theta),|\theta|\leq\delta\) is RES on
  \(\mathcal{X}_1\) by \Cref{thm:mpc:robust:exp}. Our next aim is to find such a
  \(\delta>0\). %
}%

Let \(|x|\leq 2\), \(\theta\in\real\), \(x^+\defas f(x,\kappa_1(x),\theta)\),
and \(\hat x^+\defas\hat f(x,\kappa_1(x))\). Then \(x^+=\hat x^+ +
\theta\kappa_1(x)\), and we have
\[
  |x^+| \leq |\hat x^+| + |\theta||\kappa_1(x)| \leq 1 + |\theta|
\]
for all \(\theta\in\real\). But this means \(|x^+|\leq 2\) for all
\(|\theta|\leq 1\), so \(\mathcal{X}_1\) is RPI for
\(x^+=f(x,\kappa_1(x),\theta),|\theta|\leq 1\). Continuity of \(f\), \(\ell\),
\(V_f\), and \(\kappa_1\) implies continuity of \(V_1^0(f_c(\cdot,\cdot))\), at
least for all \(|x|\leq 2\) and \(|\theta|\leq 1\) on which the function is
well-defined. Then, by \Cref{prop:cont:2}, there exists \(\sigma\in\calKinf\)
such that, if \(|\theta|\leq 1\), we have \(|V_1^0(x^+)-V_1^0(\hat
x^+)|\leq\sigma(|\theta|)\), and therefore \(V_1^0(x^+) \leq V_1^0(\hat x^+) +
|V_1^0(x^+)-V_1^0(\hat x^+)| \leq V_1^0(x) - x^2 + \sigma(|\theta|)\). Finally,
\(x^+=f(x,\kappa_1(x),\theta),|\theta|\leq 1\) is RES in
\(\mathcal{X}_1=[-2,2]\) by \Cref{thm:lyap:robust}.

Next, we aim to show the MPC is inherently strongly stabilizing via
\Cref{assum:lyap:exp,thm:mpc:mismatch:exp}. Consider the candidate Lyapunov
function \(V(x)\defas x^2\) for all \(|x|\leq 2\) and \(V(x)\defas\infty\)
otherwise, and let \(\rho\geq 4\), \(\mathcal{S} \defas \textnormal{lev}_\rho V
= [-2,2] = \mathcal{X}_1\), and \(\delta_0\defas 1\). If we can show
\Cref{assum:lyap:exp}(a,b) hold with these ingredients, then
\Cref{assum:lyap:exp} will hold for all \(\rho>0\). \Cref{assum:lyap:exp}(a) and
\cref{eq:lyap:exp:a} are already satisfied with \(|\theta|\leq\delta_0=1\), and
\(\mathcal{S}\) is RPI, but it remains to construct the bound
\cref{eq:lyap:exp:b}. Throughout this derivation, let \(x^+\defas
f(x,\kappa_1(x),\theta)\) and \(\hat x^+\defas\hat f(x,\kappa_1(x))\).

First, suppose \(|x|\leq x^*\) and \(|\theta|\leq 1\). Then the controller does
not saturate, i.e., \(\kappa_1(x)=-0.8x-0.4\gamma(x)\), and we have in the
nominal case \(\hat x^+ = 0.2x + 0.1\gamma(x)\), \(|\hat x^+|\leq 0.3|x|\), and
\begin{equation}\label{eq:example:sin:1}
  V(\hat x^+) - V(x) = |\hat x^+|^2 - |x|^2 \leq -0.91|x|^2.
\end{equation}
Next, consider the identity
\begin{equation}\label{eq:example:sin:2}
  y^2 - z^2 = 2z(y-z) + (y-z)^2
\end{equation}
for all \(y,z\in\real\). We have \(x^+ = (0.2 - 0.8\theta)x +
(0.1-0.4\theta)\gamma(x)\), so \(|x^+-\hat x^+| = |0.8\theta x +
0.4\theta\gamma(x)| \leq 1.2|\theta||x|\), and \cref{eq:example:sin:2} implies
\begin{equation}\label{eq:example:sin:3}
  |V(x^+) - V(\hat x^+)| \leq 0.72|\theta||x|^2 + 1.44|\theta|^2|x|^2.
\end{equation}

Next, suppose \(x^*<x\leq 2\) and \(|\theta|\leq 1\). Then the controller always
saturates, i.e., \(\kappa_1(x)=-1\). Since \(\gamma(\tilde x)\leq 0\) for all
\(1\leq\tilde x\leq 2\), we have \(0\leq 0.5x+0.5\gamma(x)\leq 0.5x\leq 1\) and
\(\hat x^+ = x + 0.5\gamma(x) - 1 \leq 0.5x\). Moreover, \(x-1>x^*-1>0\), so
\(\hat x^+ = x + 0.5\gamma(x) - 1 > 0.5\gamma(x) \geq -0.5x\). Then we have
\(|\hat x^+|\leq 0.5|x|\) and
\begin{equation}\label{eq:example:sin:4}
  V(\hat x^+) - V(x) = |\hat x^+|^2 - |x|^2 \leq -0.75|x|^2.
\end{equation}
Moreover, \(|x^+-\hat x^+|=|\theta|\) and \cref{eq:example:sin:2} implies
\begin{equation}\label{eq:example:sin:5}
  |V(x^+) - V(\hat x^+)| \leq (1/x^*)|\theta||x|^2 + (1/x^*)^2|\theta|^2|x|^2
\end{equation}
where we have used the fact that \(|x|/x^*>1\). By symmetry,
\cref{eq:example:sin:4,eq:example:sin:5} also hold for \(-2\leq x<-x^*\).

Combining
\cref{eq:example:sin:1,eq:example:sin:3,eq:example:sin:4,eq:example:sin:5}, we
have
\begin{equation*}
  V(x^+) \leq V(x) -a_3|x|^2 + \sigma_V(|\theta|)|x|^2
\end{equation*}
for all \(x\in\mathcal{X}_N\), where \(a_3\defas 0.75\) and \(\sigma_V(t) \defas
\max\set{0.72t + 1.44t^2, (2/x^*)t + (1/x^*)^2t^2}\) and \Cref{assum:lyap:exp}
is satisfied. Finally, by \Cref{thm:mpc:mismatch:exp} (and its proof), the
closed-loop system \(x^+=f(x,\kappa_1(x),\theta),|\theta|\leq\delta\) is SES in
\(\mathcal{X}_N=[-2,2]\) for any \(\delta\in(0,\sigma_V^{-1}(a_3))\). Thus,
it suffices to take \(|\theta|\leq\delta=0.5\) since
\[
  \sigma_V(0.5) = \max\set{ 0.72, 0.3809\ldots } = 0.72 < 0.75 = a_3.
\]
}{}

\subsection{Upright pendulum}\label[appendix]{app:example:pendulum}
Consider the plant \cref{eq:plant:pendulum} and MPC defined in
\Cref{ssec:example:pendulum}. It is noted in the main text that
\Cref{assum:cont,assum:cons,assum:quad,assum:steady-state,assum:diff} are
automatically satisfied. To design \(P_f\) and show \Cref{assum:stabilizability}
holds, consider the linearization
\begin{equation}\label{eq:pendulum:lin}
  x^+ = \underbrace{\begin{bmatrix} 1 & 0.1 \\ 0.1 & 1 \end{bmatrix}}_{\asdef A}x +
  \underbrace{\begin{bmatrix} 0 \\ 5 \end{bmatrix}}_{\asdef B}u
\end{equation}
and the feedback gain \(K\defas\begin{bmatrix} 2 & 2 \end{bmatrix}\), which
stabilizes \cref{eq:pendulum:lin} because \(A_K\defas A-BK=\begin{bsmallmatrix}
  1 & 0.1 \\ -0.9 & 0 \end{bsmallmatrix}\) has eigenvalues of \(0.9\) and
\(0.1\). Numerically solving the Lyapunov equation
\[
  A_K^\top P_fA_K - P_f = -2Q_K
\]
where \(Q_K\defas Q+K^\top RK = \begin{bsmallmatrix} 5 & 4 \\ 4 &
  5 \end{bsmallmatrix}\), we have a unique positive definite solution \(P_f
\defas
\begin{bsmallmatrix} 31.133\ldots & 10.196\ldots \\
  10.196\ldots & 10.311\ldots \end{bsmallmatrix}\). Using the inequality \(|\sin
 x_1-x_1|\leq (1/6)|x_1|^3\) for all \(x_1\in\real\), we have
\begin{align*}
  |V_f&(\hat f(x,-Kx)) - V_f(A_Kx)| \\
      &= 2x^\top A_K^\top P_f\begin{bmatrix} 0 \\ \Delta(\sin x_1-x_1) \end{bmatrix}
        + [P_f]_{22}\Delta^2(\sin x_1-x_1)^2 \\
      &\leq b|x|^4 + a|x|^6
\end{align*}
for all \(x\in\real^2\), where \(a\defas [P_f]_{22}\Delta^2/36 =
2.8643\ldots\times 10^{-3}\) and
\(b\defas \Delta|A_K^\top P_f\begin{bsmallmatrix} 0 \\
                               1 \end{bsmallmatrix}|/3 = 0.045675\ldots\).
 Moreover, \(\underline{\sigma}(Q_K)=1\), so
\begin{align*}
  V_f&(\hat f(x,-Kx)) - V_f(x) + \ell(x,-Kx) \\
  \ifthenelse{\boolean{LongVersion}}{%
     &= |A_Kx|_{P_f}^2 - |x|_{P_f}^2 + |x|_{Q_K}^2 + V_f(\hat f(x,-Kx)) - V_f(A_Kx) \\
     &= -|x|_{Q_K}^2 + V_f(\hat f(x,-Kx)) - V_f(A_Kx) \\
  }{}
     &\leq -[1 - b|x|^2 - a|x|^4]|x|^2
\end{align*}
for all \(x\in\real^2\). The polynomial inside the brackets has roots at
\(x_*=-1.0231\ldots\) and \(x^*=0.9774\ldots\) and is positive in between.
Recall \(c_f\defas\underline{\sigma}(P_f)/8\). Then
\(\underline{\sigma}(P_f)|x|^2 \leq V_f(x) \leq c_f =
\underline{\sigma}(P_f)/8\) implies \(|x| \leq 1/(2\sqrt{2}) < x^*\) and \(|u| =
|Kx| = 2(|x_1| + |x_2|) \leq 2\sqrt{2}|x| \leq 1\), so
\Cref{assum:stabilizability} is satisfied with \(\kappa_f(x)\defas -Kx =
-2x_1-2x_2\), and \(P_f\) and \(\mathbb{X}_f\) as defined.

\bibliographystyle{IEEEtranSN}
\bibliography{paper_twcccreport}

\end{document}